\documentclass[preprint,authoryear,12pt]{elsarticle}
\usepackage{amsmath,amssymb,epsfig,amsfonts,epsfig,graphicx}
\usepackage{epsfig}
\usepackage{epstopdf}

\usepackage{amssymb}
\usepackage{amsmath} \usepackage{stmaryrd}
\usepackage{epic} \usepackage{eepic}
\usepackage{latexsym}

\usepackage{fancyhdr}

\usepackage{mathtools}

\usepackage{url}
\usepackage{morefloats}

\renewcommand{\rho}{\varrho}
\renewcommand{\phi}{\varphi}

\newcommand{\beq}{\begin{equation}}
\newcommand{\eeq}{\end{equation}}
\newcommand{\bea}{\begin{eqnarray}}
\newcommand{\eea}{\end{eqnarray}}
\newcommand{\beas}{\begin{eqnarray*}}
\newcommand{\eeas}{\end{eqnarray*}}

\newcommand{\Images}{y}  %"y"=include images, "n"=don't include images

\newcommand{\mbs}[1]{\mathbf{#1}}

\def\bV{{\mbs{V}}}

  \def\fb{{\mbs{f}}}
\def\bg{{\mbs{g}}}  
  
 \def\bn{{\mbs{n}}}

\def\bv{{\mbs{v}}}

\usepackage{color}

\newcommand{\cFF}{{}}

\makeatletter
\def\ps@pprintTitle{%
\let\@oddhead\@empty
\let\@evenhead\@empty
\let\@oddfoot\@empty
\let\@evenfoot\@oddfoot
}
\makeatother

\begin{document}

\begin{frontmatter}

\title{On the mechanical modeling of the extreme softening/stiffening response of axially loaded tensegrity prisms}

\author{F.~Fraternali}
\ead{f.fraternali@unisa.it}
%\address{Department of Civil Engineering, University of Salerno,84084 Fisciano(SA), Italy}

\author{G.~Carpentieri}
\ead{gcarpentieri@unisa.it}
%\address{Department of Civil Engineering, University of Salerno,84084 Fisciano(SA), Italy}

\author{A.~Amendola}
\ead{adamendola@gmail.com}
\address{Department of Civil Engineering, University of Salerno,84084 Fisciano(SA), Italy}

\begin{abstract}
We study the geometrically nonlinear behavior of uniformly compressed tensegrity prisms, through fully elastic and rigid--elastic models. The presented models predict a variety of mechanical behaviors in the regime of large displacements, including an extreme stiffening-type  response, already known in the literature, and a newly discovered, extreme softening behavior. The latter may lead to a snap buckling event producing an axial collapse of the structure. The switching from one mechanical regime to another depends on the aspect ratio of the structure, the magnitude of the applied prestress, and the material properties of the constituent elements. 
We discuss potential acoustic applications of such behaviors, which are related to the design and manufacture of tensegrity lattices and innovative phononic crystals.
\end{abstract}


\end{frontmatter}

\section{Introduction}\label{into}

The category of `Extremal Materials' has been introduced in \cite{MC95} to define unconventional materials that alternately show very soft and very stiff deformation modes (unimode, bimode, trimode, quadramode and pentamode materials, depending on the number of soft modes). Such a definition applies to a variety of composite materials,  structural foams, pin-jointed trusses; cellular materials with re-entrant cells; {\cFF{rigid rotational elements}}: chiral lattices; etc., which feature special mechanical properties, such as, e.g.: auxetic deformation modes; negative compressibility; negative stiffness phases; high composite stiffness and damping,  to name just a few examples (cf.\cite{Lake87,Milton92, Milton2002,KBS12,Spadoni:2012, Nicolaou:2012, Milton:2013, Kochmann:2014}, and references therein). Extremal materials are well suited to manufacture composites with enhanced  toughness and shear strength (auxetic fiber reinforced composite); artificial blood vessels; energy absorption tools; and intelligent materials (cf. \cite{Liu:2006}). Rapid prototyping techniques for the manufacturing of materials with nearly pentamode behavior, and bistable elements with negative stiffness have been recently presented in \cite{KBS12} and  \cite{Kashdan:2012}, respectively.

From the acoustic point of view, extremal materials can be employed to manufacture nonlinear periodic lattices and phononic crystals, i.e., periodic arrays of particles/units, freestanding or embedded in in fluid or solid matrices with contrast in mass density and/or elastic moduli. Such 
artificial materials may feature a variety of unusual 
acoustic behaviors, which include: spectral band-gaps; sound attenuation;  negative effective mass density; negative elastic moduli; negative effective refraction index; energy trapping; sound focusing; wave steering and directional wave propagation (cf., e.g., \cite{Liu:2000, Li:2004,Ruzzene:2005,Daraio:2006,EZ06, FXXMWCX06, Gonella:2008,Lu:2009, Zhang:2009, Bigoni:2013, Casadei:2013}, and the references therein).  Particularly interesting is the use of geometrical nonlinearities for the in situ tuning of phononic crystals \citep{Bertoldi:2008,Wang:2013};
{\cFF{
pattern transformation by elastic instability \citep{Howon2}; as well as the optimal design of auxetic composites  \citep{KochVent2013}, and soft metamaterials incorporating fluids, gels and soft solid phases \citep{Brunet:2013}}}. It is worth noting that `extremal'  periodic lattices support solitary wave dynamics, which in particular feature atomic scale localization of traveling pulses in the presence of extremely stiff deformation modes (‘locking’ behavior, cf. \cite{Friesecke:2002,FSD12}), and rarefaction pulses in the presence of elastic softening (\cite{Nesterenko:2001, HN12, Herbold:2013}).

This paper presents a mechanical study of the compressive behavior of tensegrity prisms featuring large displacements, varying aspect ratios, prestress states, and material properties. We focus on the response of such structures under uniform axial loading, showing that they can feature extreme stiffening or, alternatively, extreme softening behavior, depending on suitable design variables. Interestingly, such a variegated mechanical response is a consequence of purely geometric nonlinearities.
By extending the tensegrity prism models already in the literature \citep{Oppenheim:2000, FSD12}, we assume that the bases and bars of the tensegrity prism may feature  either elastic or rigid behavior. The presented models lead us to recover the extreme stiffening-type response in the presence of rigid bases already studied in \cite{Oppenheim:2000, FSD12}. In addition, we discover a new, extreme softening-type response. The latter is associated with a snap buckling phenomenon eventually leading to the complete axial collapse of the structure. We validate our theoretical and numerical results through comparisons with an experimental study on the quasi-static compression of physical prism models \citep{prot}. 
The extreme hard/soft behaviors of tensegrity prisms can be usefully exploited to manufacture  periodic lattices and acoustic metamaterials supporting special types of solitary waves. Such waves may feature extreme compact support, in correspondence with a stiffening response  of the unit cells (`atomic scale localization,' cf. \cite{Friesecke:2002,FSD12}); or alternatively rarefaction pulses, when instead the unit cells exhibit  a softening-type behavior \citep{Nesterenko:2001, HN12,Herbold:2013}. 
Tensegrity  lattices can also be employed to manufacture highly anisotropic composite metamaterials, which include soft and hard units and are designed to show special wave-steering and stop-band properties \citep{Ruzzene:2005, Casadei:2013}.
The structure of this paper is as follows: 
in Section \ref{model}, we formulate a geometrically nonlinear model of a regular minimal tensegrity prism.
Next, we present a collection of numerical results referring to tensegirity prisms with different aspect ratios, prestress states, and material properties (Section \ref{results}). In Section \ref{experiment}, we validate such results against compression tests on physical tensegrity prism models. We end in Section \ref{conclusions} by drawing the main conclusions of the present study, and discussing future applications of tensegrity structures for the manufacture of innovative periodic lattices and acoustic metamaterials.

%-----------------------------------------------------------------------------------------
\section{Geometrically nonlinear model of an axially loaded tensegrity prism}\label{model}

\graphicspath{ {figures/} }
 
\begin{figure}[hbt] \begin{center}
\includegraphics[width=13cm]{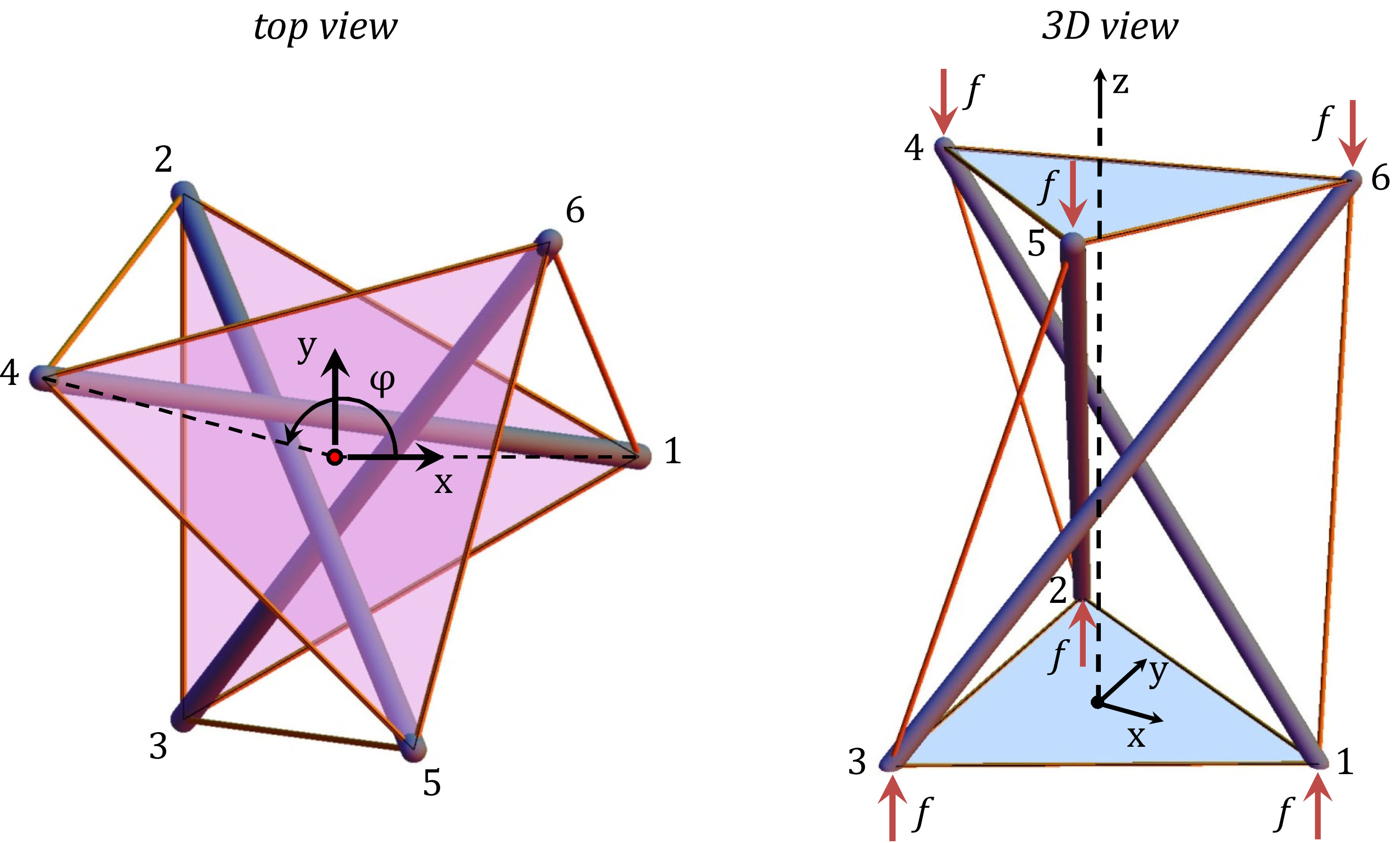}
\caption{Reference configuration of a minimal regular tensegrity prism.}
 \label{prism_notation}
\end{center}
\end{figure}

Let us consider an arbitrary configuration of a \textit{regular minimal tensegrity prism} 
\citep{Skelton2010}, which consists of two sets of \textit{horizontal strings}: $1-2-3$  (\textit{top strings}) and $4-5-6$ (\textit{bottom strings}); three \textit{cross strings}: 1-6, 2-4, and 3-5; and three \textit{bars}: 1-4, 2-5, and 3-6 (Fig. \ref{prism_notation}). 
The horizontal strings form two equilateral triangles with side length $\ell$, which are
rotated with respect to each other by an arbitrary \textit{angle of twist} $\varphi$. 
On introducing the Cartesian frame $\{O,x,y,z\}$ depicted in Fig. \ref{prism_notation}, which has the origin at the center of mass of the bottom base, we obtain the following expressions of the nodal coordinate vectors

\bea
\label{nodes}
\bn_1
 & = &  
\left[ \begin{array}{c}
\frac{\ell}{\sqrt{3}} \\
\\
 0 \\
\\
0 \end{array} \right],
\ \bn_2 =  
\left[ \begin{array}{c}
-\frac{\ell}{2 \sqrt{3}} \\
\\
\frac{\ell}{2} \\
\\
0 \end{array} \right],
\ \bn_3  = 
\left[ \begin{array}{c}
-\frac{\ell}{2 \sqrt{3}} \\
\\
-\frac{\ell}{2} \\
\\
0 \end{array} \right], 
\ \bn_4   =  
\left[ \begin{array}{c}
\frac{\ell  \cos (\phi )}{\sqrt{3}} \\
  \\
\frac{\ell  \sin (\phi )}{\sqrt{3}}\\
 \\
h
\end{array} \right] , 
\nonumber \\
\nonumber \\
 \bn_5  & = &  
\left[ \begin{array}{c}
-\frac{1}{2} \ell  \sin (\phi )-\frac{\ell  \cos (\phi )}{2 \sqrt{3}} \\
  \\
\frac{1}{2} \ell  \cos (\phi ) - \frac{\ell  \sin (\phi )}{2 \sqrt{3}} \\
 \\
h
\end{array} \right] , 
\label{n6}
\ \bn_6  =  
\left[ \begin{array}{c}
\frac{1}{2} \ell  \sin (\phi )-\frac{\ell  \cos (\phi )}{2 \sqrt{3}} \\
  \\
- \frac{\ell  \sin (\phi )}{2 \sqrt{3}}-\frac{1}{2} \ell  \cos (\phi )\\
 \\
h
\end{array} \right] \label{nccords}
\eea

\noindent with $h$ denoting the \textit{prism height}. 
The bars 1-4, 2-5, and 3-6 have the same length $b$, which is easily computed by

\bea
\label{Lh}
b & = & \sqrt{h^2-\frac{2}{3} \ell ^2 \cos (\phi )+\frac{2 \ell ^2}{3}}
\eea

\noindent while the cross strings 1-6, 2-4, and 3-5 have equal lengths $s$ given by
 
\bea
\label{sh}
s & = & \frac{\sqrt{3 h^2-\sqrt{3} \ell ^2 \sin (\phi )+\ell ^2 \cos (\phi )+2 \ell ^2}}{\sqrt{3}}
\eea

\begin{figure}[hbt] \begin{center}
\includegraphics[width=15cm]{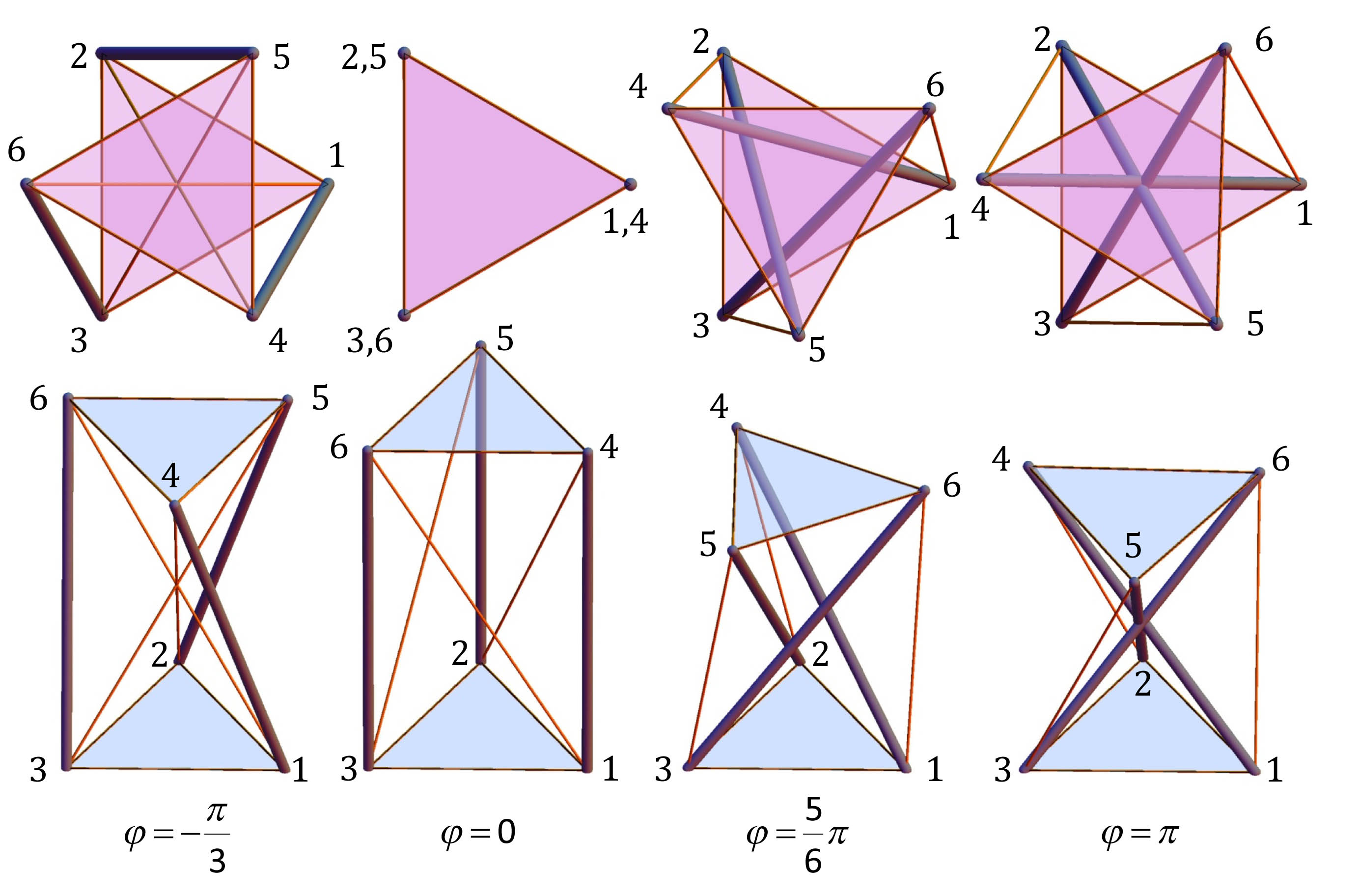}
\caption{Sequence of configurations corresponding to feasible values of the twisting angle $\varphi$.}
 \label{prism_configurations}
\end{center}
\end{figure}

We assume that the prism is loaded in the $z$ direction by three equal forces (each of magnitude $f=F/3$) in correspondence with the bottom base $1,2,3$, and three forces of equal magnitude but opposite direction in correspondence with the top base $4,5,6$ (Fig. \ref{prism_notation}). 
Under such a uniform axial loading, it is easy to recognize that the deformation of the prism maintains its top and bottom bases parallel to each other, and simultaneously changes the angle of twist $\varphi$ and the height $h$.
The \textit{geometrically feasible} configurations are obtained by letting $\varphi$ vary between $\varphi=-\pi/3$ (cross-strings touching each other), and $\varphi= \pi$ (bars touching each other), as shown in Fig.  \ref{prism_configurations}.
Hereafter, we refer to the configuration with the bars touching each other as the \textit{`locking'} configuration of the prism.
Let us consider the equilibrium equations associated with an arbitrary node of the prism, which set to zero the summation of all the forces acting on the given node in the current configuration. It is an easy task to show that such equations can be written as it follows

\bea
\frac{1}{6} \ell  \left(2 \sqrt{3} (x_{1}+3 x_{2}-x_{3})+\sqrt{3} (x_{1}+2 x_{3}) \cos (\phi
   )-3 x_{1} \sin (\phi )\right)
   & = &  0  \nonumber  \\
\frac{1}{6} \ell  \left(\sqrt{3} (x_{1}+2 x_{3}) \sin (\phi )+3 x_{1} \cos (\phi )\right)
   & = & 0 \label{equilibrium} \\
 h (x_{3}-x_{1})-\frac{F}{3}  & = &  0
  \nonumber 
\eea

\noindent where $x_{1}$, $x_2$ and $x_3$ are the forces per unit length (i.e, the \textit{force densities}) acting in  the cross-string,  base-strings, and bar attached to the current node, respectively.  
Such force densities are assumed positive if the strings are stretched, and the bars are compressed.
We say that the prism occupies a proper \textit{tensegrity placement} if one has: $x_1 \ge 0, x_2 \ge 0$ (i.e., the strings are either in tension or, at most, slack).
It is not difficult to verify that the system of equations (\ref{equilibrium}) admits the following general solution  

\bea
x_1 & = & 
-\frac{2 F \sin (\phi )}{3 \sqrt{3} h \left(\sqrt{3} \sin (\phi )+\cos (\phi )\right)}
  \nonumber \\
x_2 & = & 
-\frac{F \left(\sin ^2(\phi )-\sqrt{3} \sin (\phi )+\cos ^2(\phi )-\cos (\phi )\right)}{9 h \left(\sqrt{3} \sin
   (\phi )+\cos (\phi )\right)}\label{xsol} \\
x_3 & = & 
\frac{F}{3 h}-\frac{2 F \sin (\phi )}{3 \sqrt{3} h \left(\sqrt{3} \sin (\phi )+\cos (\phi )\right)}
  \nonumber 
\eea

Restricting our attention to the geometrically feasible configurations ($\varphi \in [-\pi/3, \pi ]$), we note that the solution (\ref{xsol}) becomes indeterminate when either  $\varphi=-\pi/6$, or $\varphi=5\pi/6$, that is, when the quantity $\sqrt{3} \sin (\phi )+\cos (\phi )$ is zero. This means that the configurations corresponding to such values of $\varphi$ may exhibit nontrivial states of self-stress, i.e., nonzero force densities in the prism members for $F=0$ (\textit{prestressable} configurations). 
By solving the first two equations (\ref{equilibrium}) for $x_2$ and $x_3$, we characterize the self-stress states of the prism by

\bea
\varphi=-\frac{\pi}{6}: & x_2 = -\frac{x_1}{\sqrt{3}}, & x_3 = x_1
 \label{pi6} 
 \eea

\bea
\varphi= \frac{5}{6} \pi: & x_2 = \frac{x_1}{\sqrt{3}}, & x_3 = x_1
 \label{56pi}
 \eea
 
\noindent for arbitrary $x_1$. 
Eqs. (\ref{pi6}) and (\ref{56pi}) show that a nontrivial state of self-stress compatible with an effective {tensegrity placement} is possible only for $\varphi=5\pi/6$. As a matter of fact, Eq.  (\ref{pi6}) highlights that $x_1$ and $x_2$ have opposite signs for $\varphi=-\pi/6$, which implies that the prism is either unstressed  ($x_1=x_2=x_3=0$), or has some strings stretched and the others compressed in such a configuration.
In  contrast, Eq.  (\ref{56pi}) reveals that $x_1$ and $x_2$ have equal signs for $\varphi=5\pi/6$. 
{\cFF{The prism is loaded in compression for $\theta >0 $, and in tension for $\theta <0$, where $\theta = \varphi - 5 \pi/6$ (cf. Section \ref{results}, and \cite{Oppenheim:2000, FSD12})}}.
By manipulating Eqs. (\ref{nccords}) and (\ref{xsol}), we detect that all the cross strings are vertical and carry force densities $x_1 = f/h$, for $\varphi = 2/3 \pi$ ($\theta= -\pi/6$). In the same configuration, the base strings and the bars carry zero forces  ($x_2=x_3=0$).
We take as a \textit{reference} the configuration of the prism such that $\varphi=\varphi_0=5\pi/6$,
and let $s_0$, $\ell_0$ and $b_0$ denote the lengths of the cross-strings, base-strings and bars in such a configuration, respectively.
By inserting $\ell_0$ and $s_0$ into Eqs. (\ref{Lh}) and (\ref{sh}), we can easily compute the reference values of the prism height and bar length as follows

\bea
h_0 & = & \sqrt{s _0^2+\frac{1}{3} \left(\sqrt{3}-2\right) \ell _0^2}, \ \ \ \ \ \
b_0 \ = \ \sqrt{s_0^2+\frac{2 \ell _0^2}{\sqrt{3}}}
\label{h0s0}
\eea

\subsection{Fully elastic model} \label{fullelresp}

A \textit{fully elastic model} is obtained by describing all the prism members (bars and strings) as linear springs characterized by the following constitutive laws \citep{Skelton2010} 

\bea
x_1  & = & \frac{1}{s} \ k_1 \ (s - s_N), \ \ \
x_2  \ = \  \frac{1}{\ell} \ k_2 \ (\ell - \ell_N), \ \ \
x_3  \ = \ - \  \frac{1}{b} \ k_3 \ (b- b_N)
\label{xel}
 \eea
 
 \noindent where $k_1$, $k_2$ and $k_3$ are spring constants, and $s_N$, $\ell_N$ and $b_N$ are the \textit{rest lengths} (or \textit{natural lengths}) of cross-strings, base-strings and bars, respectively.
Upon neglecting the change of the cross-section areas of all members during the prism deformation, we compute the spring constants as follows \citep{Skelton2010} 

\bea
k_1 & = & \frac{E_1 A_1}{s_N}, \ \ \
k_2 \ = \ \frac{E_2 A_2}{\ell_N}, \ \  \
k_3 \ = \ \frac{E_3 A_3}{b_N}
\label{k123}
\eea
 
\noindent where $E_1$, $E_2$, $E_3$, and $A_1$, $A_2$, $A_3$ are the elastic moduli and the cross-section areas of the cross-strings, base-strings and bars, respectively.

\subsubsection{Reference configuration} \label{reference}

Hereafter, we assume that  $\ell_N$ and $s_N$ are given, and that the cross-string \textit{prestrain} is prescribed, i.e., the quantity

\bea
p_0 & = {(s_0 - s_N)}/{s_N}
\label{prestrain}
\eea

\noindent  In line with the above assumptions, we compute the reference length of the cross-strings ($s_0$), and the reference value of the force density in such members ($x_1^{(0)}$) through

\bea
s_0 & =  & {s_N}({1 + p_0})
\label{s0} \\
& & \nonumber \\
x_1^{(0)}  & = & \frac{1}{s_0} \ k_1 \ (s_0 - s_N) 
\ = \ \frac{A_1 E_1}{s_N} \ \frac{p_0}{1+p_0},
\label{x10}
\eea

Using (\ref{56pi}), (\ref{xel}) and (\ref{x10}), we are led to the following reference values of the force densities in the base strings ($x_2^{(0)}$)  and bars ($x_3^{(0)}$)

\bea
x_2^{(0)}  & = &  \frac{1}{\ell_0} \ k_2 \ (\ell_0 - \ell_N)
\ = \ \frac{A_1 E_1}{\sqrt{3} s_N} \ \frac{p_0}{1+p_0}
\label{x20} \\
& & \nonumber \\
x_3^{(0)}  & = & - \  \frac{1}{b_0} \ k_3 \ (b_0- b_N) 
\ = \ \frac{A_1 E_1}{s_N} \ \frac{p_0}{1+p_0}
\label{x30}
\eea

Eq. (\ref{x20}) can be solved for $\ell_0$, yielding

\bea
\ell_0 & = & 
\frac{3 A_2 E_2 \left(p_0+1\right) s_N \ell _N}{p_0 \left(3 A_2 E_2 s
   _N-\sqrt{3} A_1 E_1 \ell _N\right)+3 A_2 E_2 s_N}
\label{el0}
\eea

On the other hand, the substitution of (\ref{s0}) and (\ref{el0}) into (\ref{h0s0})$_2$ gives

\bea
b_0 & = & 
\eta \ s_0 \ = \
\eta \  \left(1 + p_0\right) \ s_N
\label{b0}
\eea

\noindent where

\bea
\eta & = & 
\sqrt{
\frac{6 \sqrt{3} A_2^2 E_2^2 \ell _N^2}{\left(p_0 \left(3 A_2 E_2 s_N-\sqrt{3}
   A_1 E_1 \ell _N\right)+3 A_2 E_2 s_N\right){}^2}+1
}   
\label{eta}
\eea

\noindent By solving Eq. (\ref{x30}) for $b_N$ and employing (\ref{b0}), we finally obtain

\bea
b_N & = & 
\frac{\eta A_3 E_3}{A_3 E_3 - \eta A_1 E_1} \  \left(1 + p_0\right) \ s_N
\label{bN}
\eea

\subsubsection{The elastic problem} \label{elproblem}

The substitution of Eqns. (\ref{xel}) into (\ref{equilibrium}) leads us to the following elastic problem

\bea
g_1 & = & 
\frac{1}{6} \ell  \left(k_3 \ 4 \sin ^2\left(\frac{\phi }{2}\right) \left(\sqrt{3}-\frac{3 b_N}{\sqrt{3 h^2-2 \ell ^2
   \cos (\phi )+2 \ell ^2}}\right) \right. \nonumber \\
& & \ \ \  \ \ \
\left .+  \  k_1 \ \left(-3 \sin (\phi )+\sqrt{3} \cos (\phi )+2 \sqrt{3}\right) 
\ + \ k_2 \ \frac{6 \sqrt{3} \left(\ell -\ell _N\right)}{\ell }
\right.
 \nonumber \\
 & & \ \ \  \ \ \
  \left.   - \ k_1 \ \frac{3 s_N \left(-\sqrt{3} \sin
   (\phi )+\cos (\phi )+2\right)}
   {\sqrt{3 h^2-\sqrt{3} \ell ^2 \sin (\phi )+\ell ^2 \cos (\phi )+2 \ell
   ^2}} \right)
   \ = \ 0
\label{elsys1}
\eea

\bea
g_2 & = & 
\frac{1}{6} \ell  \left(k_3 \ 2 \sin (\phi ) \left(\frac{3 b_N}{\sqrt{3 h^2-2 \ell ^2 \cos (\phi )+2 \ell
   ^2}}-\sqrt{3} \right) \right. \nonumber \\
& & \ \ \  \ \ \
\left .+  \  k_1 \ \left(\sqrt{3} \sin (\phi )+3 \cos (\phi )\right)
\right.
 \nonumber \\
 & & \ \ \  \ \ \
  \left.   - \ k_1 \ \frac{3 s_N \left(\sin (\phi )+\sqrt{3} \cos (\phi) \right)}
   {\sqrt{3 h^2-\sqrt{3} \ell ^2 \sin (\phi )+\ell ^2 \cos (\phi )+2 \ell
   ^2}} \right)
   \ = \ 0
\label{elsys2}
\eea

\bea
g_3 & = &
-f \ + \
   k_3 \ h \left(\frac{b_N}{\sqrt{h^2-\frac{2}{3} \ell ^2 \cos (\phi )+\frac{2 \ell ^2}{3}}}-1\right)
 \nonumber \\
 & & \ \ \  
   + \ k_1 \ h 
   \left(\frac{\sqrt{3} s_N}{\sqrt{3 h^2-\sqrt{3} \ell ^2 \sin (\phi )+\ell ^2 \cos (\phi )+2 \ell
   ^2}}-1\right)
   \ = \ 0
\label{elsys3}
\eea

%\noindent which can be solved for $h$, $\ell$, $\varphi$ and $f$, in correspondence with a given initial point, when any of these variables is given (\textit{independent coordinate}). 

{\cFF{
\subsubsection{Path-following method} \label{rateproblem}

We formulate a path-following approach to the nonlinear problem (\ref{elsys1})--(\ref{elsys3}), by introducing the following  `extended system' \citep{Riks84,Wrisimo90,frabuck13}
\bea
\label{g}
\tilde{\bg}
 & = &  
\left[ \begin{array}{c}
\bg(\bv,f) \\
\\
\psi(\bv,f) \end{array} \right] 
 = 
0
\eea

\noindent where we set $\bv =  [\ell, \varphi, h ]^T$, $\bg =  [g_1, g_2, g_3 ]^T$, and let $\psi({\bv},f) = 0$ denote a constraint equation characterizing the given loading condition. In the case of a displacement control loading, we in particular assume

\bea
\psi & = & v_k - c  =  0,
\label{psi1}
\eea

\noindent letting $v_k$ coincide with $v_1 \equiv \ell$ (\textit{base edge control}); $v_2 \equiv \varphi$ (\textit{twist control}) or $v_3 \equiv h$ (\textit{height control}), and letting $c$ denote a given constant.
The Newton--Raphson linearization of (\ref{g}) at a given starting point ($\bar \bv$, $\bar f$) leads us to the incremental problem

\begin{equation}
\begin{bmatrix}
\nabla_{\bv} \bg & \nabla_{f} \bg \\
\\
\nabla_{\bv} \psi^T & \nabla_{f} \psi \\
\end{bmatrix}
\begin{bmatrix}
\Delta \bv \\
\\
\Delta f \\
\end{bmatrix}
= -
\begin{bmatrix}
{\bar \bg} \\
\\
\bar \psi 
\end{bmatrix}
\label{incremental}
\end{equation}

\noindent where we set ${\bar \bg}  = \bg(\bar \bv, \bar f)$; ${\bar \psi}  = \psi(\bar \bv, \bar f)$; and 

\beq
\nabla_{\bv} \bg =
\begin{bmatrix}
\frac{\partial g_1}{\partial v_1} & \frac{\partial g_1}{\partial v_2}& \frac{\partial g_1}{\partial v_3} \\
\\
\frac{\partial g_2}{\partial v_1} & \frac{\partial g_2}{\partial v_2}& \frac{\partial g_2}{\partial v_3} \\
\\
\frac{\partial g_3}{\partial v_1} & \frac{\partial g_3}{\partial v_2}& \frac{\partial g_3}{\partial v_3} 
\end{bmatrix},
\ \ 
\nabla_{f} \bg =\left[ \begin{array}{c}
\frac{\partial g_1}{\partial f} \\
\\
\frac{\partial g_2}{\partial f} \\
\\
\frac{\partial g_3}{\partial f} \end{array} \right],
\ \
\nabla_{\bv} \psi =\left[ \begin{array}{c}
\frac{\partial \Psi}{\partial \bv_1} \\
\\
\frac{\partial \Psi}{\partial \bv_1} \\
\\
\frac{\partial \Psi}{\partial \bv_1} \end{array} \right]
\eeq

We now introduce the notations $\bV:=\nabla_{\bv} \bg$ and $\fb := \nabla_{f} \bg = [0,0,-1]^T$, and assume that $\bV$ is invertible at ($\bv = \bar \bv$, $f = \bar f$). The incremental problem (\ref{incremental}) is solved by first computing the partial solutions

\beq
\Delta \bv_f \ = \ - \bV^{-1} \fb \ = \ 
\begin{bmatrix}
V^{-1}_{13} \\
V^{-1}_{23} \\
V^{-1}_{33}
\end{bmatrix},
\ \ \ \ \
\Delta \bv_g \ = \ -\bV^{-1} \bar{\bg},
\label{a}
\eeq

\noindent and next the updates

\bea
\Delta \bv & = &  \Delta f \Delta \bv_f + \Delta \bv_g, \label{b} \\
\Delta f & = & - \ \frac {\bar{\psi} + \nabla_{\bv}\psi \cdot \Delta \bv_g} {\nabla_{f}\psi + \nabla_{\bv}\psi \cdot \Delta \bv_f}
\label{c}
\eea

Equations (\ref{b})--(\ref{c}) lead us to the new predictor  ($\bar{\bv}$ + $\Delta \bv$, $\bar f$ + $\Delta f$), which is used to reiterate the updates (\ref{b})--(\ref{c}), until the residual $\left\| \bg(\bar{\bv},\bar{f})\right\|$ gets lower than a given tolerance.
Once a new equilibrium point is obtained, the value of constant $c$ in Eqn. (\ref{psi1}) is updated and the path-following procedure is continued.
The explicit expression for the $\bV$ matrix is given in Appendix.

Let us assume $\psi  =  h - {\bar h}$ (height control loading). By writing Eqn. (\ref{b}) in correspondence with a solution of the extended system (\ref{g}) (${\bar \bg}  = \mathbf{0}$, ${\bar \psi}  = 0$), we easily obtain $\Delta \bv_g = 0$, and 

\bea
\begin{bmatrix}
\Delta \ell \\
\Delta\varphi \\
\Delta h
\end{bmatrix} & = &   \Delta f 
\begin{bmatrix}
V^{-1}_{13} \\
V^{-1}_{23} \\
V^{-1}_{33}
\end{bmatrix}
\label{d} 
\eea

\noindent which implies

\bea
\Delta h & = & {\Delta f} \ {V^{-1}_{33}}
\ = \ \frac{\Delta F}{3} \ {V^{-1}_{33}}
\label{e}
\eea

Eqn. (\ref{e}) shows that the \textit{axial stiffness} $K_h^{el}$ of the fully elastic model is given by 

\bea
K_h^{el} =   - \frac {3}{V^{-1}_{33}}
\label{Khel}
\eea

\noindent The value of the above quantity at $\bv = \bv_0 = [\ell_0, \varphi_0, h_0]^T$ represents the axial stiffness $K_0^{el}$ of the prism in correspondence with the reference configuration, and it is not difficult to show that such a quantity is zero for $p_0=0$ (see the Appendix).
}}

\subsection{Rigid-elastic model} \label{rigelresp}

In a series of studies available in the literature, the mechanical response of tensegrity prisms has been analyzed by assuming that the bases and bars behave rigidly, while the cross strings respond as elastic springs (\textit{rigid-elastic model}, cf., e.g., \cite{Oppenheim:2000, FSD12}).
Such a modeling keeps $b$ and $\ell$ fixed ($b=b_0=\mbox{const}$, $\ell=\ell_0=\mbox{const}$), and relates $h$ to $\varphi$ through
Eq. (\ref{Lh}). Let us solve Eq. (\ref{Lh}) for $h$, obtaining the equation

\bea
\label{hL}
h & = & \sqrt{b^2 \ - \ \frac{2}{3}\ell^2(1-\cos \phi)} 
\eea

\noindent which, once inverted (for $-\pi/3 \le \varphi \le \pi$), gives

\bea
\label{phih}
\varphi & = & \arccos{\left( 1-\frac{b^2  - h^2}{2 a^2} \right)}
\eea

\noindent where $a = \ell / \sqrt{3}$ denoted the radius of the circumference circumscribed to the base triangles. The response of the rigid--elastic model is easily modeled  by substituting 
(\ref{xel})$_1$ into the equilibrium equations (\ref{equilibrium}), and solving the resulting system of algebraic equations with respect to $F$, $x_2$, and $x_3$, for given $h$ (or $\varphi$). It is not difficult to verify that such an approach leads to the same constitutive law given in \cite{Oppenheim:2000, FSD12}, that is  

\bea
{F} & = &  3 k_1 \ (s - s_N) \ 
\frac{h}{2 s} \ \left(3+ \frac{\sqrt{3} \left(2 a^2+h^2-b^2\right)}{a^2 \sqrt{-\frac{\left(h^2-b^2\right) \left(4
   a^2+h^2-b^2\right)}{a^4}}}\right) \nonumber \\
& & \nonumber \\
& = & 
\frac{k_1 \csc (\phi ) \left(3 \sin (\phi )+\sqrt{3} \cos (\phi )\right) \sqrt{3 b^2+2 \ell ^2 \cos (\phi )-2 \ell
   ^2}}{2 \sqrt{3
   b^2-\sqrt{3} \ell ^2 \sin (\phi )+3 \ell ^2 \cos (\phi )}} \nonumber \\
& & \nonumber \\
& & \times \
\left(\sqrt{9 b^2-3 \sqrt{3} \ell ^2 \sin (\phi )+9 \ell ^2 \cos (\phi )}-3 s_N\right) 
\label{Frigel}
\eea

It is also easily shown that the rigid--elastic model predicts an infinitely stiff 
response ($F \rightarrow \infty$) for $\varphi \rightarrow \pi$  (assuming $b > 2 a$).
Once $h$ (or $\varphi$) is given, $x_1$ is computed through (\ref{xel})$_1$ and (\ref{sh}); ${F}$ is computed through (\ref{Frigel}); and $x_2$ and $x_3$ are obtained from the equilibrium equations (\ref{equilibrium}).
The differentiation of (\ref{Frigel}) with respect to $h$ gives the tangent axial stiffness of the present model (cf. the Appendix). The reference value of such a quantity ($\varphi=5/6 \pi$) is given by

\bea
K_{h_0}^{rigel} & = &  - {F}^{'}(h=h_0) 
\ = \ 12 \ \sqrt{3} \ k_1 \frac{p_0}{1+p_0} \ \left(\frac{h_0}{a}\right)^2
\label{K0rigel}
\eea

\noindent and it is immediately seen that also $K_0^{rigel}$ is zero for $p_0=0$, as well as $K_0^{el}$.

\section{Numerical results}
\label{results}

\begin{figure}[hbt] \begin{center}
\includegraphics[width=17cm]{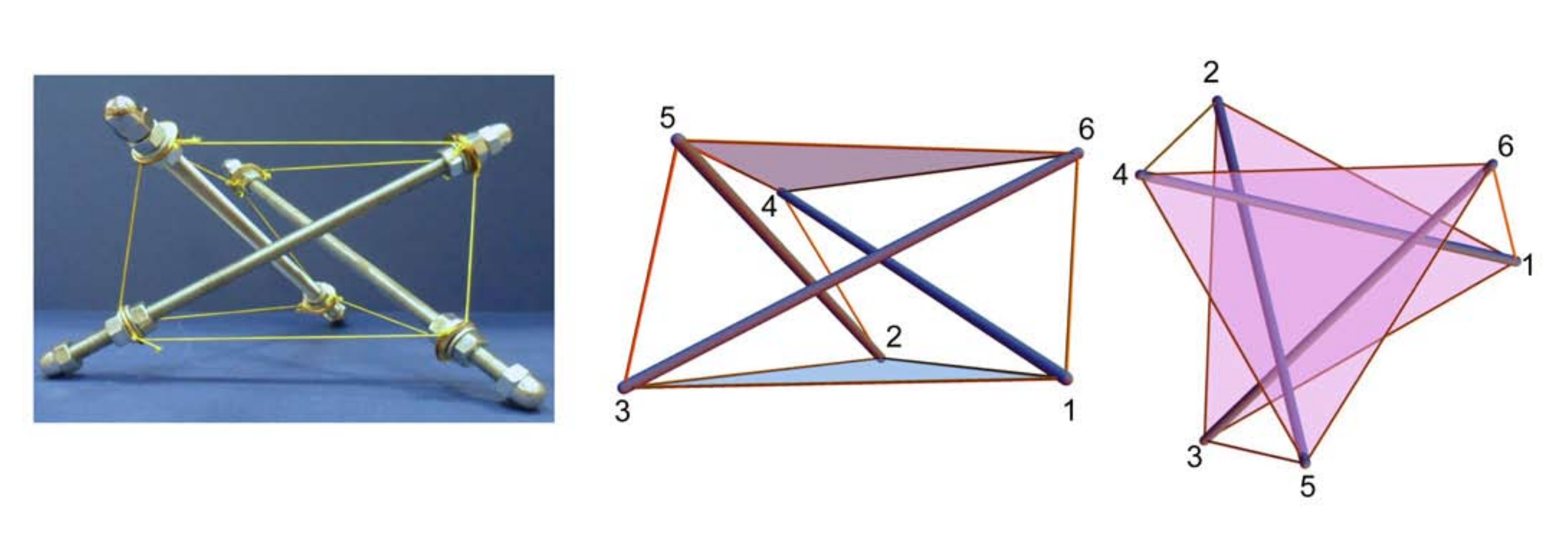}
\caption{Thick prism model. Left: photograph of a real-scale example \citep{prot}. Center and right: 3D view (center) and top view (right) of the theoretical model.}
 \label{thick_prism}
\end{center}
\end{figure}

\begin{figure}[hbt] \begin{center}
\includegraphics[height=8cm]{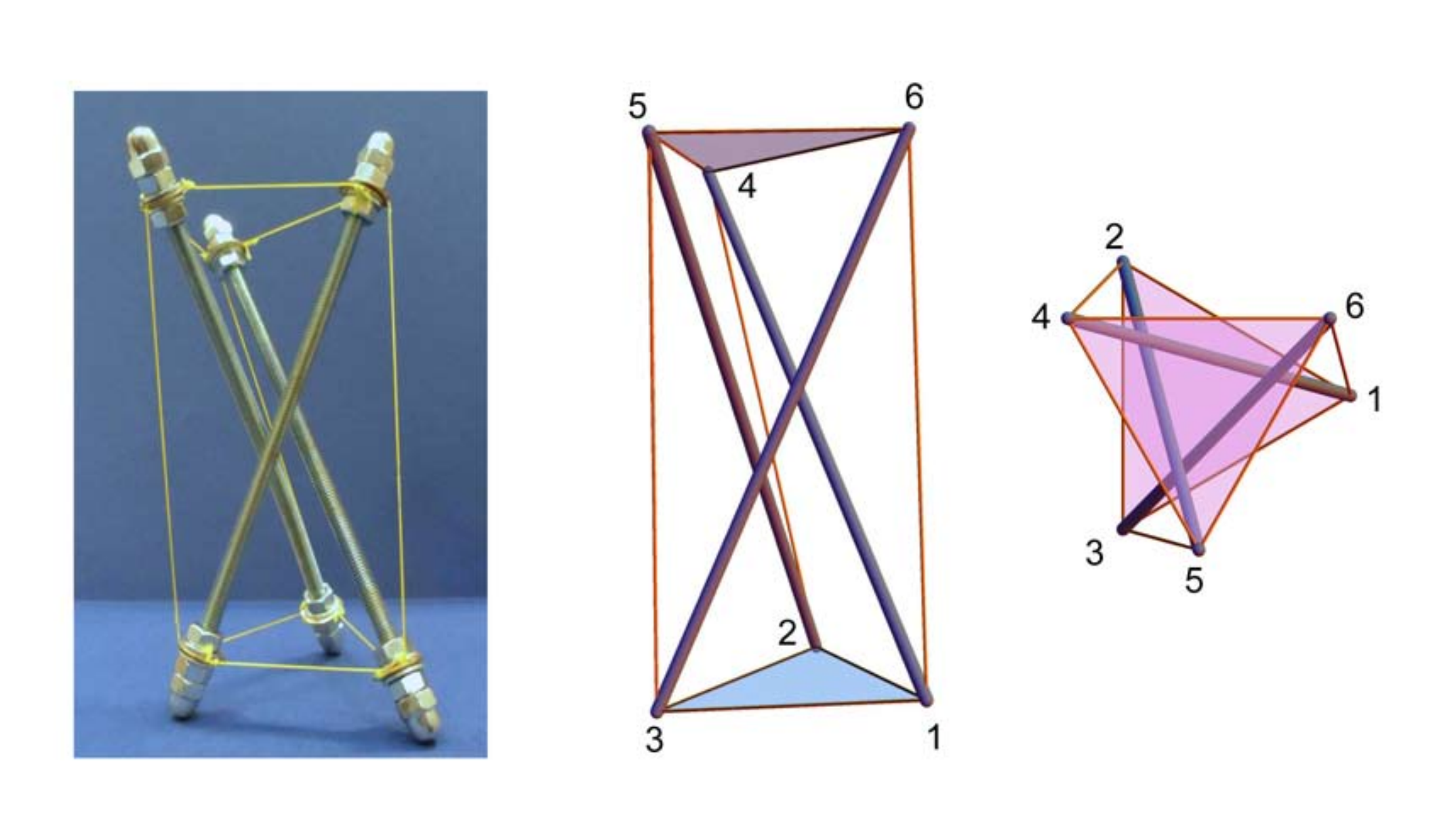}
\caption{Slender prism model. Left: photograph of a real-scale example \citep{prot}. Center and right: 3D view (center) and top view (right) of the theoretical model.}
 \label{slender_prism}
\end{center}
\end{figure}

The current section presents a collection of numerical results aimed to illustrate the main features of the mechanical models presented in Section \ref{model}. We examine the mechanical response of tensegrity prisms having the same features as the physical models studied in \cite{prot}.
Such prisms are equipped with M8 threaded bars made out of
white zinc plated grade 8.8 steel (DIN 976-1),
and strings consisting of
PowerPro\textsuperscript{\textregistered} braided Spectra\textsuperscript{\textregistered} fibers with 0.76 mm diameter (commercialized by Shimano American Corporation - Irvine CA). The properties of the employed materials are shown in Table \ref{tabmat}.
Let ${\bar A}_1$, ${\bar A}_2$, ${\bar A}_3$ and ${\bar E}_1$, ${\bar E}_2$, ${\bar E}_3$ denote the cross-sectional areas and elastic moduli of the strings and bars defined according to Table \ref{tabmat}. In  order to study the transition from the elastic to the rigid--elastic model, we hereafter study the mechanical response of elastic prisms  endowed with the following spring constants (cf. Section \ref{fullelresp}).

\bea
k_1 & = & \frac{{\bar E}_1 {\bar A}_1}{s_N}, \ \ \
k_2 \ = \ \alpha \ \frac{{\bar E}_2 {\bar A}_2}{\ell_N}, \ \  \
k_3 \ = \ \beta \ \frac{{\bar E}_3 {\bar A}_3}{b_N}
\label{k123num}
\eea

 \noindent where $\alpha$ and $\beta$ are rigidity multipliers ranging within the interval $[1, \infty]$. The case of $\alpha=\beta =1$ corresponds to the fully elastic (`el')  model of Sect. \ref{elproblem}, while the limiting case with $\alpha = \beta \rightarrow \infty$ corresponds to the rigid--elastic (`rigel') model presented in Sect. \ref{rigelresp}.
 The equilibrium configurations of the elastic prism model are numerically determined through the path-following method given in Section \ref{rateproblem}, letting the angle of twist $\varphi$ to vary within the interval 
$[2/3 \pi,  \pi )$, which corresponds to effective tensegrity placements of the structure (cf. Section \ref{model}). 
%To avoid numerical singularities in the `rigel' limit, we interrupt our calculations at $\varphi = 0.99 \pi$. 
We examine a large variety of prestrains $p_0$, and both \textit{thick}  and \textit{slender} reference configurations (cf. Figs. \ref{thick_prism} and  \ref{slender_prism}, respectively).
{\cFF{Let $\delta=h_0 - h$ denote the axial displacement of the prism from the reference configuration, and let $\varepsilon = \delta / h_0$ denote the corresponding axial strain (positive when the prism is compressed). We name \textit{stiffening} a branch of the $F-\delta$ response showing axial stiffness $K_h$ increasing with $|\delta|$ (or $|\varepsilon|$), and \textit{softening} a branch that instead shows $K_h$ decreasing with $|\delta|$  ($|\varepsilon|$). 
The axial forces carried by the cross-strings, base-strings, and bars are denoted by $N_1$, $N_2$, and $N_3$, respectively. We assume  that $N_1$ and $N_2$ are positive in tension, and that $N_3$ is instead positive in compression.
}}

\begin{table}[htbp]
\begin{center}
\medskip
\scalebox{1}{
\begin{tabular}{|c|c|c|} \hline
\text{Property} & bars & strings \\ \hline \hline
\text{area} (mm$^2$) & 36.6 & 0.45 \\ \hline
\text{mass density (kg/m$^3$}) & 7850 & 793 \\ \hline
\text{elastic modulus (GPa)}& 203.53 & 5.48 \\ \hline
%\text{yielding stress (MPa)} & 640 & $\approx$ 2000\\ \hline
\end{tabular}}
\caption{Properties of the materials employed in the numerical simulations.}
\label{tabmat}
\end{center}
\end{table}

\subsection{Thick prisms} \label{thick prism}

\begin{figure}[hbt]
\unitlength1cm
\begin{picture}(11,13)
\if\Images y\put(2,6.5){\psfig{figure=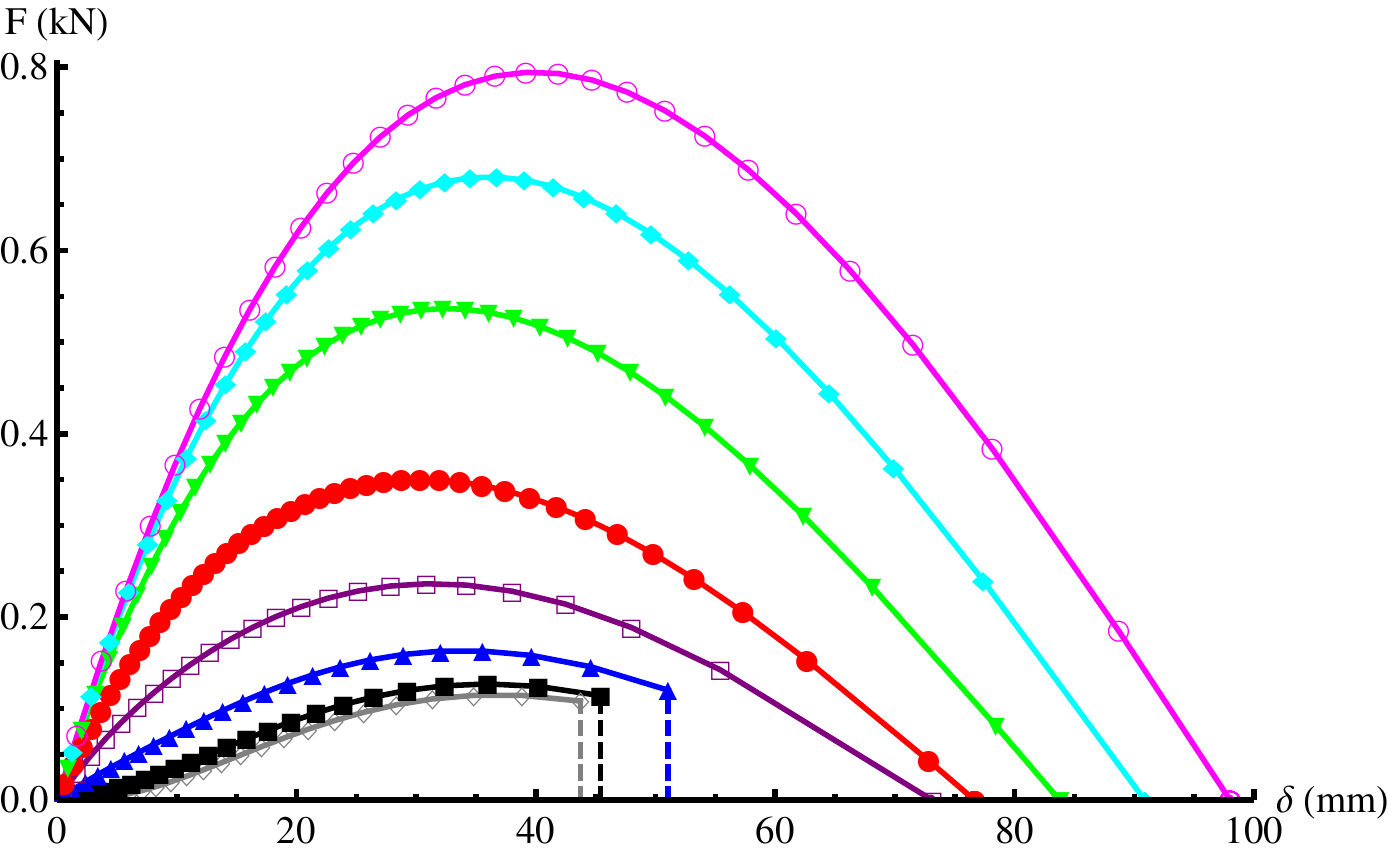,width=10cm}}\fi
\if\Images y\put(2,0.5){\psfig{figure=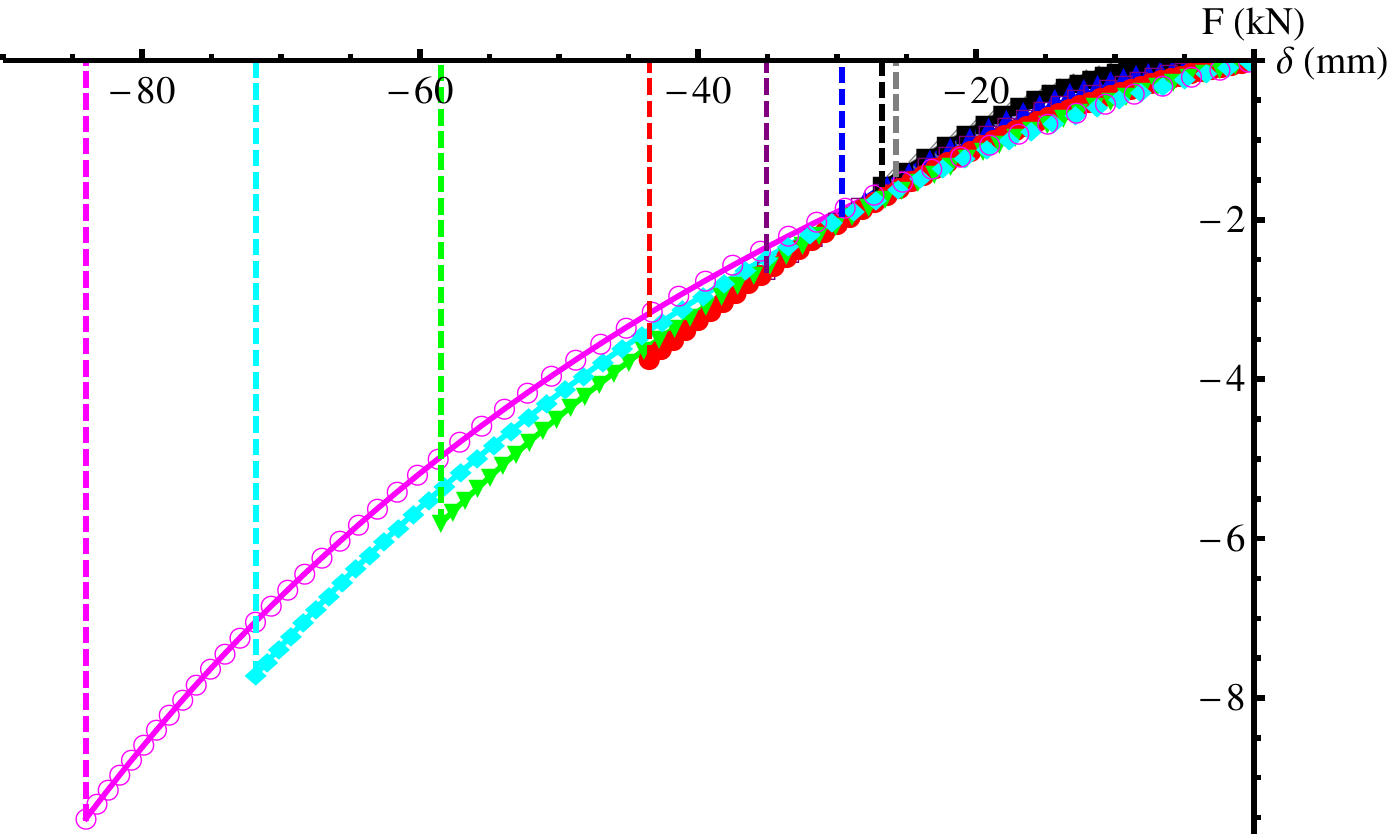,width=10cm}}\fi
\if\Images y\put(1.5,0){\psfig{figure=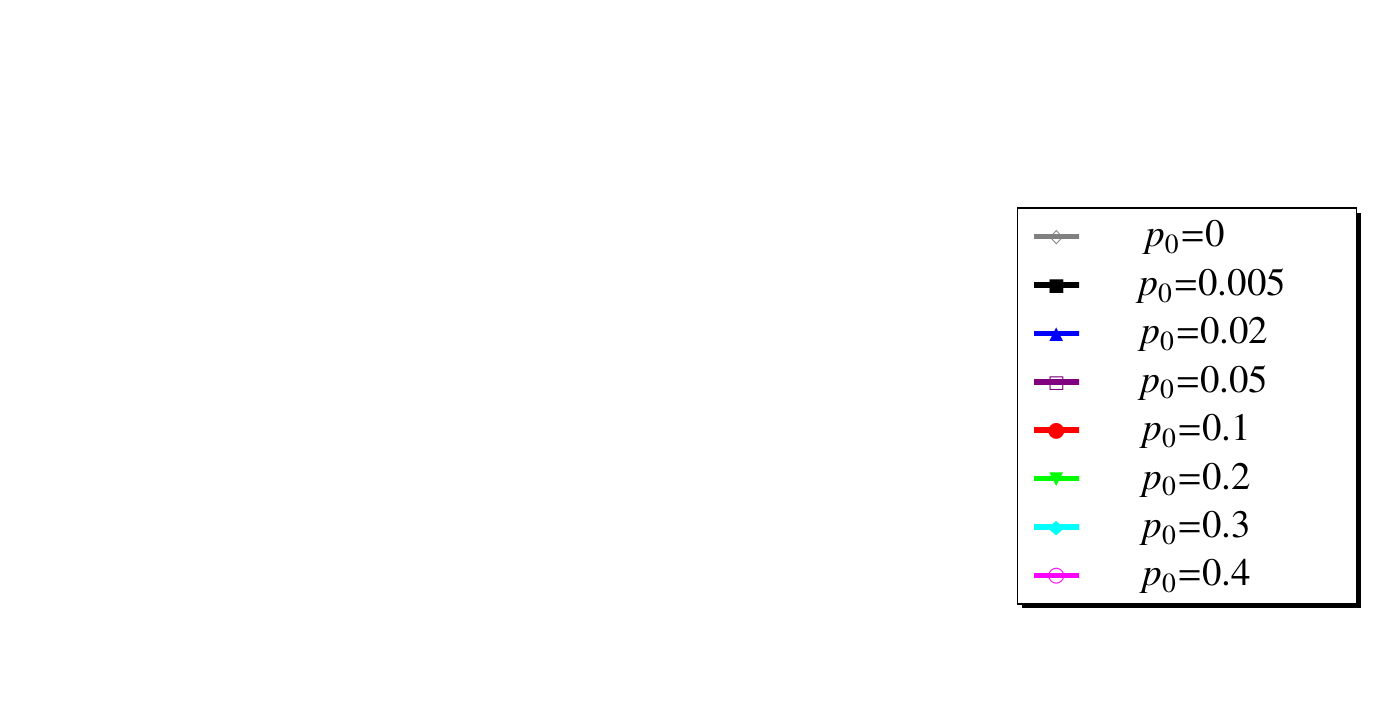,width=9cm}}\fi
\end{picture}
\caption{$F$--$\delta$ curves of the thick prism model, when loaded in compression (top), and tension (bottom), for $\alpha=\beta=1$ and different values of $p_0$.
}
\label{p-variab-system1}
\end{figure}

\begin{figure}[hbt]
\unitlength1cm
\begin{picture}(11,13)
\if\Images y\put(2,6.5){\psfig{figure=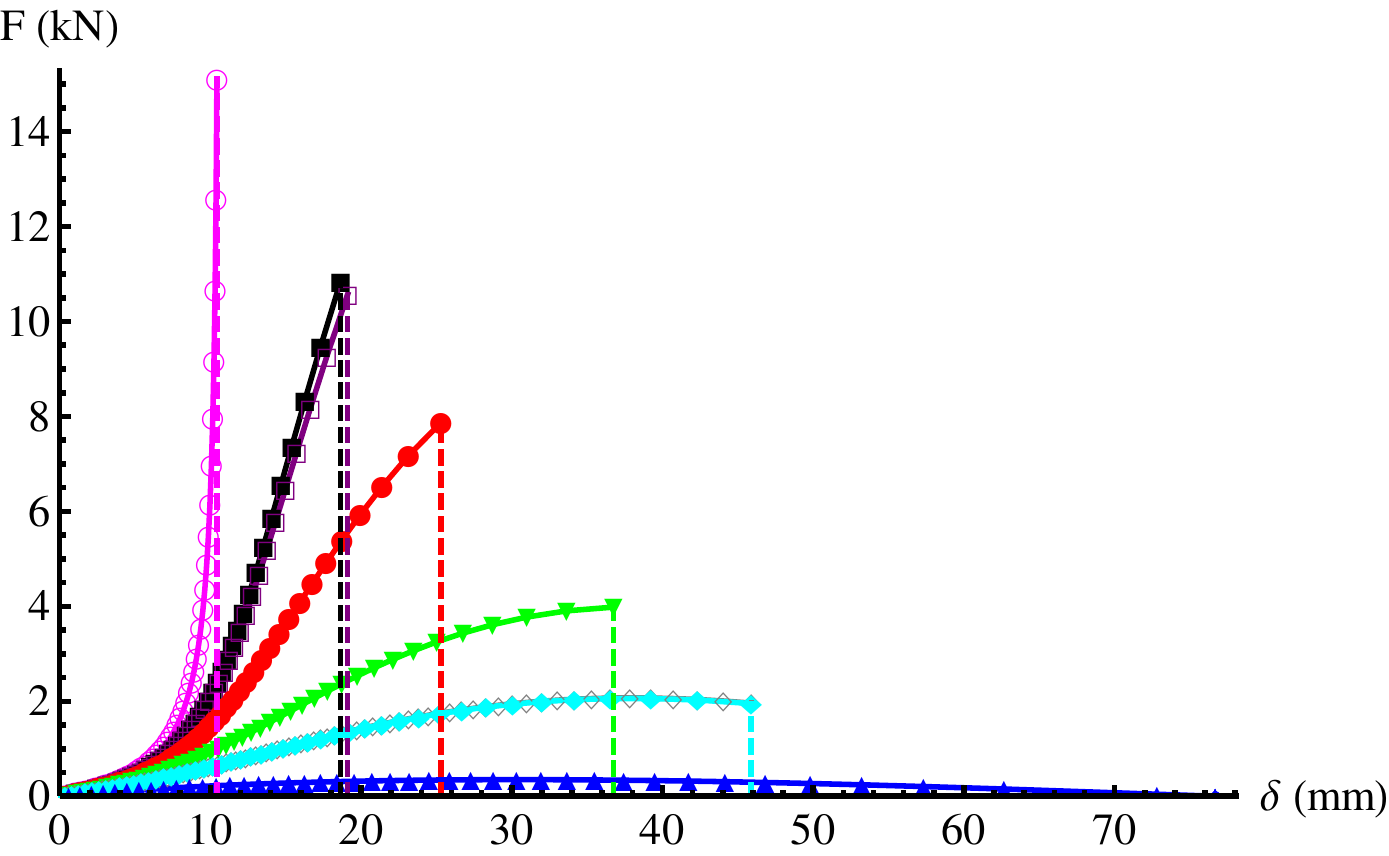,width=10cm}}\fi
\if\Images y\put(2,0.25){\psfig{figure=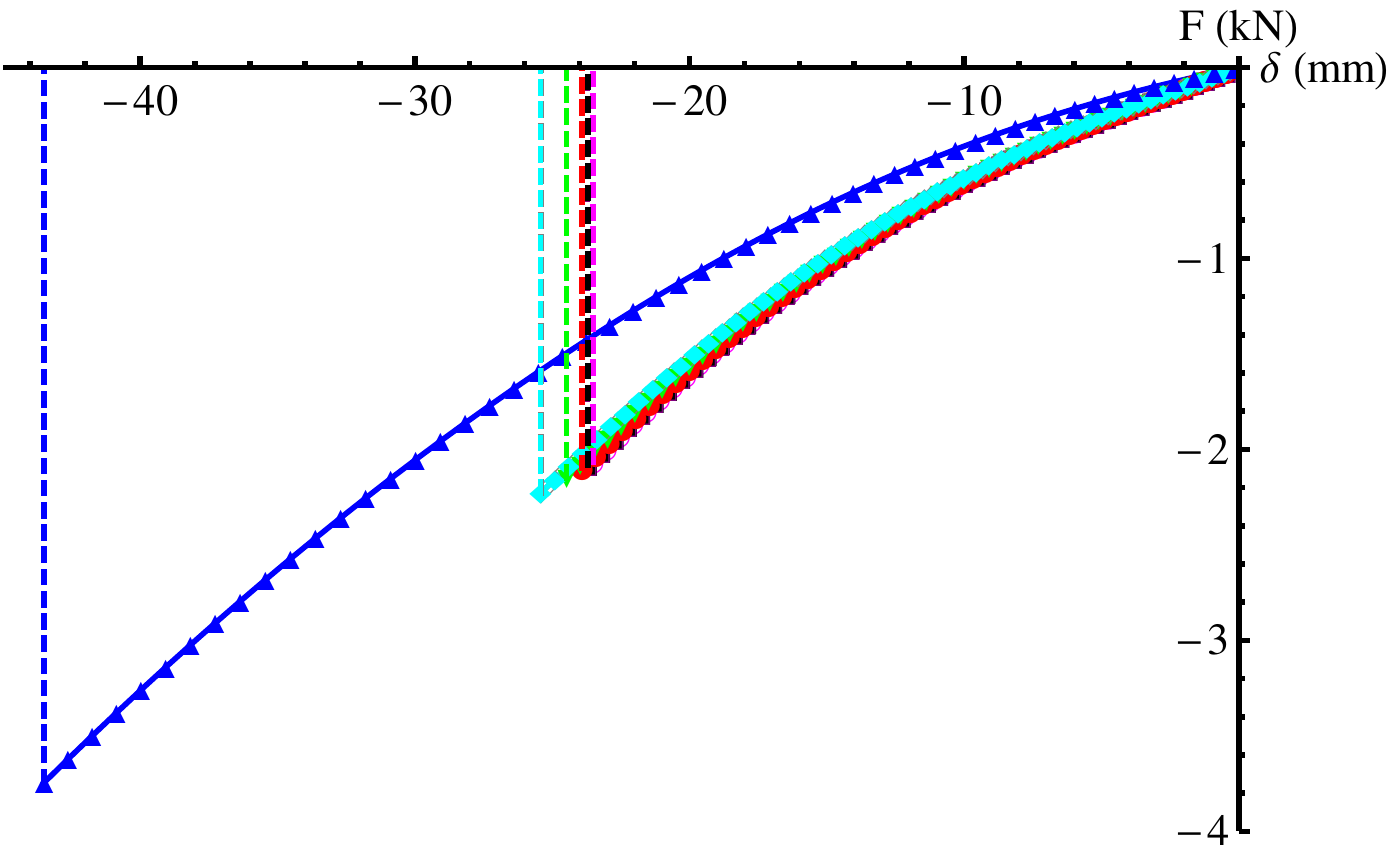,width=10cm}}\fi
\if\Images y\put(0.5,8.0){\psfig{figure=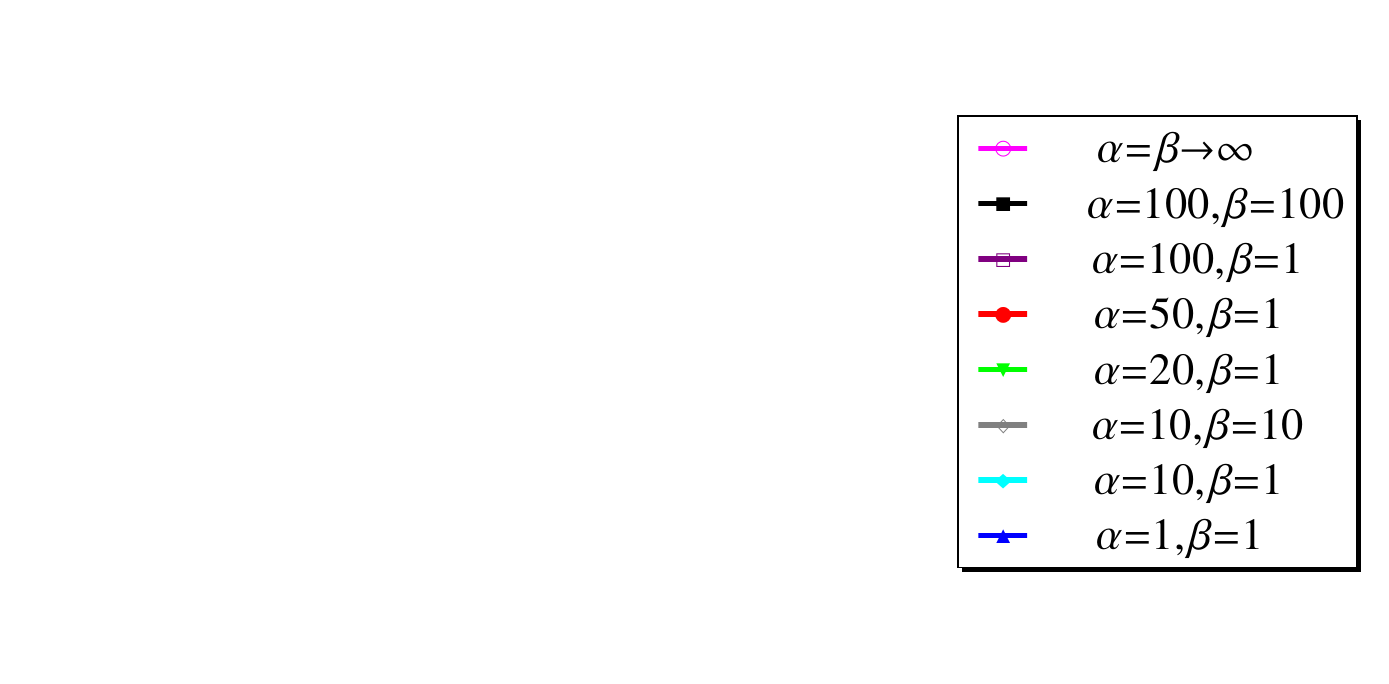,width=10cm}}\fi
\end{picture}
\caption{$F$--$\delta$ curves of the thick prism model, when loaded in compression (top), and tension (bottom), for $p_0$ = 0.1 and different values of $\alpha$ and $\beta$.
}
\label{Fr_rigidezza_variab}
\end{figure}

\begin{figure}[!hbt]
\unitlength1cm
\begin{picture}(11,15)
\if\Images y\put(0,10){\psfig{figure=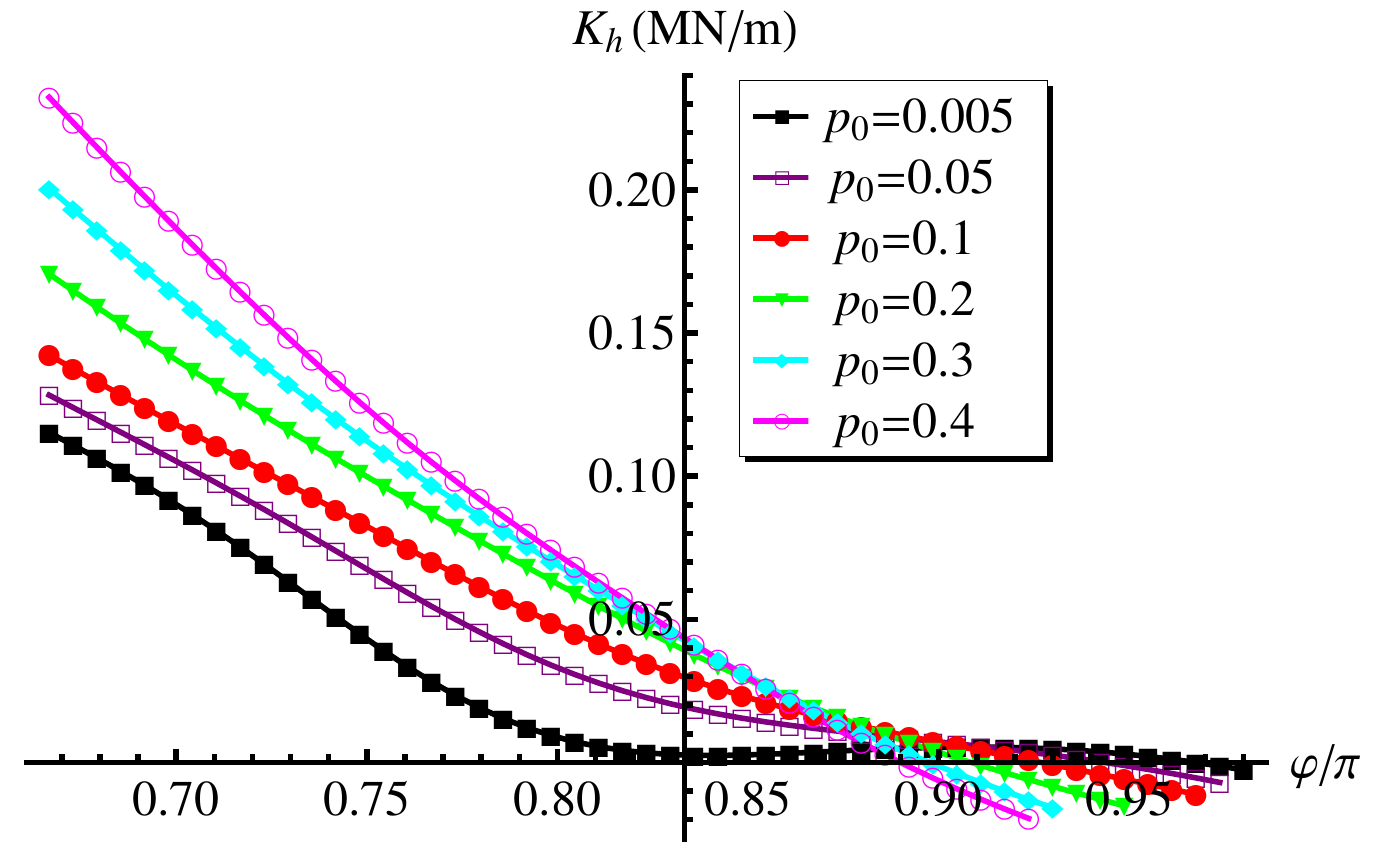,width=7.5cm}}\fi
\if\Images y\put(8.5,10){\psfig{figure=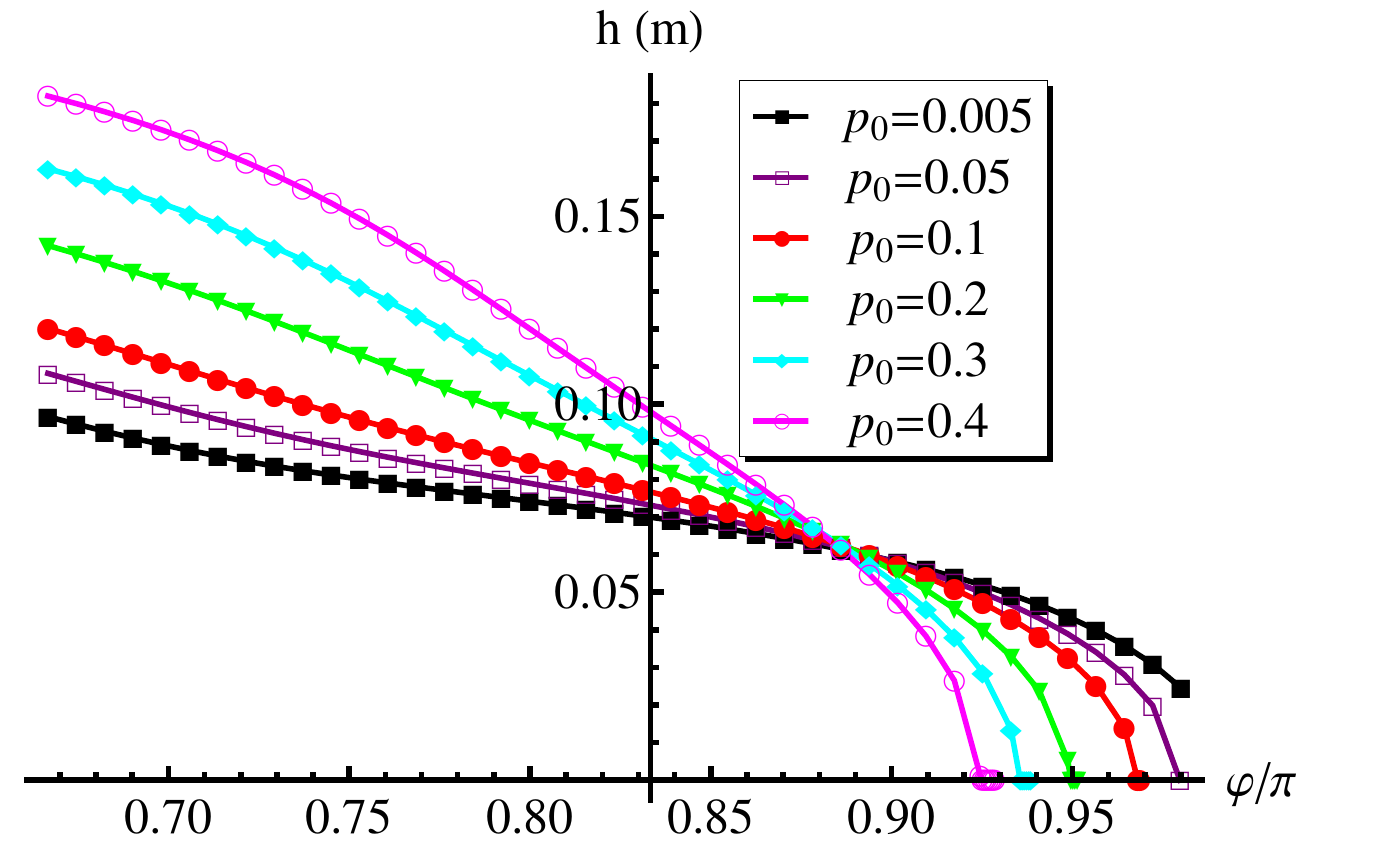,width=7.5cm}}\fi
\if\Images y\put(0.2,5){\psfig{figure=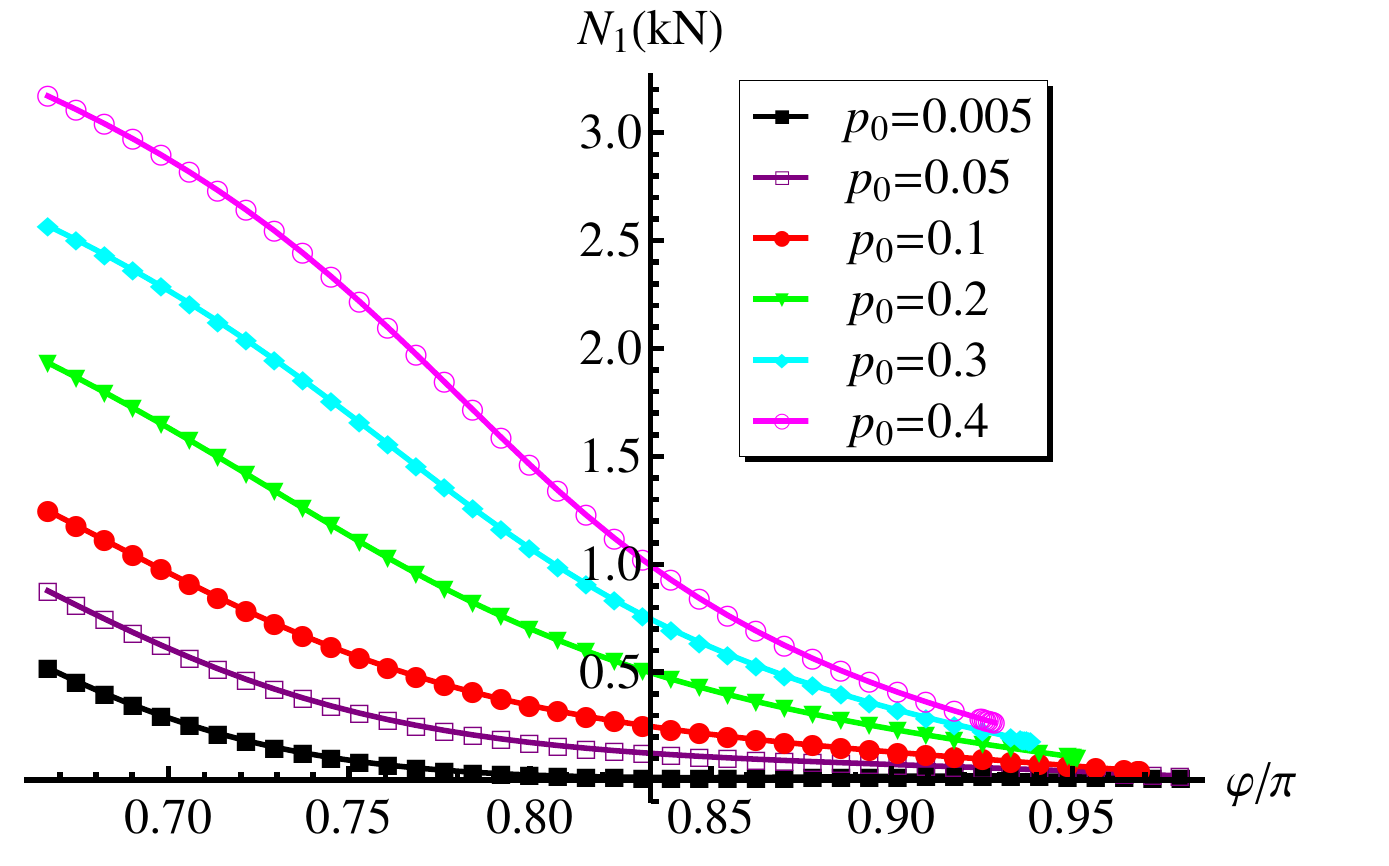,width=7.5cm}}\fi
\if\Images y\put(8.6,5){\psfig{figure=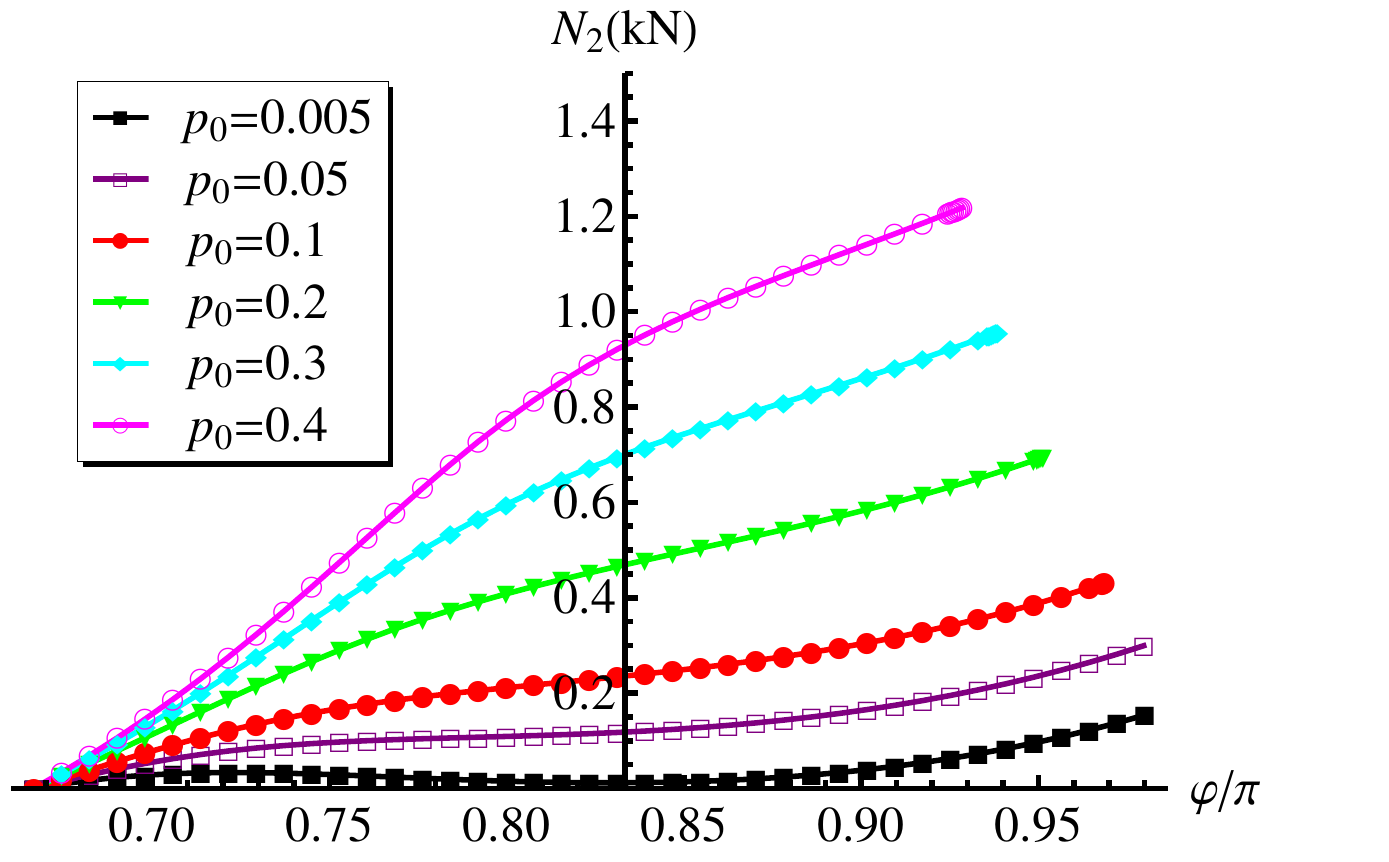,width=7.5cm}}\fi
\if\Images y\put(4.25,0){\psfig{figure=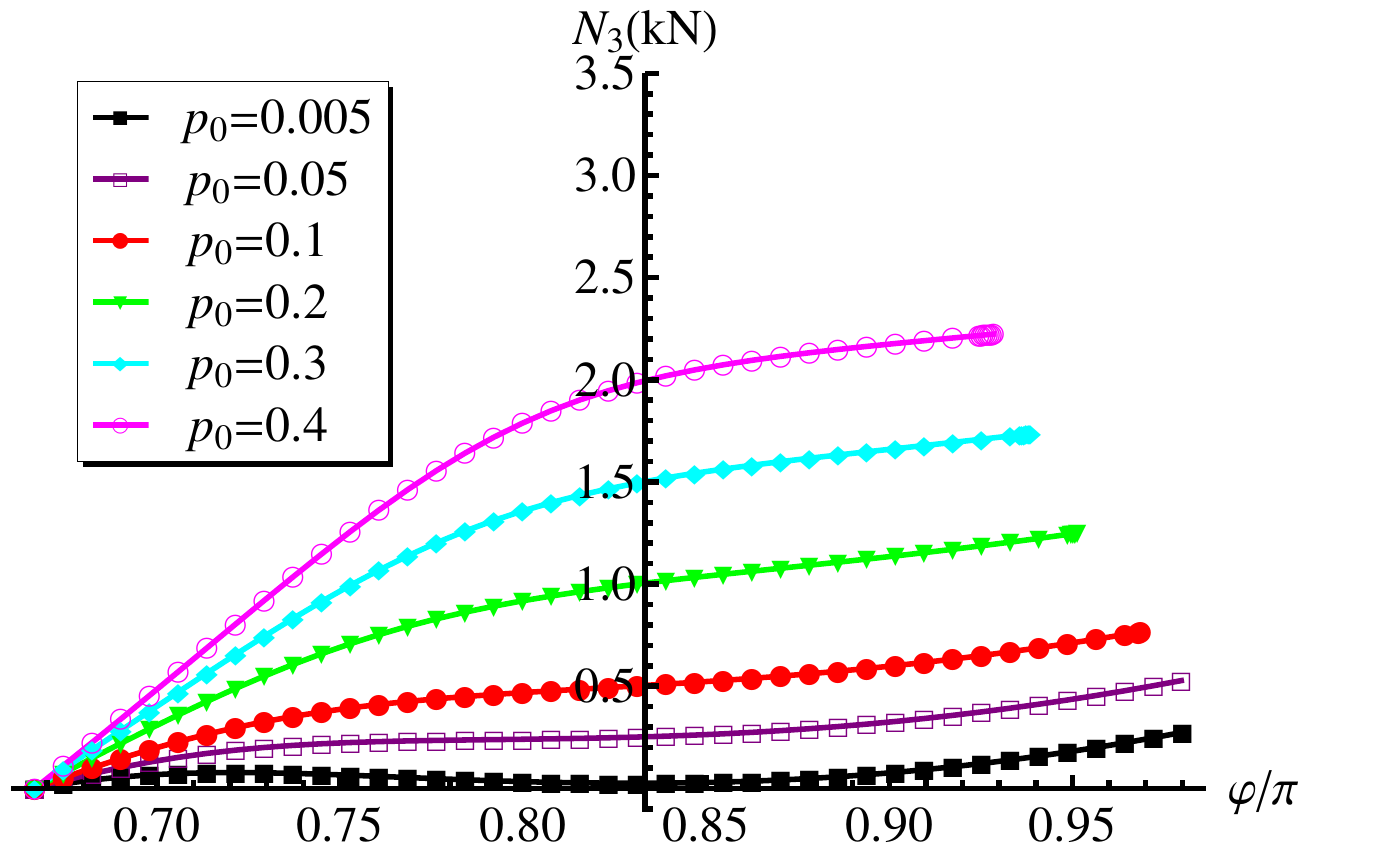,width=7.5cm}}\fi
\end{picture}
\caption{$K_h$ vs. $\varphi$, $h$ vs. $\varphi$, and $N_1,N_2,N_3$ vs. $\varphi$ curves of the thick prism model for $\alpha$ = $\beta$ = 1, and different values of $p_0$.
}
\label{NK-phi-p0-thick-panel}
\end{figure}

\begin{figure}[!hbt]
\unitlength1cm
\begin{picture}(11,21)
\if\Images y\put(-0.1,16.5){\psfig{figure=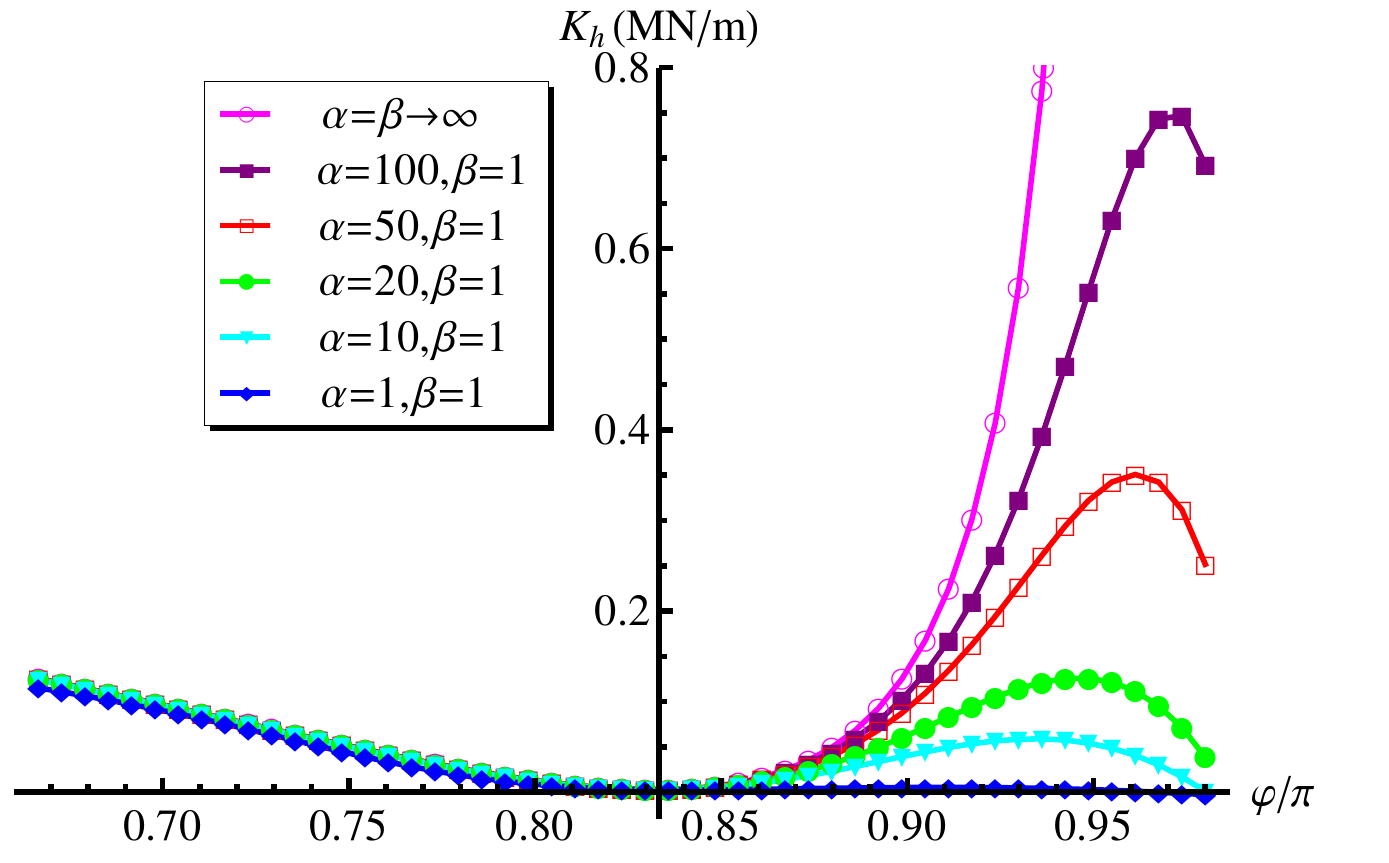,width=8.5cm}}\fi
\if\Images y\put(7.9,16.5){\psfig{figure=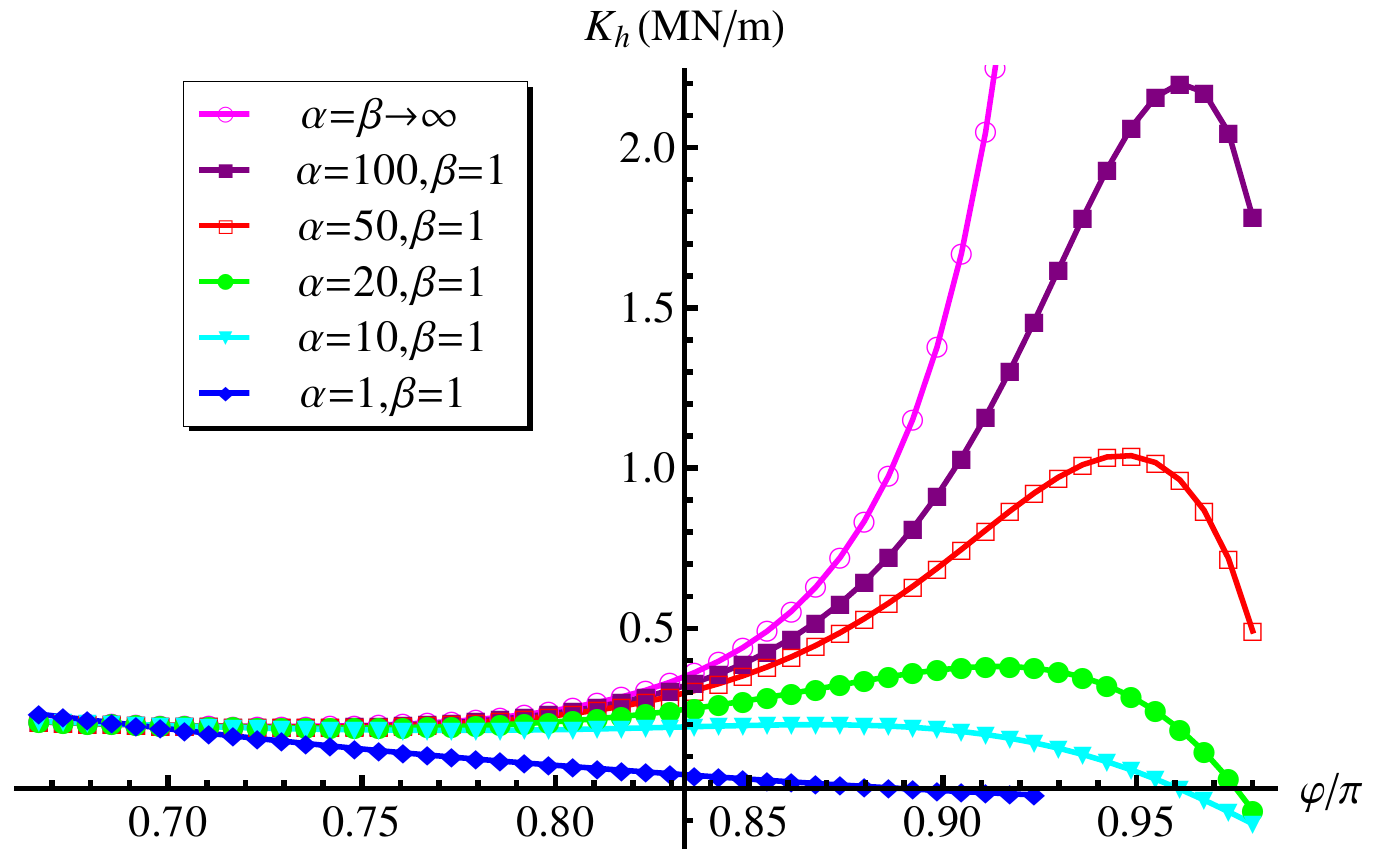,width=8.5cm}}\fi
\if\Images y\put(0,11){\psfig{figure=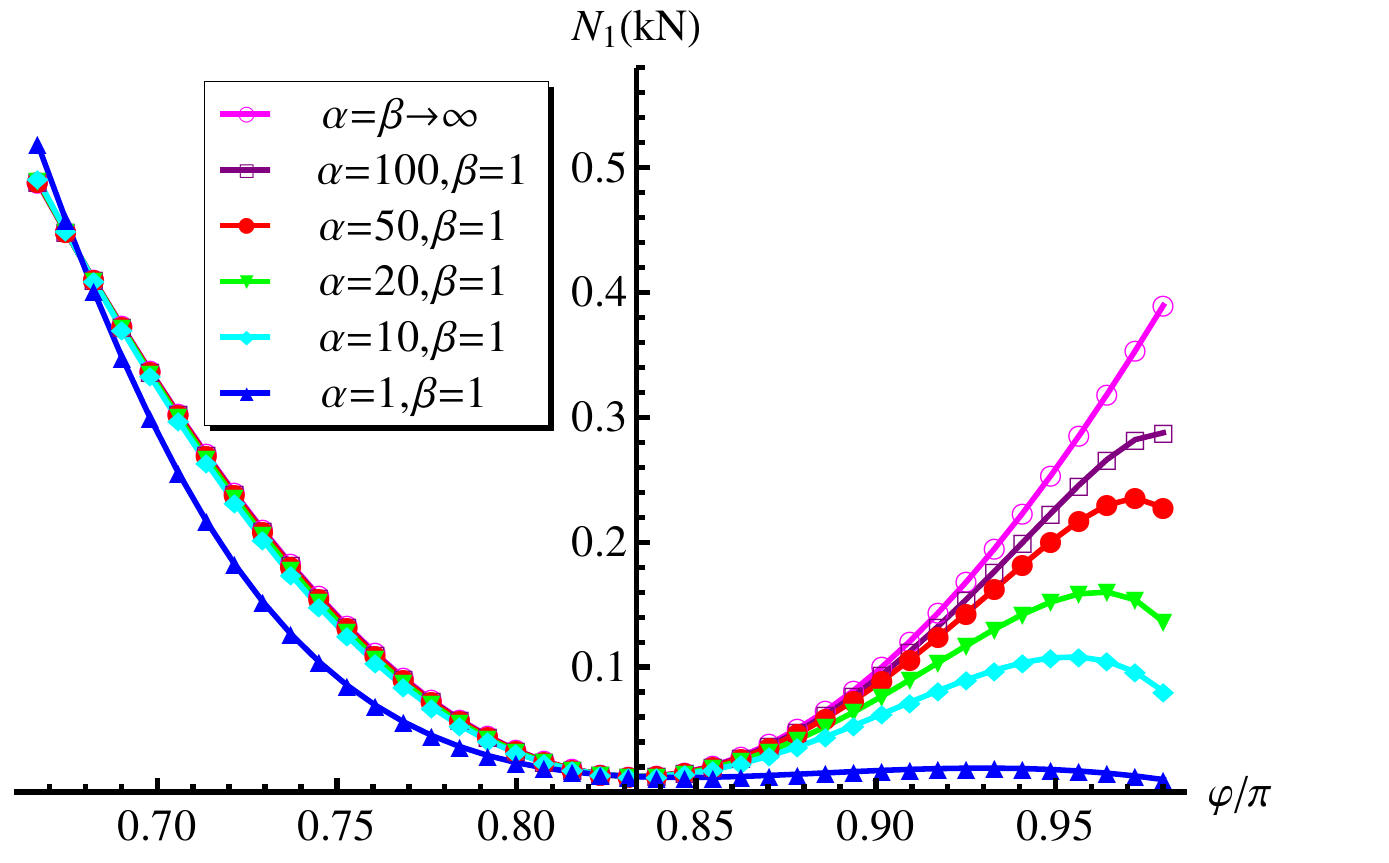,width=8.5cm}}\fi
\if\Images y\put(8,11){\psfig{figure=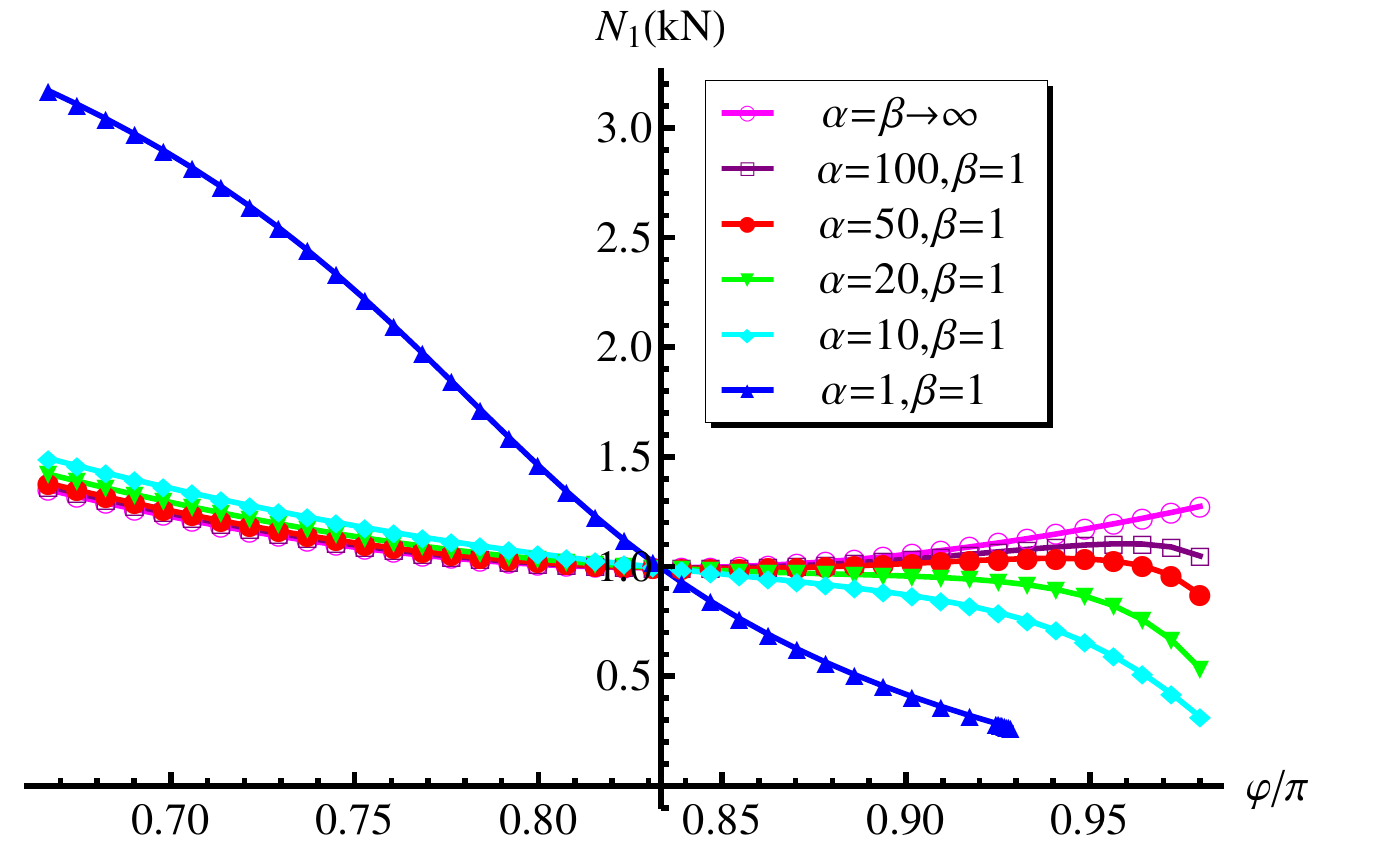,width=8.5cm}}\fi
\if\Images y\put(-0.1,5.5){\psfig{figure=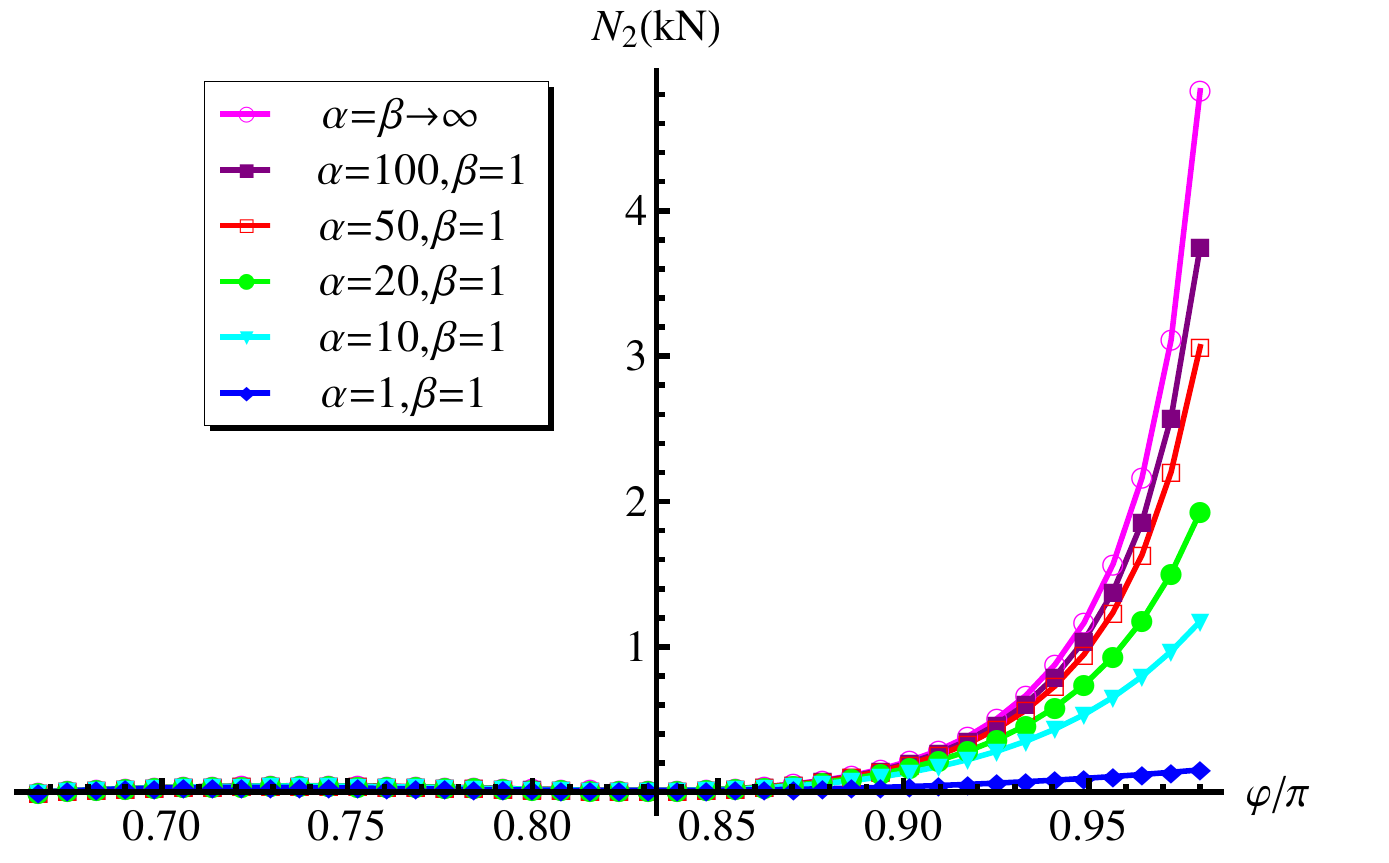,width=8.5cm}}\fi
\if\Images y\put(8,5.5){\psfig{figure=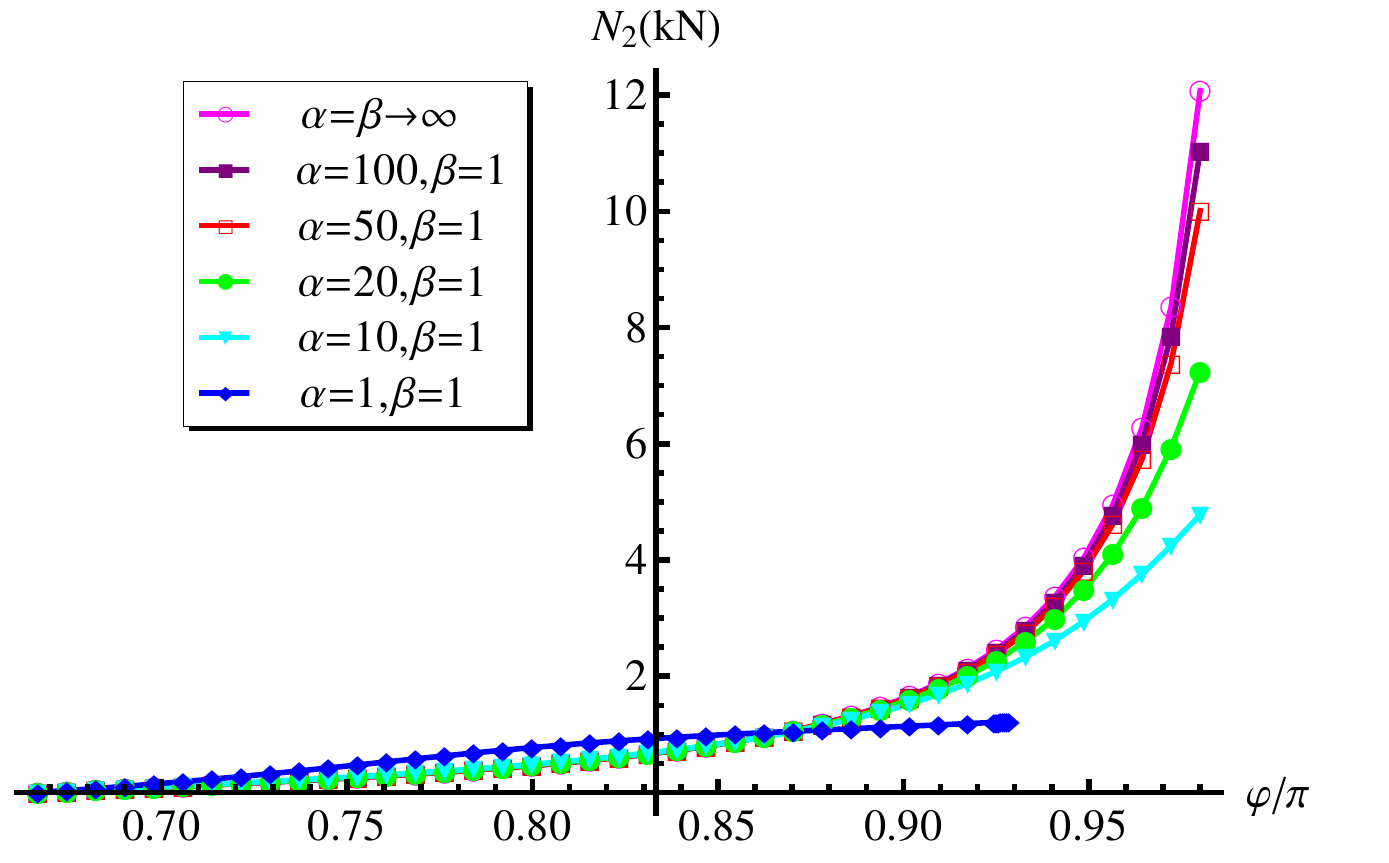,width=8.5cm}}\fi
\if\Images y\put(-0.1,0){\psfig{figure=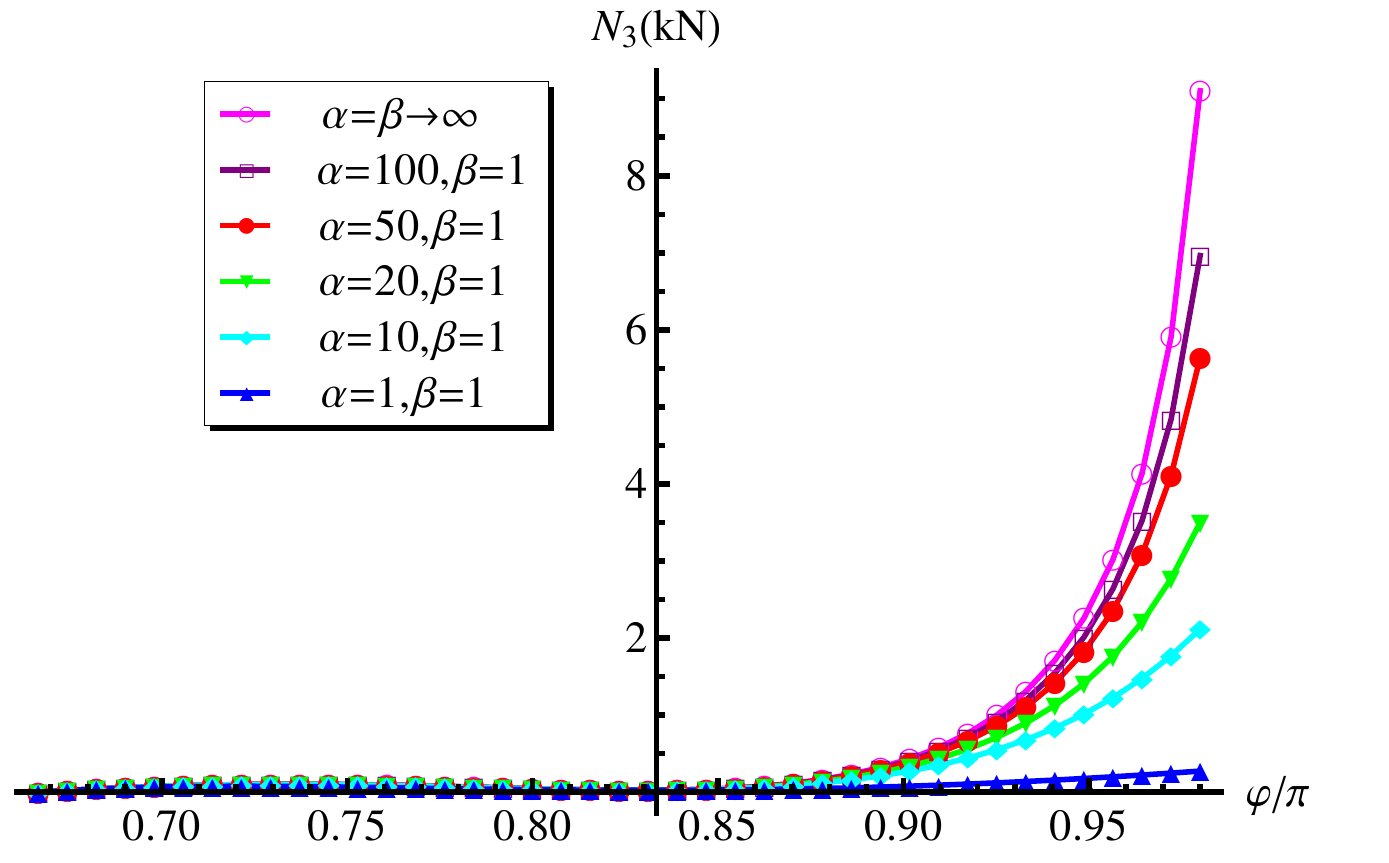,width=8.5cm}}\fi
\if\Images y\put(8,0){\psfig{figure=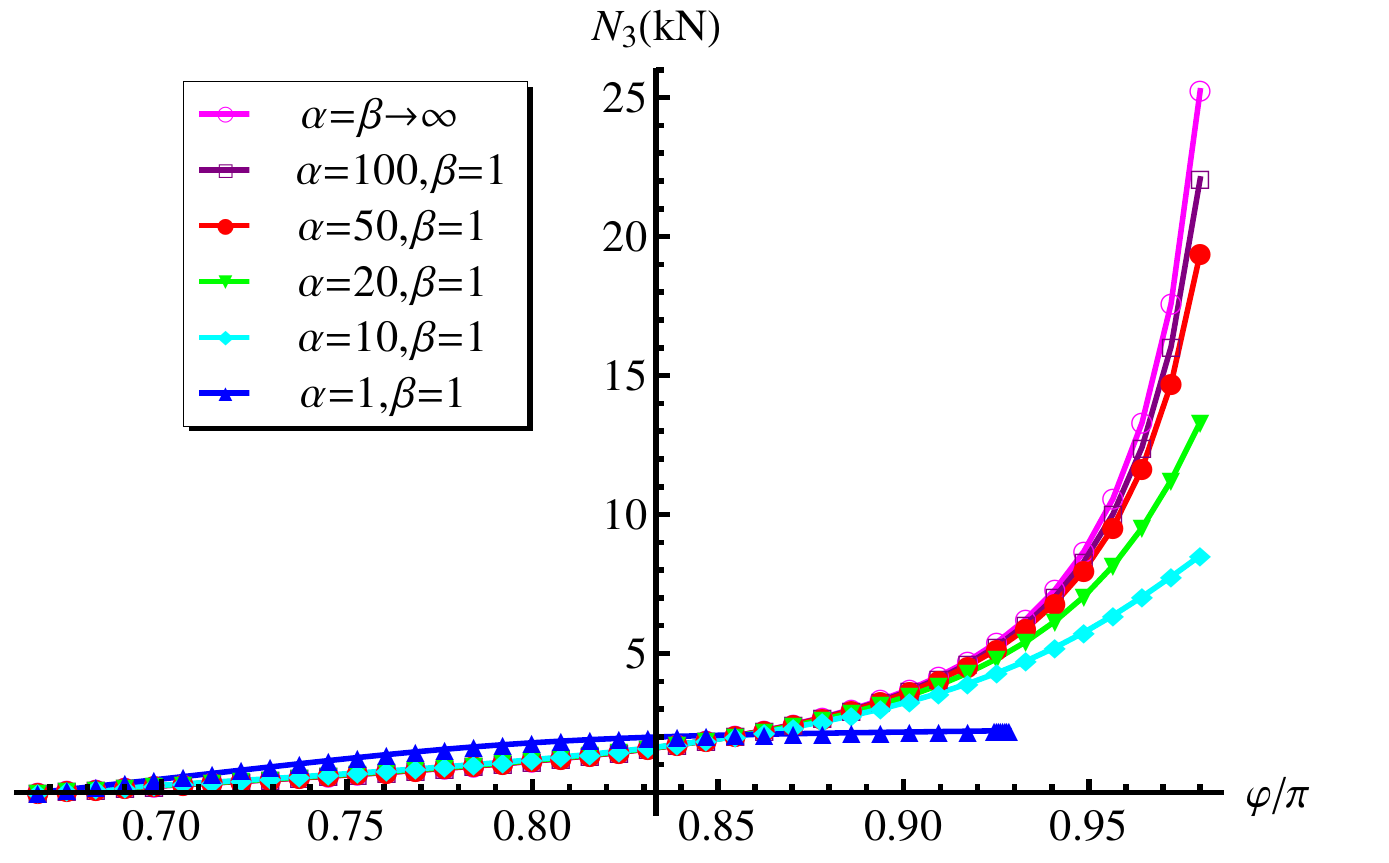,width=8.5cm}}\fi
\end{picture}
\caption{$K_h$ vs. $\varphi$ and $N_1,N_2,N_3$ vs. $\varphi$ curves of the thick prism model for $p_0$ = 0.005 (left), $p_0$ = 0.4 (right), and different values of $\alpha$ and $\beta$.
}
\label{NK-phi-ab-thick-panel}
\end{figure}

\begin{figure}[hbt]
\unitlength1cm
\begin{picture}(11,11)
\if\Images y\put(0,0){\psfig{figure=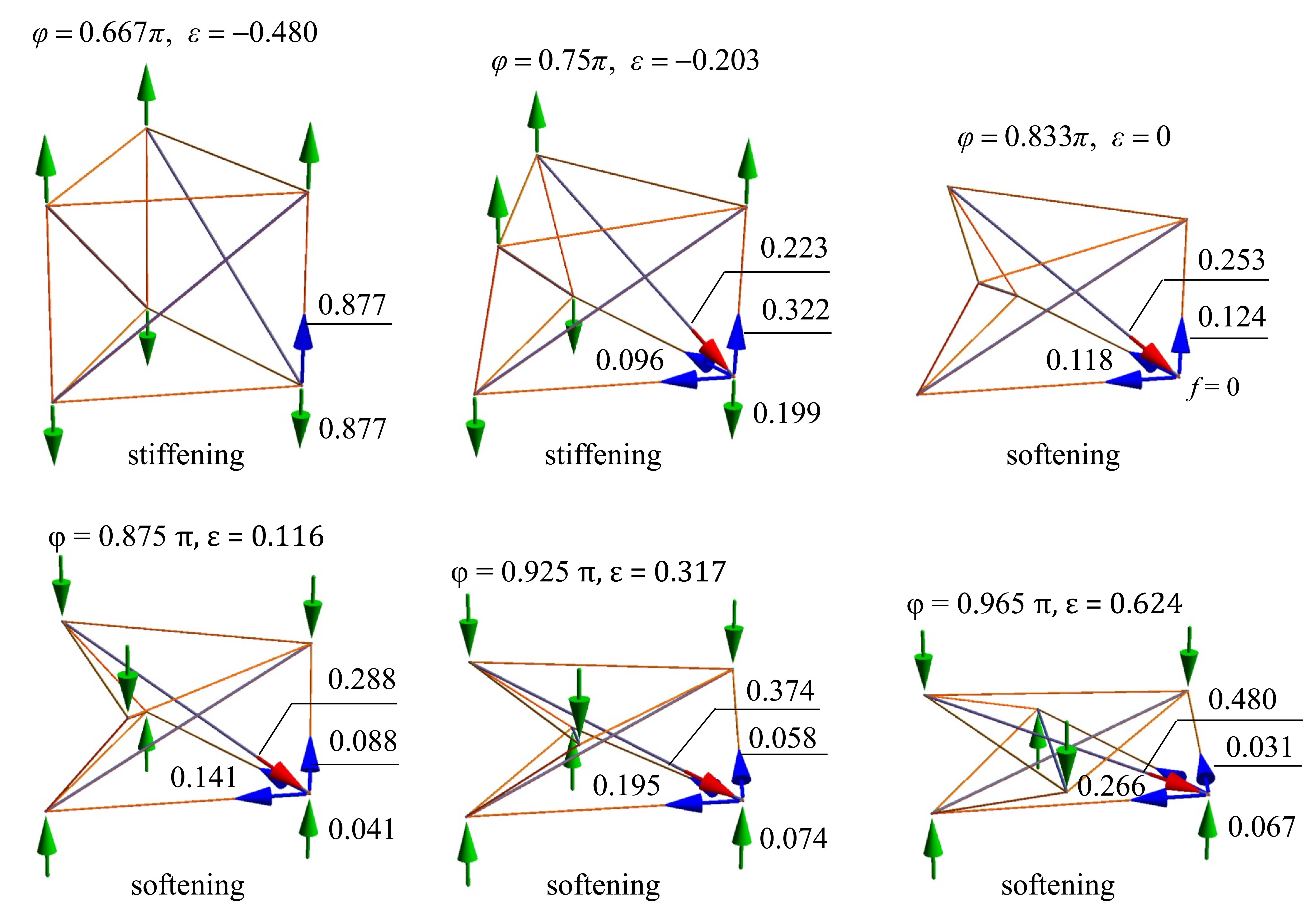,height=11cm}}\fi
\end{picture}
\caption{{\cFF{Member forces (kN) in different configurations of the thick prism, for $\alpha$ = $\beta$ = 1, and $p_0$ = 0.05.}}}
\label{Force_thick}
\end{figure}

We examine `thick'  prisms featuring: $s_N = 0.08$ m, $\ell_N = 0.132$ m, and reference lengths $s_0$, $\ell_0$, $b_0$, and $h_0$ variable with the cross string prestrain $p_0$ (cf. Section \ref{reference}). Table \ref{Sys1geom} shows noticeable values of such variables and $K_{h_0}$, for different prestrains $p_0$; the fully elastic model; and the rigid--elastic model. It is seen that $h_0$ is always smaller than $\ell_0$ in the present case, which justifies the name `thick' given to the prisms under consideration.  The difference between $K_{h_0}^{el}$ and $K_{h_0}^{rigel}$ grows with the prestrain $p_0$, being zero for $p_0=0$ ($K_{h_0}^{el}=K_{h_0}^{rigel}=0$). 
{\cFF{
Fig. \ref{p-variab-system1} shows the force $F$ vs. $\delta$ curves of the `el'  samples for different values of $p_0$. Fig. \ref{Fr_rigidezza_variab} provides the same curves for different values of the stiffness multipliers $\alpha$ and $\beta$, and $p_0=0.1$.  Finally, Figs. \ref{NK-phi-p0-thick-panel} and \ref{NK-phi-ab-thick-panel} illustrate the variations with the angle of twist $\varphi$ of the axial stiffness $K_h$; the prism height $h$; and the axial forces $N_1$, $N_2$ and $N_3$.
In the `el'  case, the results in Figs. \ref{p-variab-system1} and \ref{NK-phi-p0-thick-panel} highlight that the compressive response for $p_0 \le 0.005$ initially features a stiffening branch, next a softening branch, and finally an unstable phase (\textit{strain softening}: $F$ decreasing with $\delta$), as the axial strain $\varepsilon$ increases. 
When $p_0$ grows above 0.005, the initial stiffening branch disappears, and the compressive response is always softening. The final unstable branch is associated with the snap buckling of the prism to the completely collapsed configuration featuring zero height $h$ (cf. Fig. \ref{Force_thick}). Such a collapse event can fully take place when $p_0 \ge 0.05$, but is instead prevented by prism locking for lower values of $p_0$ (Figs. \ref{p-variab-system1} and \ref{NK-phi-p0-thick-panel}).
It is worth noting that the maximum compression displacement  $\delta^{max}$ of the current prism model increases with $p_0$. Overall, we conclude that the compressive response of the thick prism is markedly different from that of the rigid--elastic model analyzed in \cite{Oppenheim:2000, FSD12}, since the latter predicts an infinitely stiff response for $\delta \rightarrow \delta^{max}$. 
For what concerns the tensile response, we observe that the `el' model is always stiffening in tension, for any $p_0 \in [0, 0.4]$ (Figs. \ref{p-variab-system1}, \ref{NK-phi-p0-thick-panel}). We also observe that the minimum axial displacement $\delta^{min}$  (i.e. the value of $\delta$ for $\varphi = 2/3 \pi$) grows in magnitude with $p_0$.
}}

\begin{table}[hbt]
	\centering
		\scalebox{0.90}{
		\begin{tabular}{| c | c | c | c | c | c | c | c | c | c | c |}
    \hline
$\alpha$=$\beta$ & $p_0$ & $s_N$ (m) & $s_0$ (m) & $\ell_N$ (m) & $\ell_0$ (m) & $b_N$ (m) & $b_0$ (m) & $h_0$ (m) & $K_{h_0}$ (N/m) \\ \hline
1 & 0 & 0.080 & 0.0800 & 0.1320 & 0.1320 & 0.1628 & 0.1628 & 0.0696 & 0 \\ \hline
1 & 0.005 & 0.080 & 0.0804 & 0.1320 & 0.1326 & 0.1636 & 0.1636 & 0.0700 & 2595 \\ \hline
1 & 0.1 & 0.080 & 0.0880 & 0.1320 & 0.1445 & 0.1785 & 0.1785 & 0.0767 & 29720 \\ \hline
1 & 0.2 & 0.080 & 0.0960 & 0.1320 & 0.1569 & 0.1941 & 0.1940 & 0.0838 & 39257 \\ \hline
1 & 0.3 & 0.080 & 0.1040 & 0.1320 & 0.1692 & 0.2095 & 0.2095 & 0.0909 & 42601 \\ \hline
1 & 0.4 & 0.080 & 0.1120 & 0.1320 & 0.1814 & 0.2248 & 0.2248 & 0.0980 & 43465 \\ \hline
$\rightarrow \infty$ & 0 & 0.080 & 0.0800 & 0.1320 & 0.1320 & 0.1628 & 0.1628 & 0.0696 & 0 \\ \hline
$\rightarrow \infty$ & 0.005 & 0.080 & 0.0804 & 0.1320 & 0.1320 & 0.1630 & 0.1630 & 0.0701 & 2682 \\ \hline
$\rightarrow \infty$ & 0.1 & 0.080 & 0.0880 & 0.1320 & 0.1320 & 0.1669 & 0.1669 & 0.0787 & 49582 \\ \hline
$\rightarrow \infty$ & 0.2 & 0.080 & 0.0960 & 0.1320 & 0.1320 & 0.1713 & 0.1713 & 0.0875 & 92033 \\ \hline
$\rightarrow \infty$ & 0.3 & 0.080 & 0.1040 & 0.1320 & 0.1320 & 0.1759 & 0.1759 & 0.0962 & 129005 \\ \hline
$\rightarrow \infty$ & 0.4 & 0.080 & 0.1120 & 0.1320 & 0.1320 & 0.1807 & 0.1807 & 0.1048 & 161679 \\ \hline
		\end{tabular}		
		}
	\caption{Geometric variables and initial axial stiffness $K_{h_0}$ of the thick prism model, for different values of the cross-string prestrain $p_0$; the fully elastic model ($\alpha=\beta=1$); and the rigid--elastic model ($\alpha=\beta \rightarrow +\infty$).
}	
	\label{Sys1geom}
\end{table}

\newpage

{\cFF{Let us now pass to studying the response of thick prisms for different values of the rigidity multipliers $\alpha$ and $\beta$.
The $F-\delta$ curves in  Fig. \ref{Fr_rigidezza_variab} show that the response in compression of the thick prisms analyzed in this study switches from extremely soft to extremely stiff when $\alpha$ and $\beta$ grow from 1 (`el' model) to $+\infty$ (`rigel' model). 
In particular, we observe that $\alpha$ (i.e., the base rigidity) plays a more substantial role in the mechanical response of such a structure than does $\beta$ (the bar rigidity multiplier). We indeed note that the $F-\delta$ curves for $\alpha= \beta=10$  and $\alpha=\beta=100$ are not much different from those corresponding to $\alpha=10,\ \beta=1$ and $\alpha=100,\ \beta=1$, respectively. This is due to the fact that the axial stiffness of the bars is much higher than the axial stiffness of the strings (cf. Table \ref{tabmat}), which implies that the assumption of bar rigidity is more realistic than the assumption of base rigidity, in the present case. 
When $p_0=0.005$, Fig. \ref{NK-phi-ab-thick-panel} shows that the response in tension of thick prisms is always stiffening, for all the examined values of $\alpha$ and $\beta$. In contrast, for $p_0=0.4$ we observe that such a response progressively switches from stiffening to softening, as $\alpha$ and $\beta$ grow to infinity (Fig. \ref{NK-phi-ab-thick-panel}). 
Overall, we note that the stroke of the prism ($\delta^{max}-\delta^{min}$) decreases with $\alpha$ and $\beta$ (Fig. \ref {Fr_rigidezza_variab}), and increases with $p_0$ (Fig. \ref {p-variab-system1}). Conversely, the value of $K_h$ at $\delta = \delta^{max}$ increases with $\alpha$ and $\beta$ (Fig. \ref{NK-phi-ab-thick-panel}), and decreases with $p_0$ (Fig. \ref{NK-phi-p0-thick-panel}).
}}

{\cFF{
The results in Fig. \ref{NK-phi-p0-thick-panel} highlight that the softening and unstable phases of the `el'  model are associated with a progressive decrease of the force acting in the cross-strings ($N_1$). 
The decrease of  $N_1$ with $\varphi$ for $\alpha=\beta=1$ is confirmed by the results given in Fig \ref{NK-phi-ab-thick-panel}, which show that the cross-strings tend to become slack as  $\varphi$ approaches $\pi$ ($\delta \rightarrow \delta^{max}$), in the `el'  case. 
The $N_2$  vs. $\varphi$ curves of the base-strings highlight that $N_2$ grows monotonically with $\varphi$ (starting with the value $N_2=0$ at $\varphi = 2/3 \pi$), independently of $p_0$, $\alpha$ and $\beta$ (Figs. \ref{NK-phi-p0-thick-panel} and \ref{NK-phi-ab-thick-panel}). In particular, the rate of growth of $N_2$ decreases with $p_0$, and increases with $\alpha$ and $\beta$, tending to infinity for $\varphi\rightarrow \pi$ ($\delta \rightarrow \delta^{max}$), when $\alpha=\beta \rightarrow \infty$. This implies that, in real life, the base strings would yield before reaching the `locking' configuration, in the `rigel' limit.
The axial force response of the bars resembles that of the base strings, and we note that the bars tend to buckle before reaching the locking configuration in the`rigel' limit.
For $p_0 \ge 0.05$, it is worth noting that the maximum value of $\varphi$ is less than $\pi$ (cf. Figs.  \ref{NK-phi-p0-thick-panel}, \ref{NK-phi-ab-thick-panel}), since in such cases the axial collapse precedes the locking configuration $\varphi = \pi$.
}}

\subsection{Slender prisms} \label{slender prism}

\begin{figure}[hbt]
\unitlength1cm
\begin{picture}(11,13)
\if\Images y\put(2,6.05){\psfig{figure=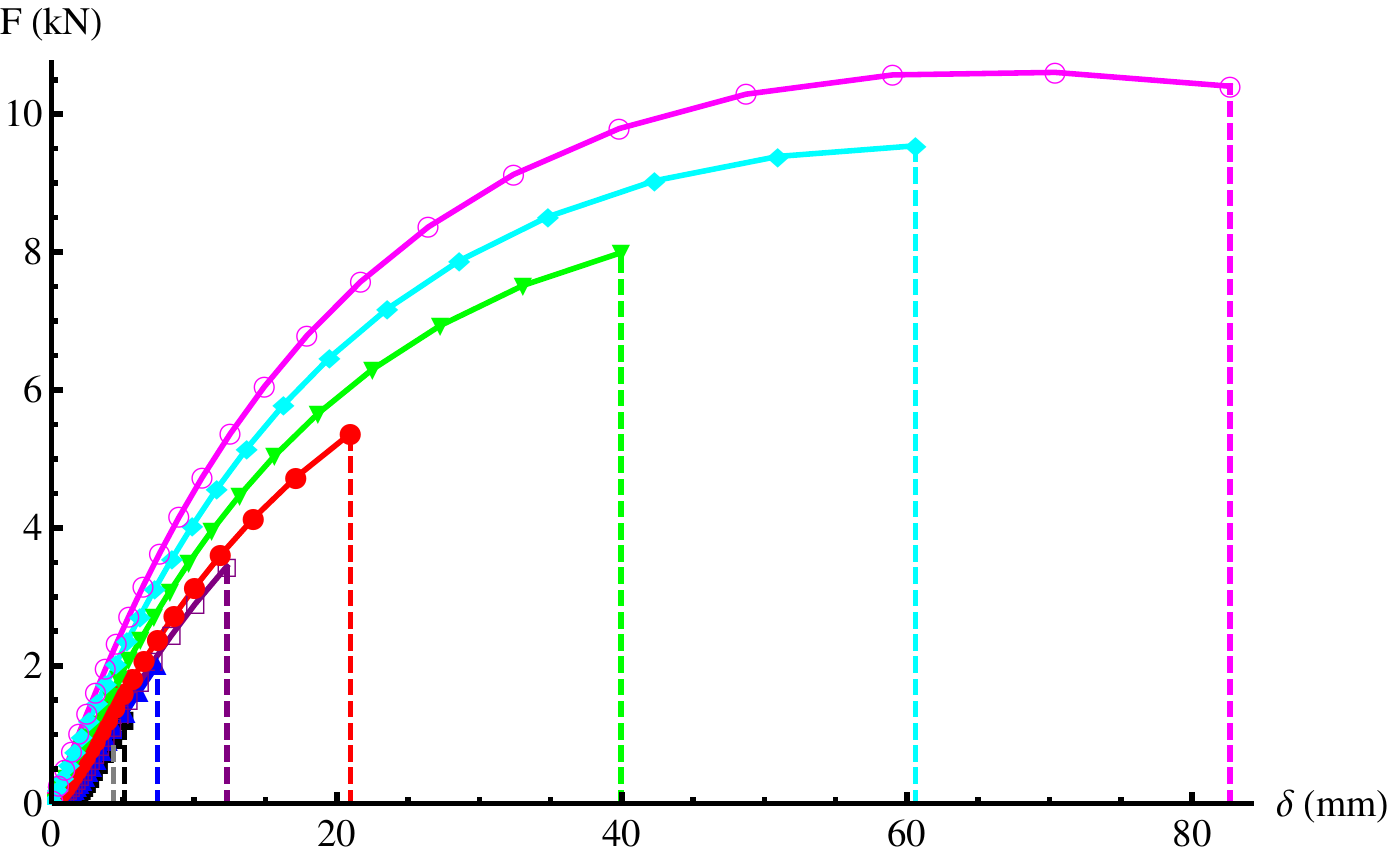,width=10cm}}\fi
\if\Images y\put(2,0){\psfig{figure=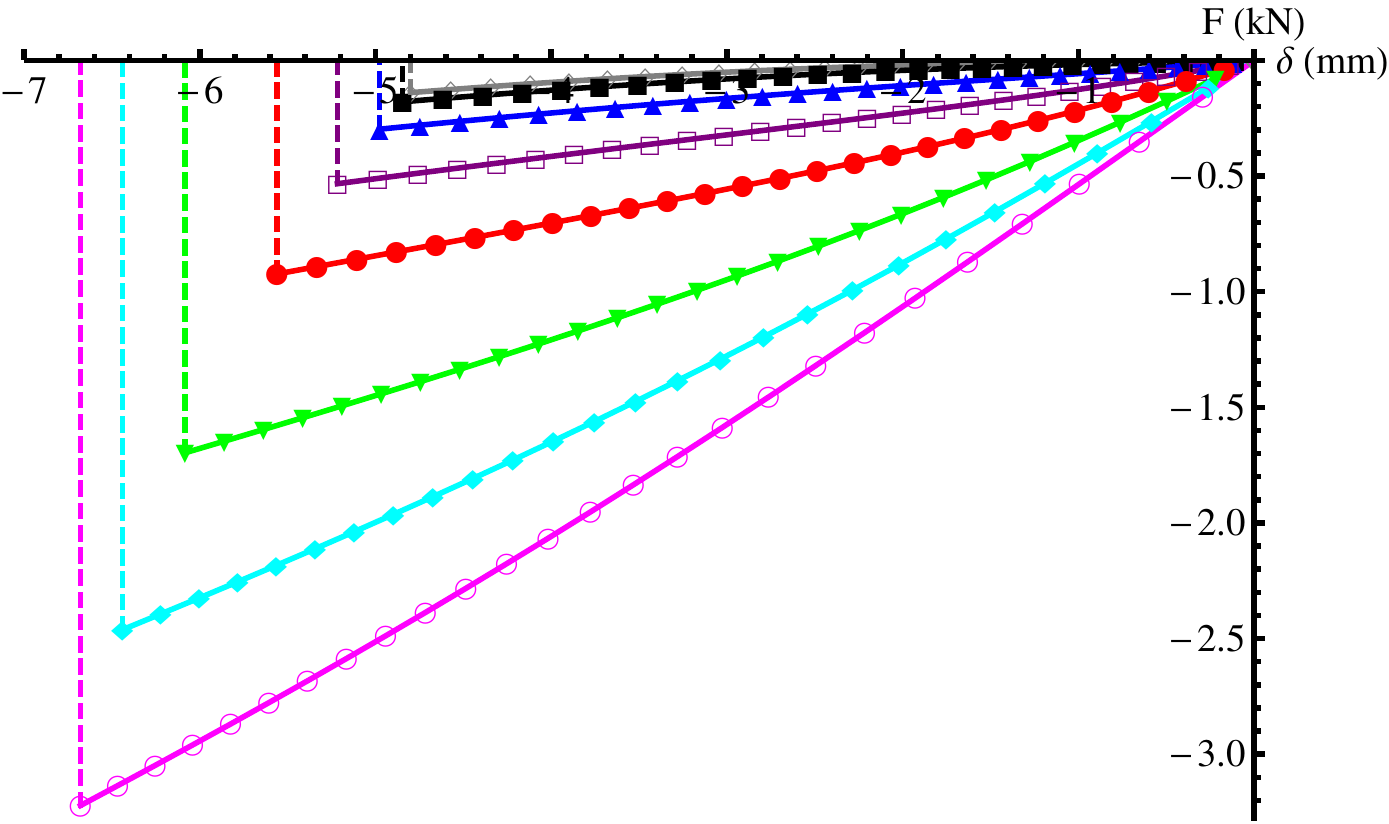,width=10cm}}\fi
\if\Images y\put(2.25,0){\psfig{figure=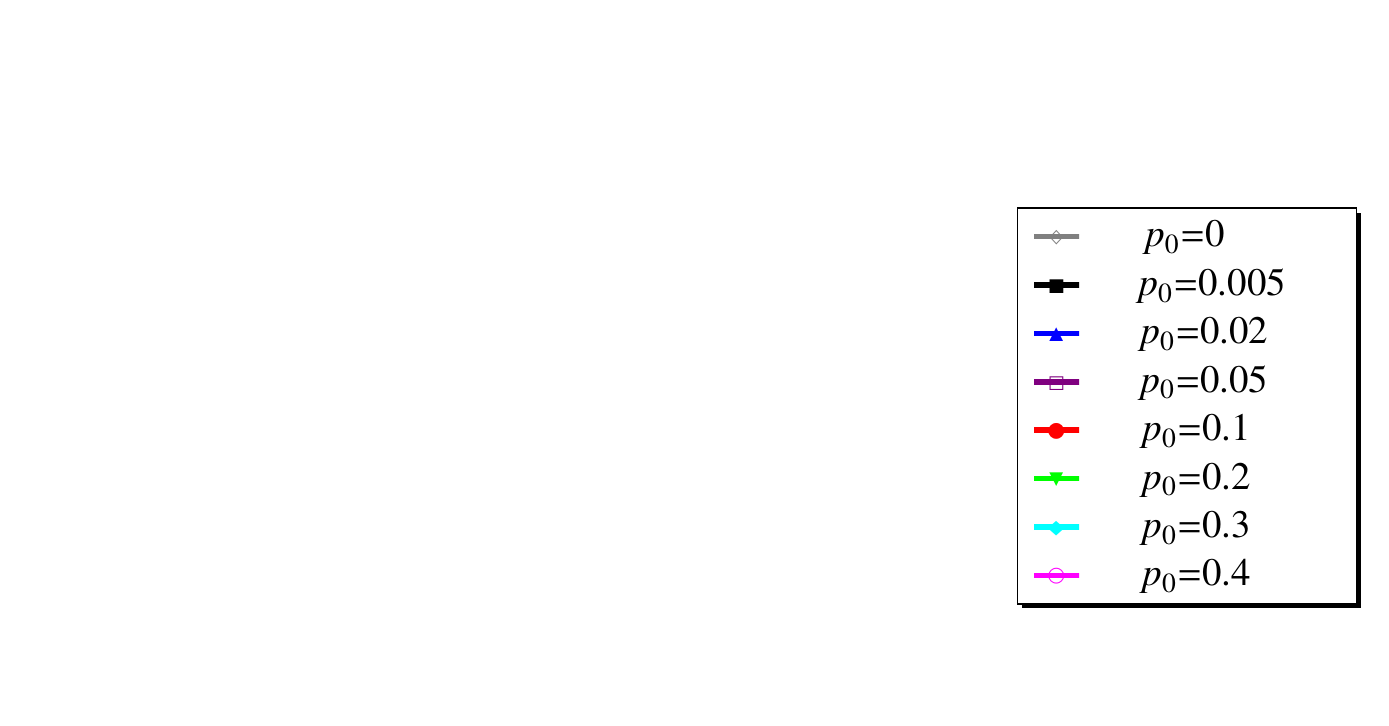,width=8cm}}\fi
\end{picture} 
\caption{$F$--$\delta$ curves of the slender prism model, when loaded in compression (top), and tension (bottom), for $\alpha=\beta=1$ and different values of $p_0$.
}
\label{p-variab-system2}
\end{figure}

\begin{figure}[hbt]
\unitlength1cm
\begin{picture}(11,12)
\if\Images y\put(2,6.5){\psfig{figure=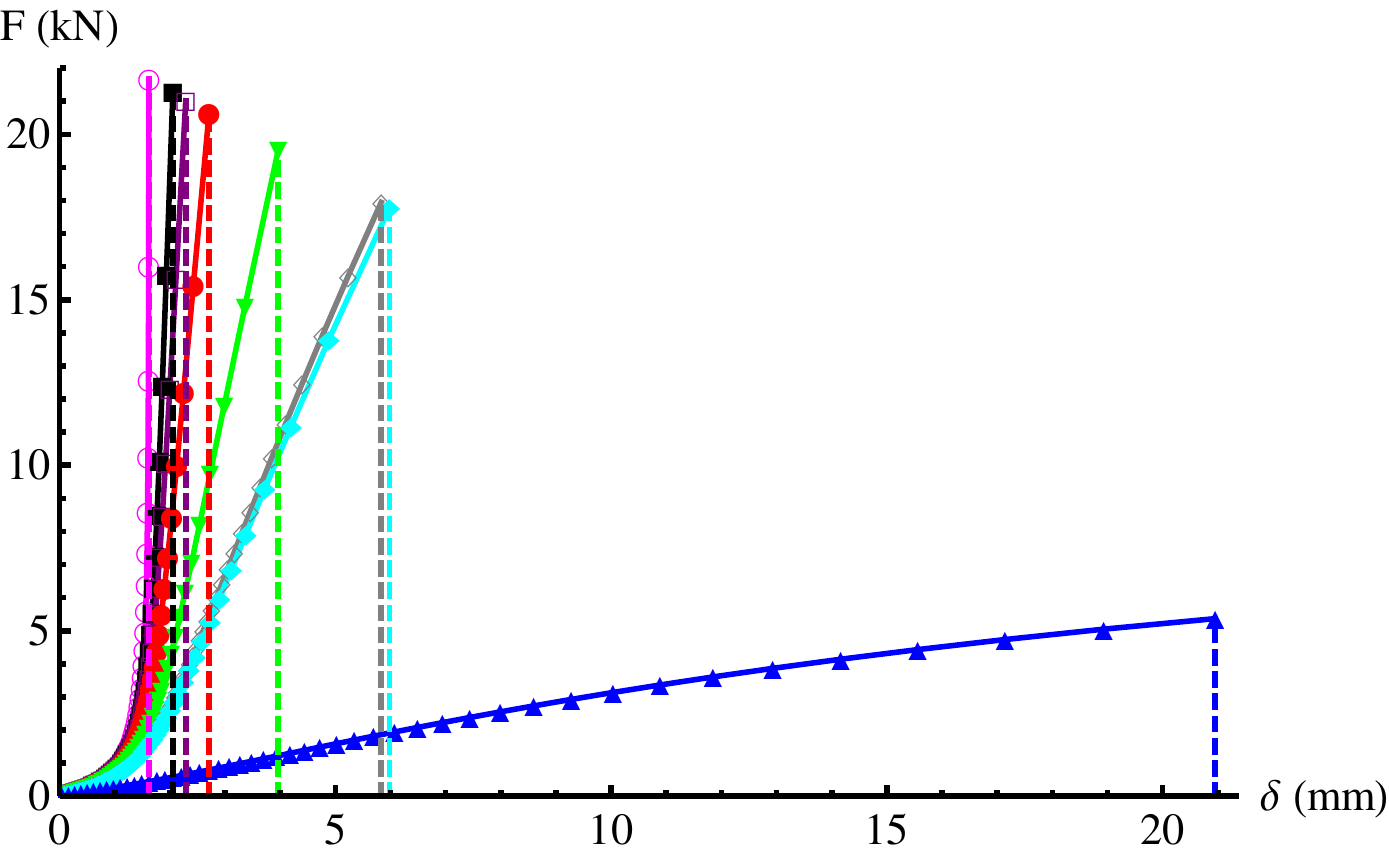,width=11cm}}\fi
\if\Images y\put(2,9){\psfig{figure=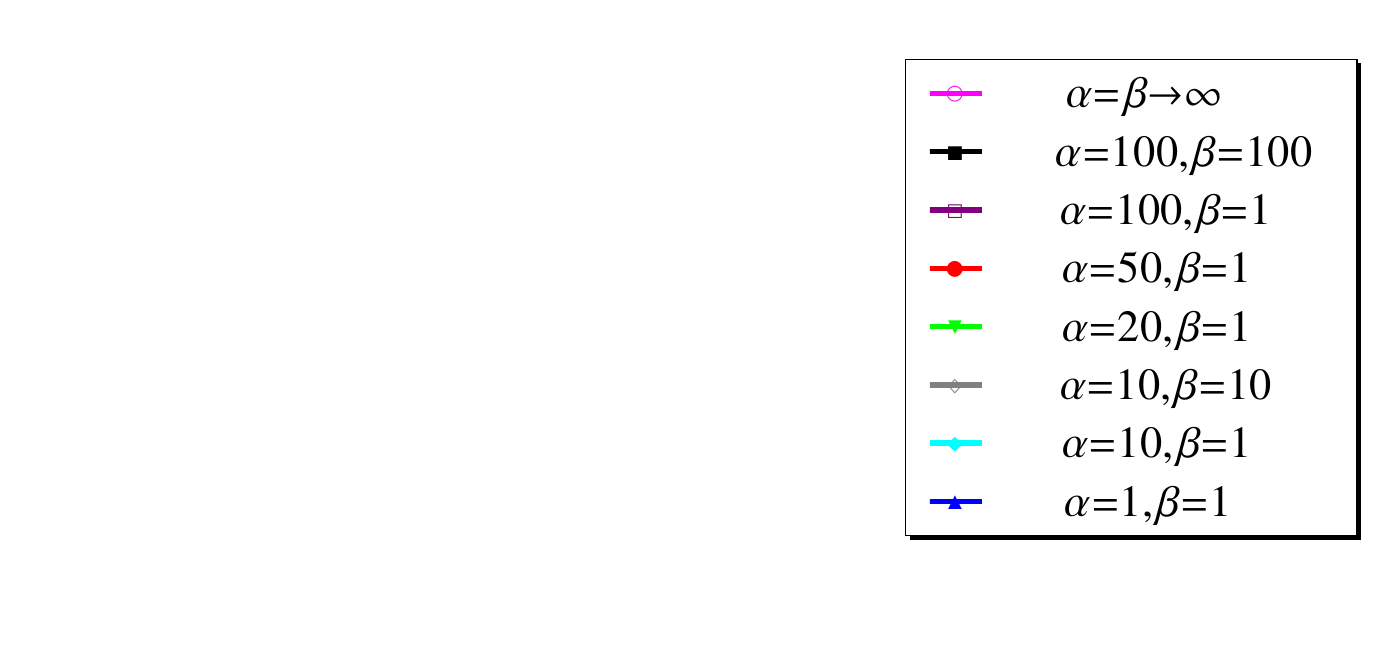,width=9cm}}\fi
\if\Images y\put(2,-0.25){\psfig{figure=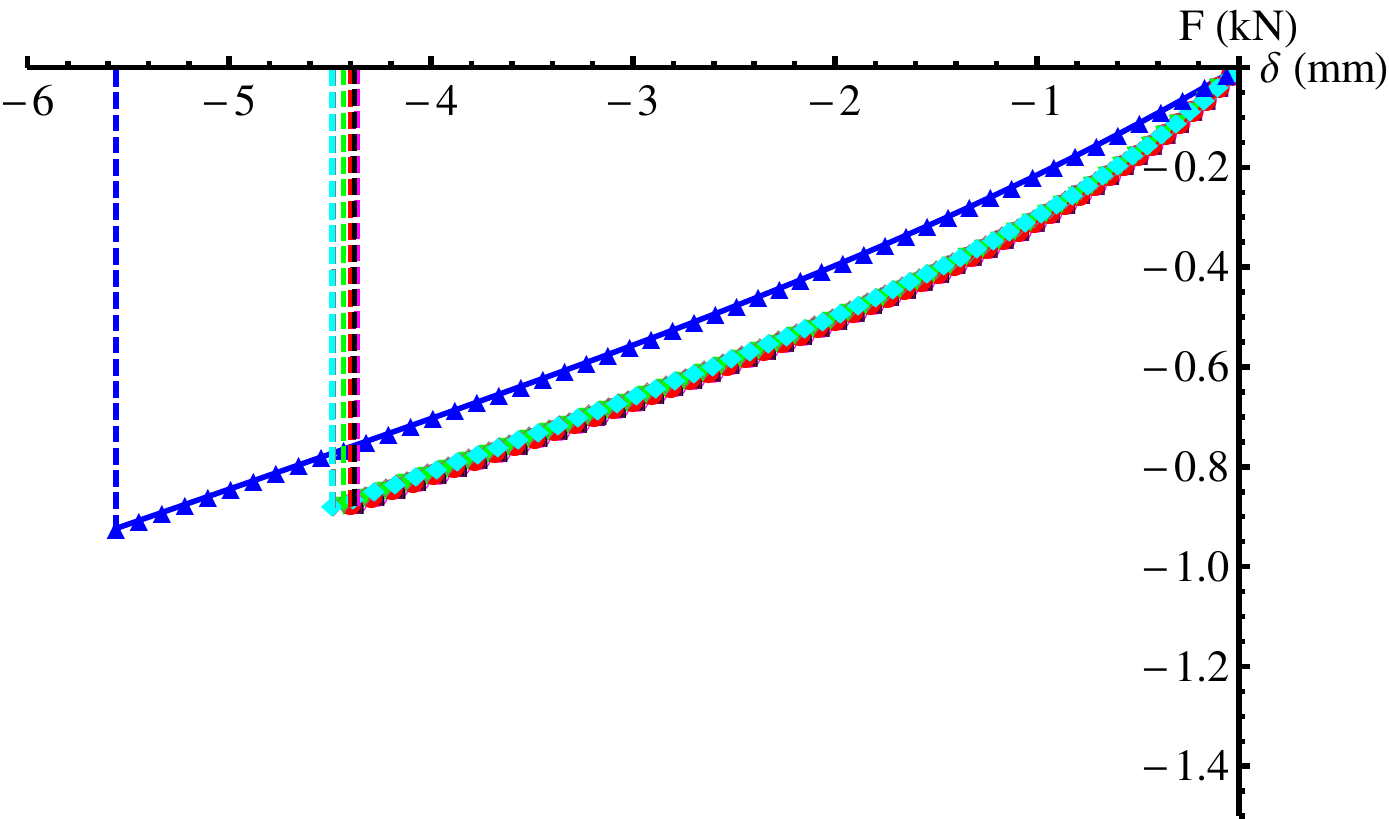,width=11cm}}\fi
\end{picture}
\caption{$F$--$\delta$ curves of the slender prism model, when loaded in compression (top), and tension (bottom), for $p_0$ = 0.1 and different values of $\alpha$ and $\beta$.
}
\label{Fr_rigidezza_variab_2}
\end{figure}

\begin{figure}[!hbt]
\unitlength1cm
\begin{picture}(11,15)
\if\Images y\put(0,10){\psfig{figure=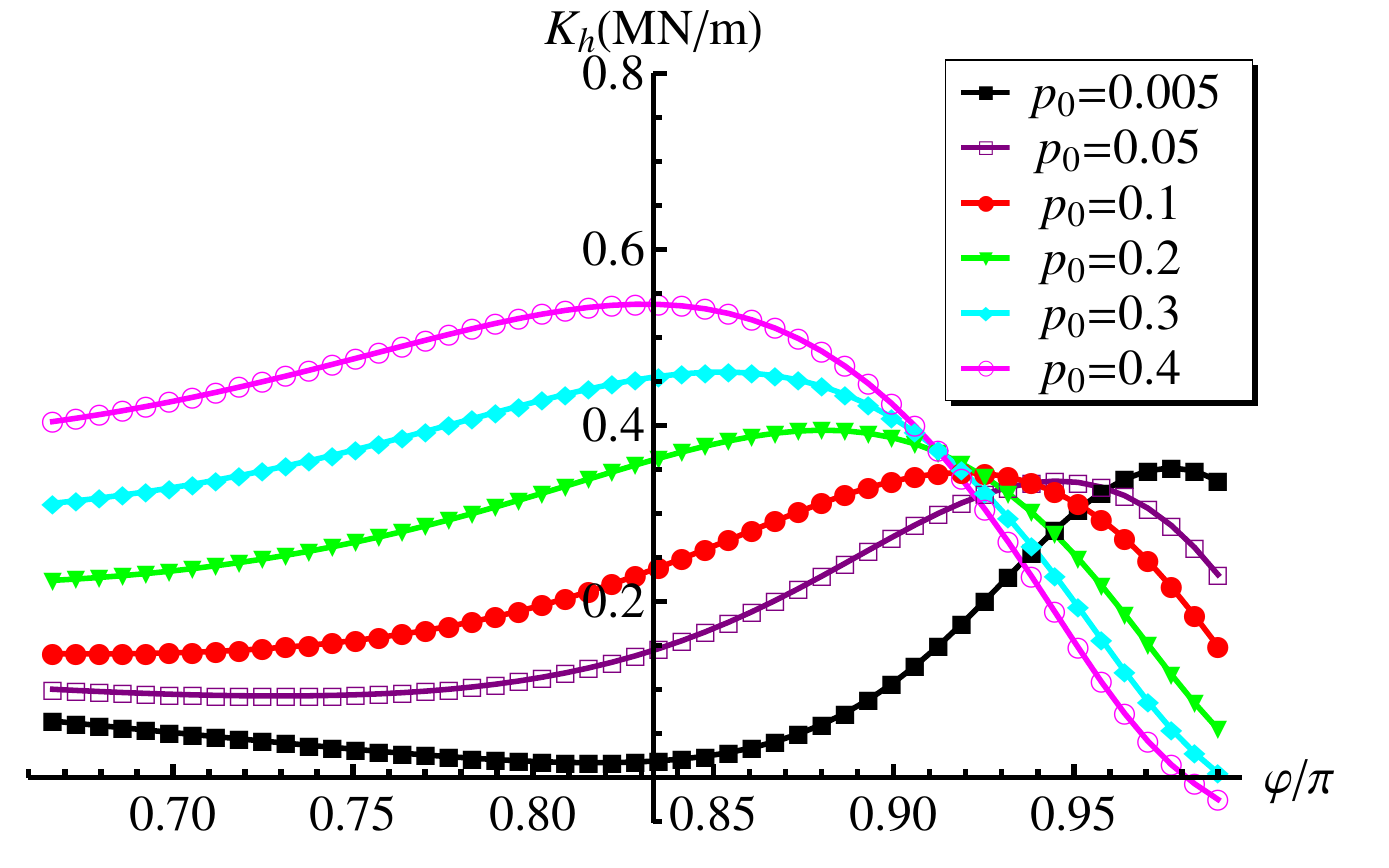,width=7.5cm}}\fi
\if\Images y\put(8.5,10){\psfig{figure=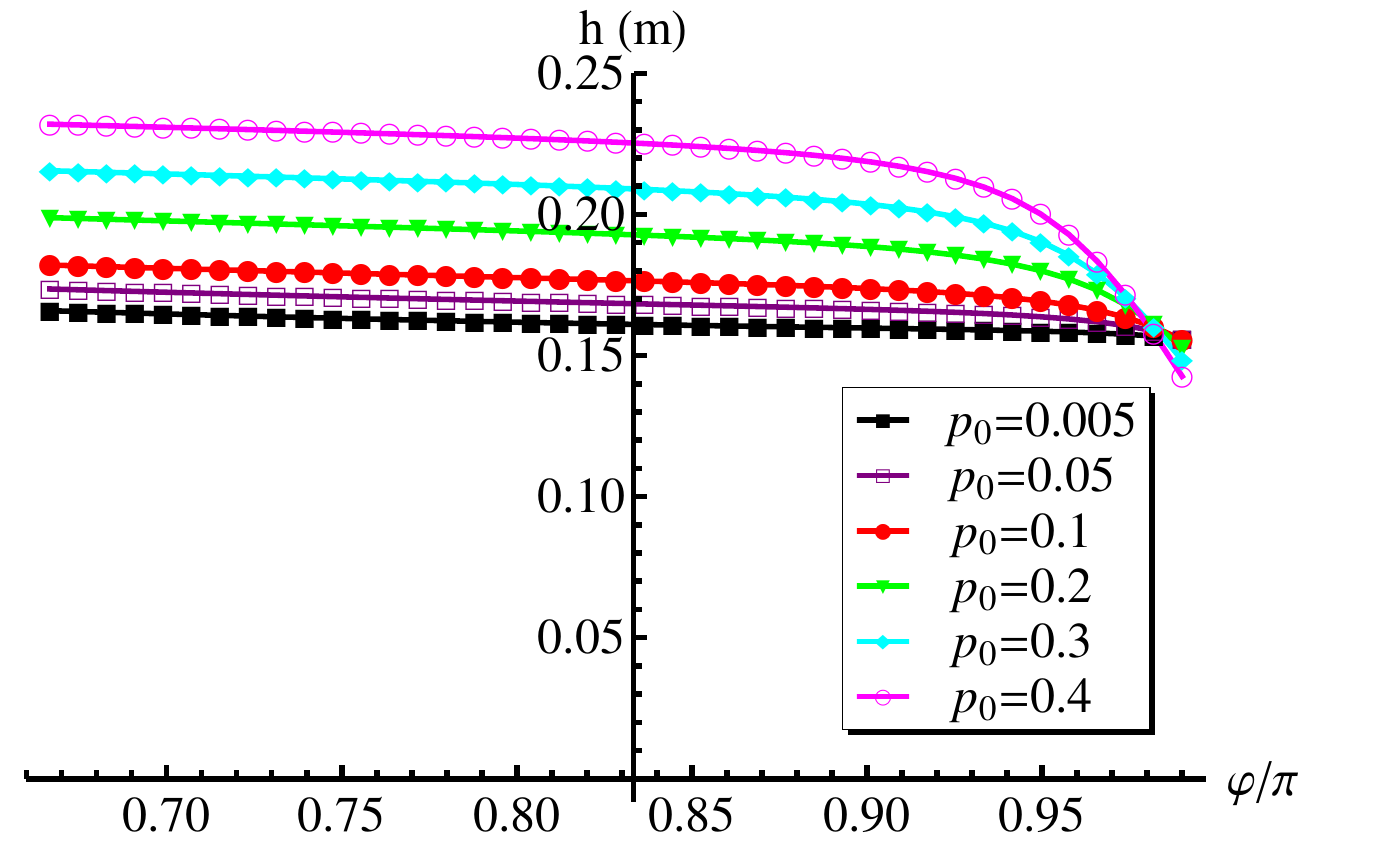,width=7.5cm}}\fi
\if\Images y\put(0.25,5){\psfig{figure=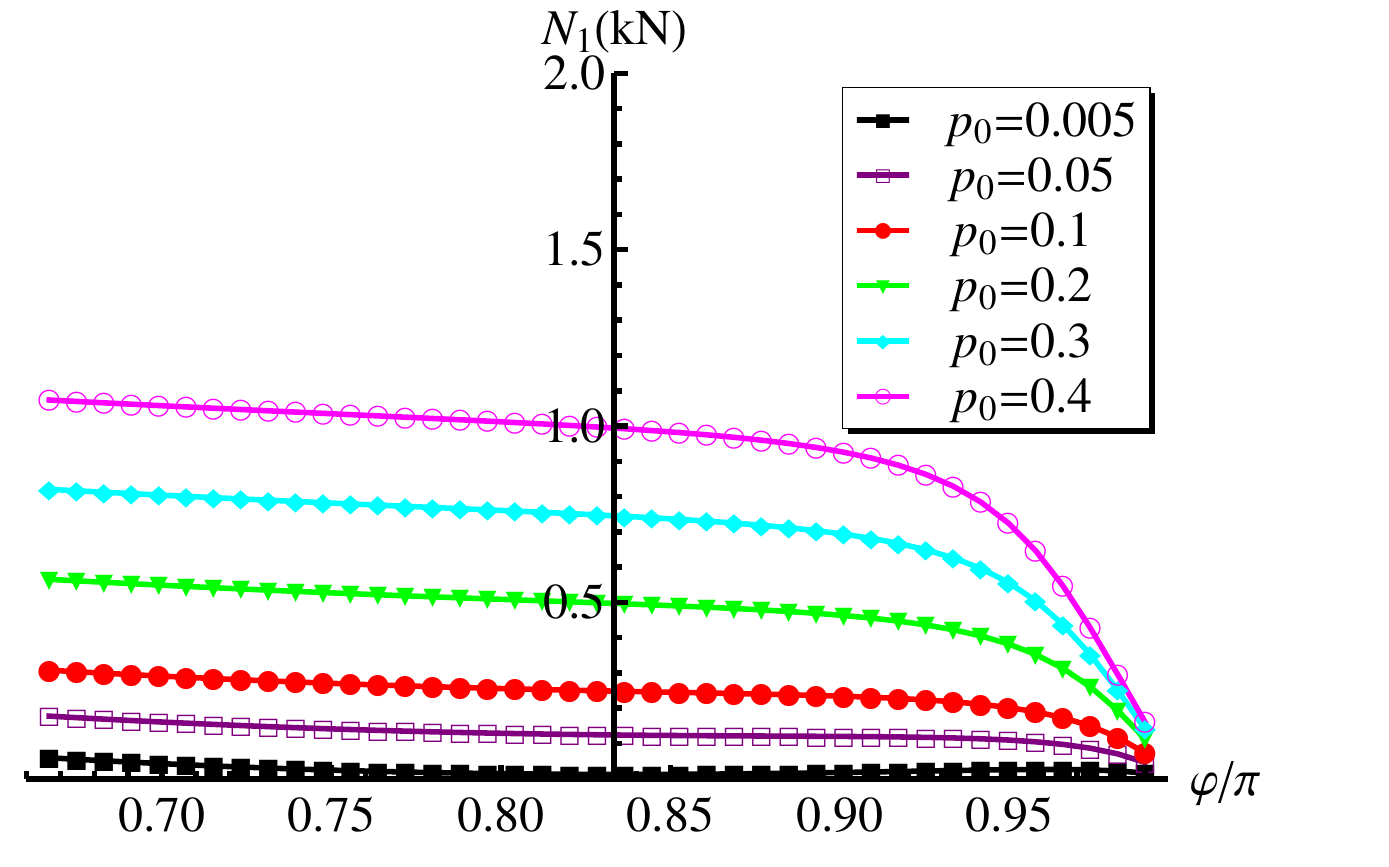,width=7.5cm}}\fi
\if\Images y\put(8.5,5){\psfig{figure=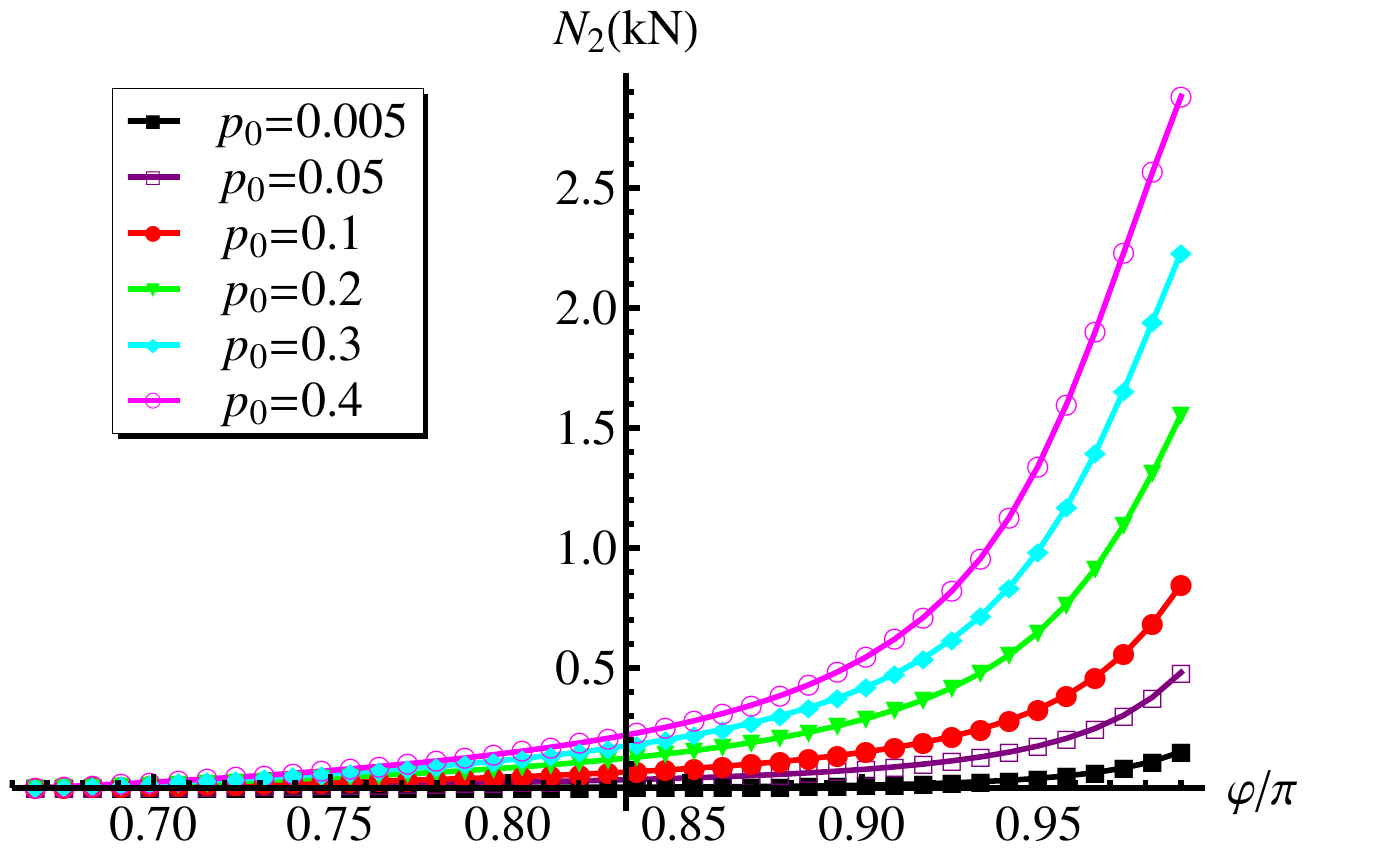,width=7.5cm}}\fi
\if\Images y\put(4.25,0){\psfig{figure=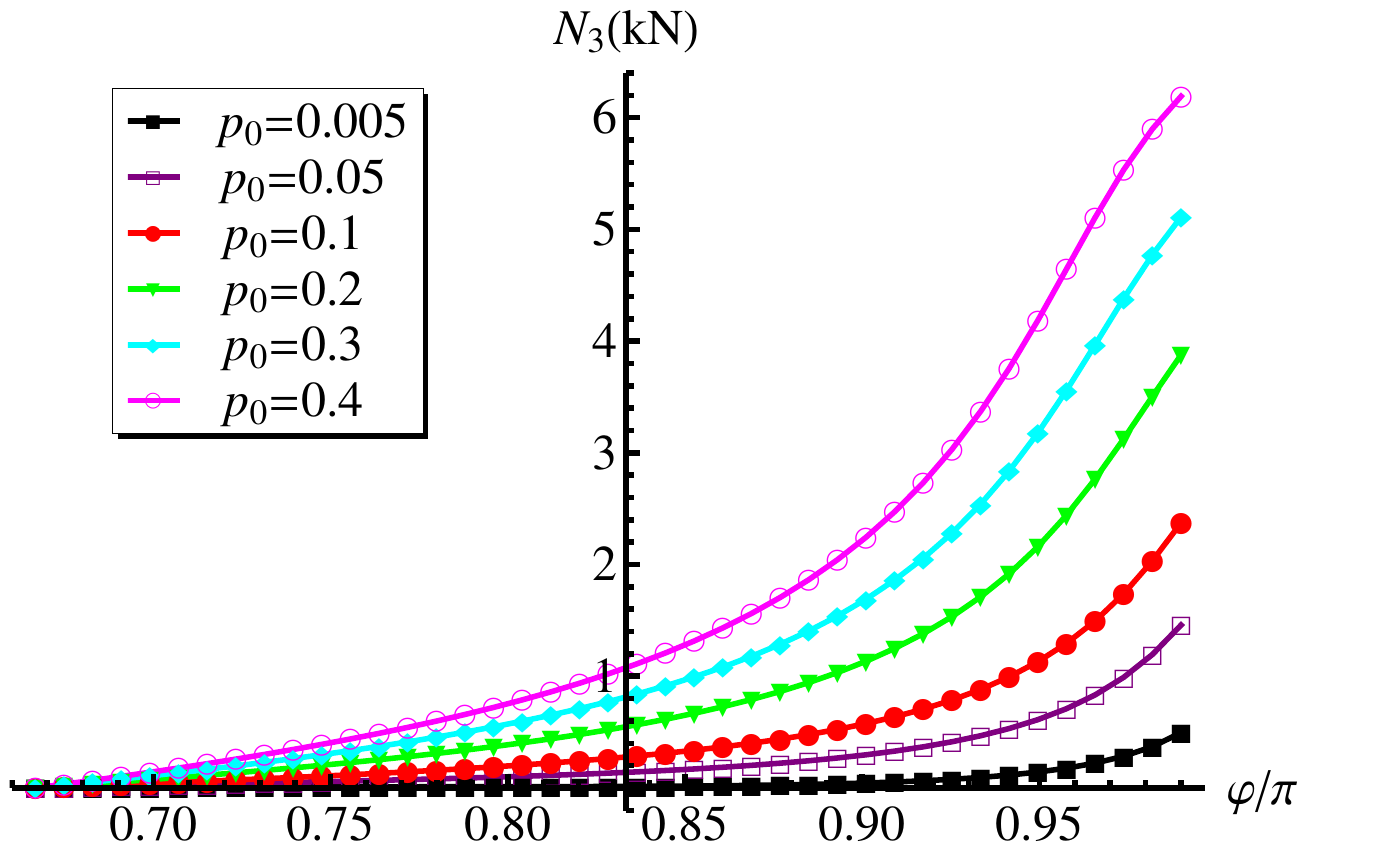,width=7.5cm}}\fi
\end{picture}
\caption{$K_h$ vs. $\varphi$, $h$ vs. $\varphi$ and $N_1,N_2,N_3$ vs. $\varphi$  curves of the slender prism model for $\alpha$ = $\beta$ = 1, and different values of $p_0$.
}
\label{NK-phi-p0-slender-panel}
\end{figure}

\begin{figure}[!hbt]
\unitlength1cm
\begin{picture}(11,21)
\if\Images y\put(0,16.5){\psfig{figure=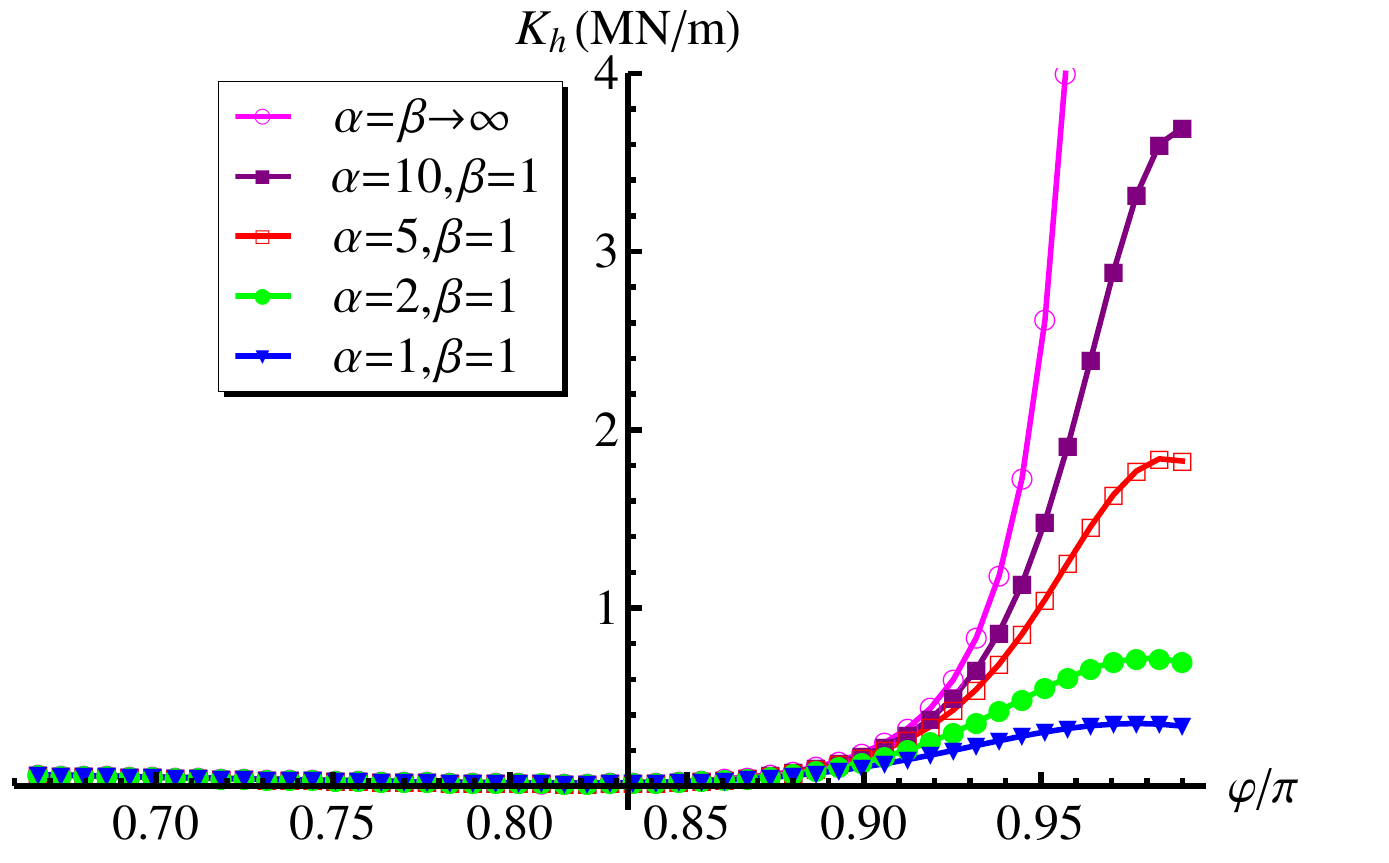,width=8.5cm}}\fi
\if\Images y\put(8,16.5){\psfig{figure=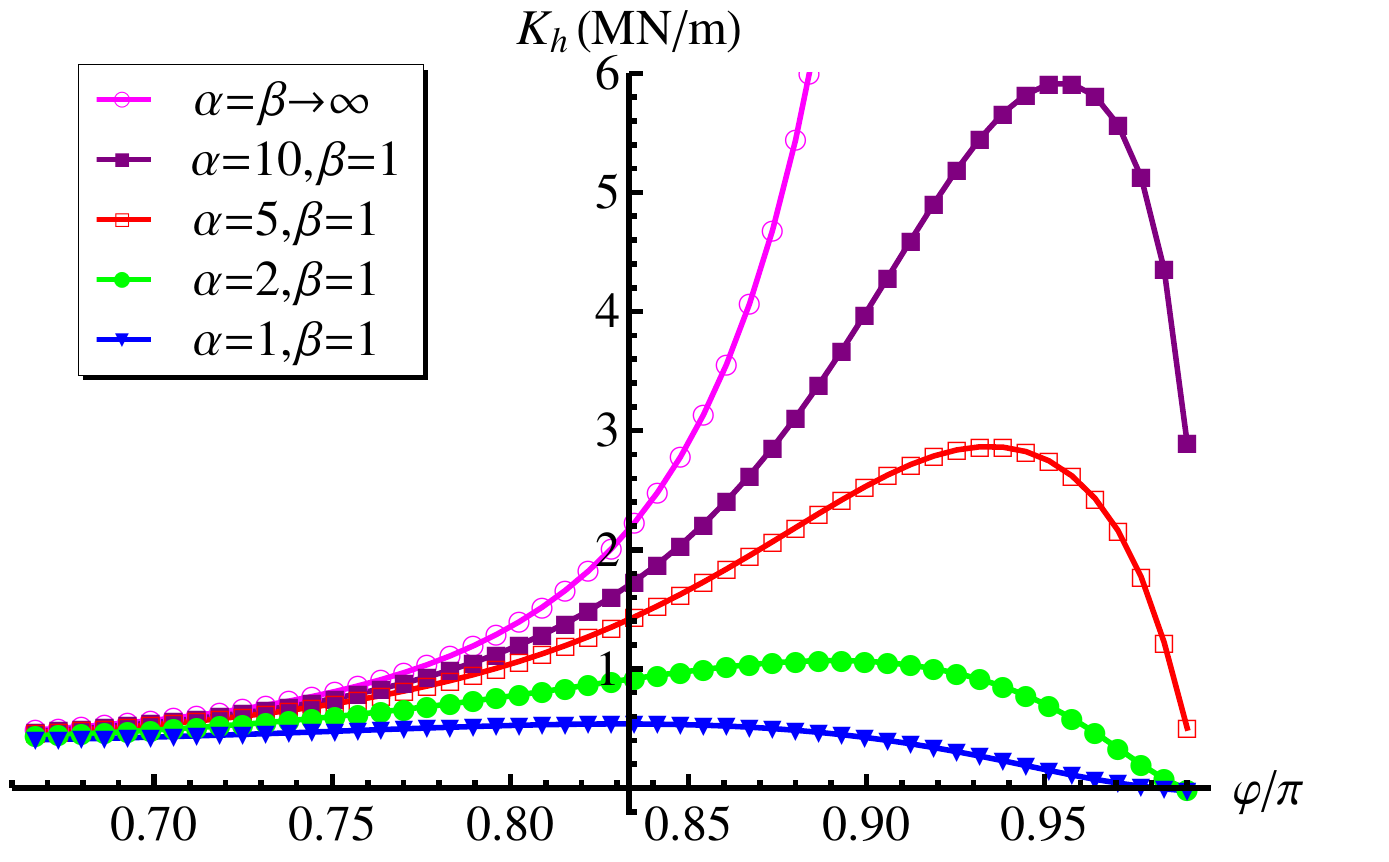,width=8.5cm}}\fi
\if\Images y\put(0.1,11){\psfig{figure=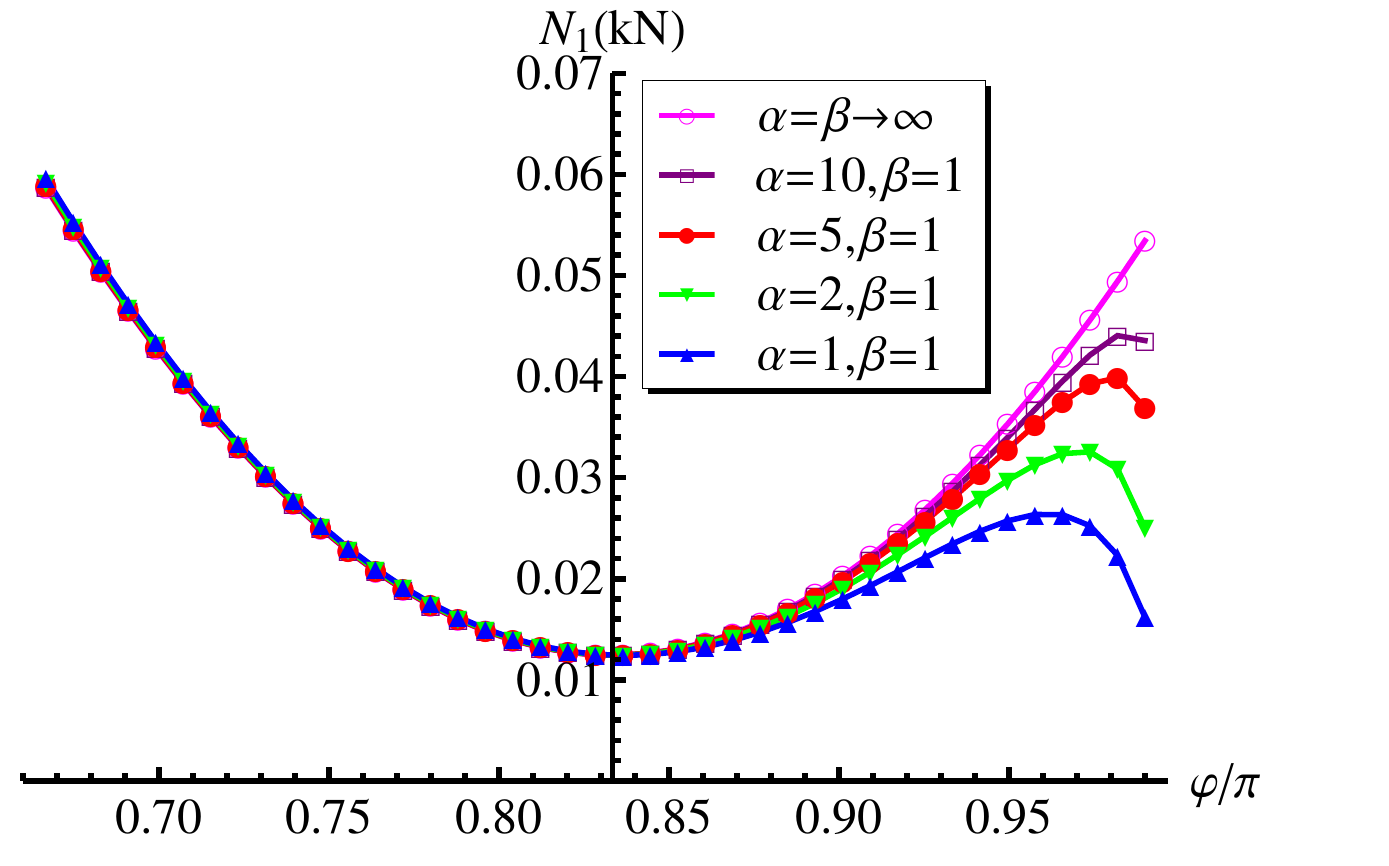,width=8.5cm}}\fi
\if\Images y\put(8,11){\psfig{figure=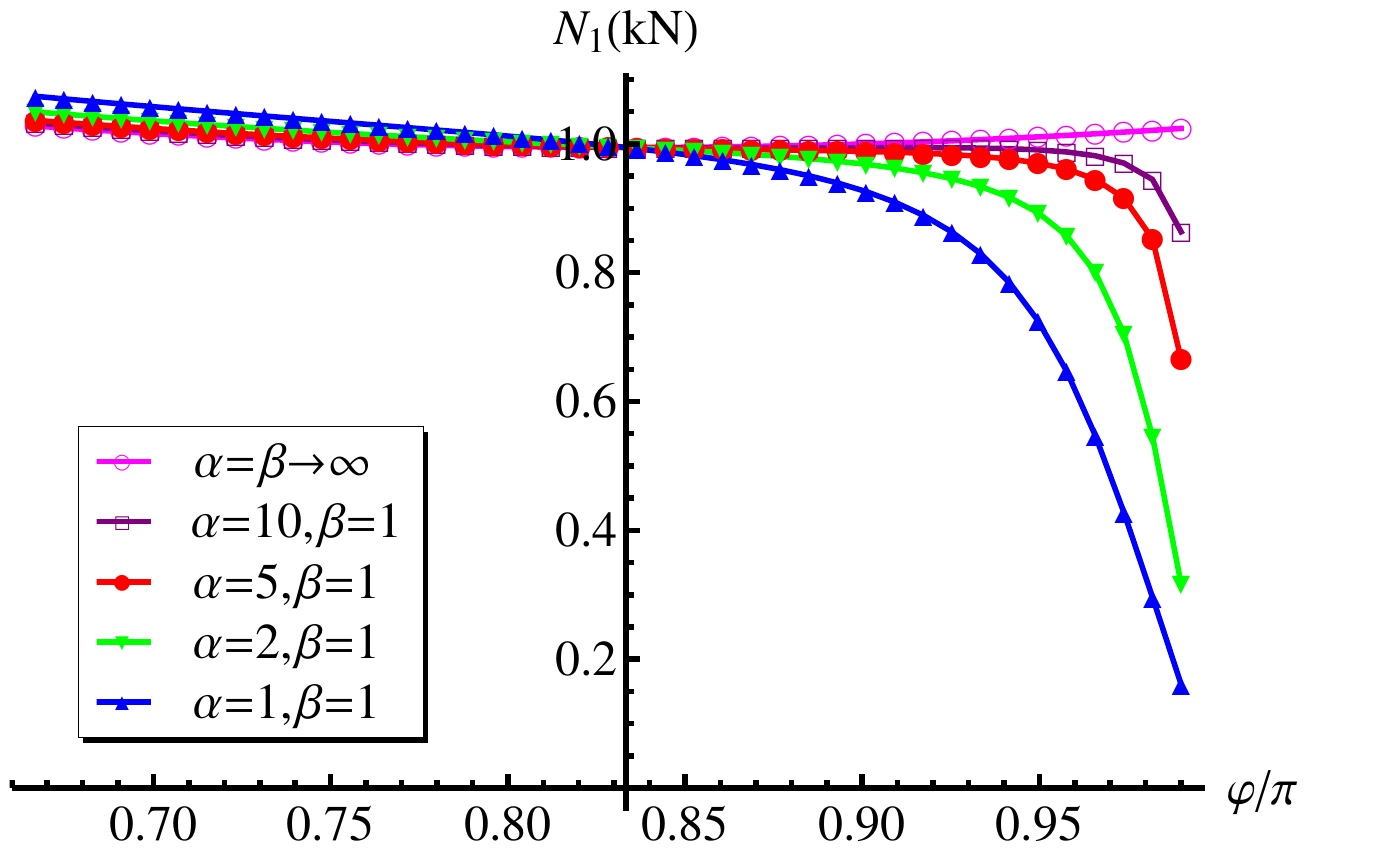,width=8.5cm}}\fi
\if\Images y\put(0,5.5){\psfig{figure=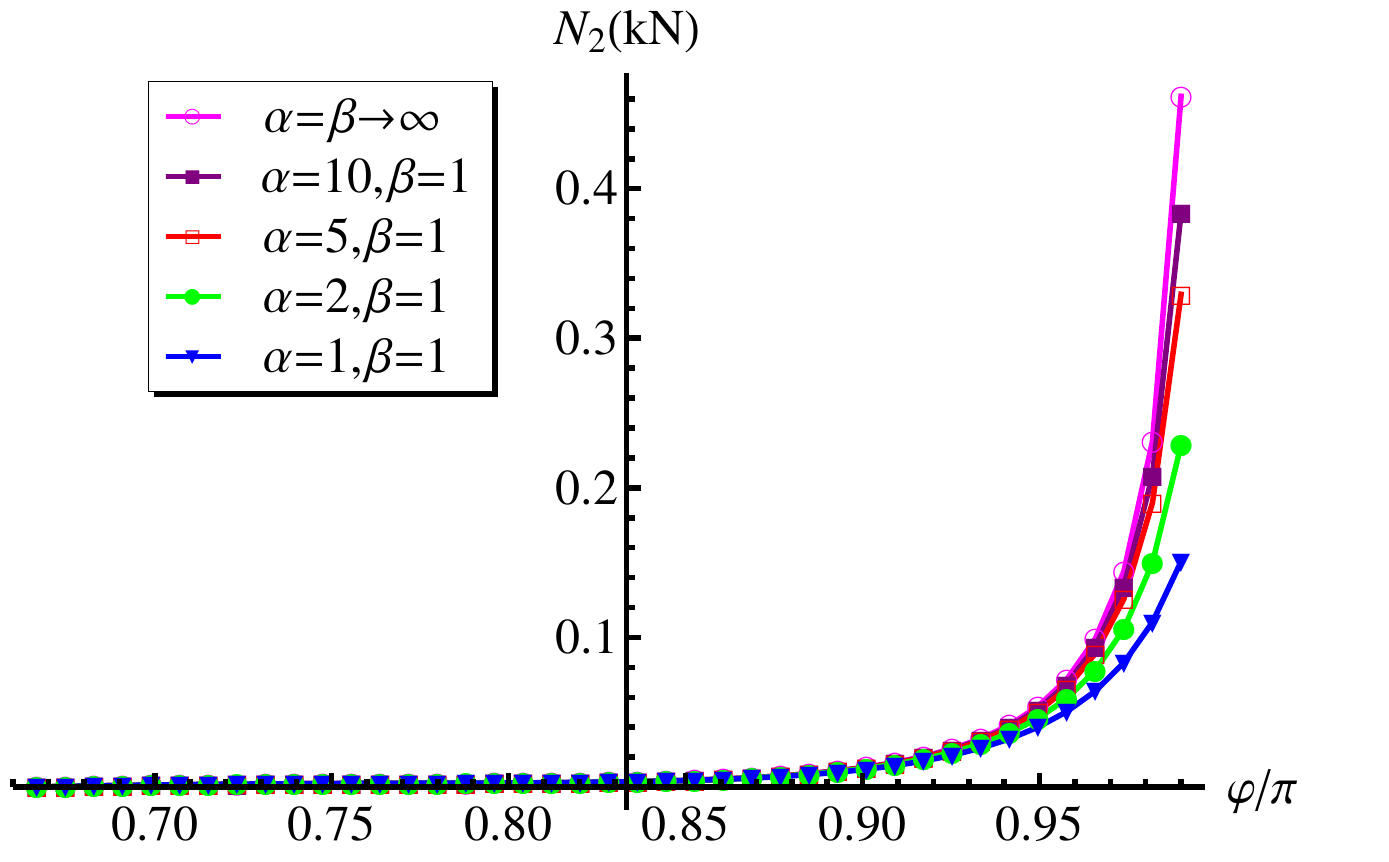,width=8.5cm}}\fi
\if\Images y\put(8,5.5){\psfig{figure=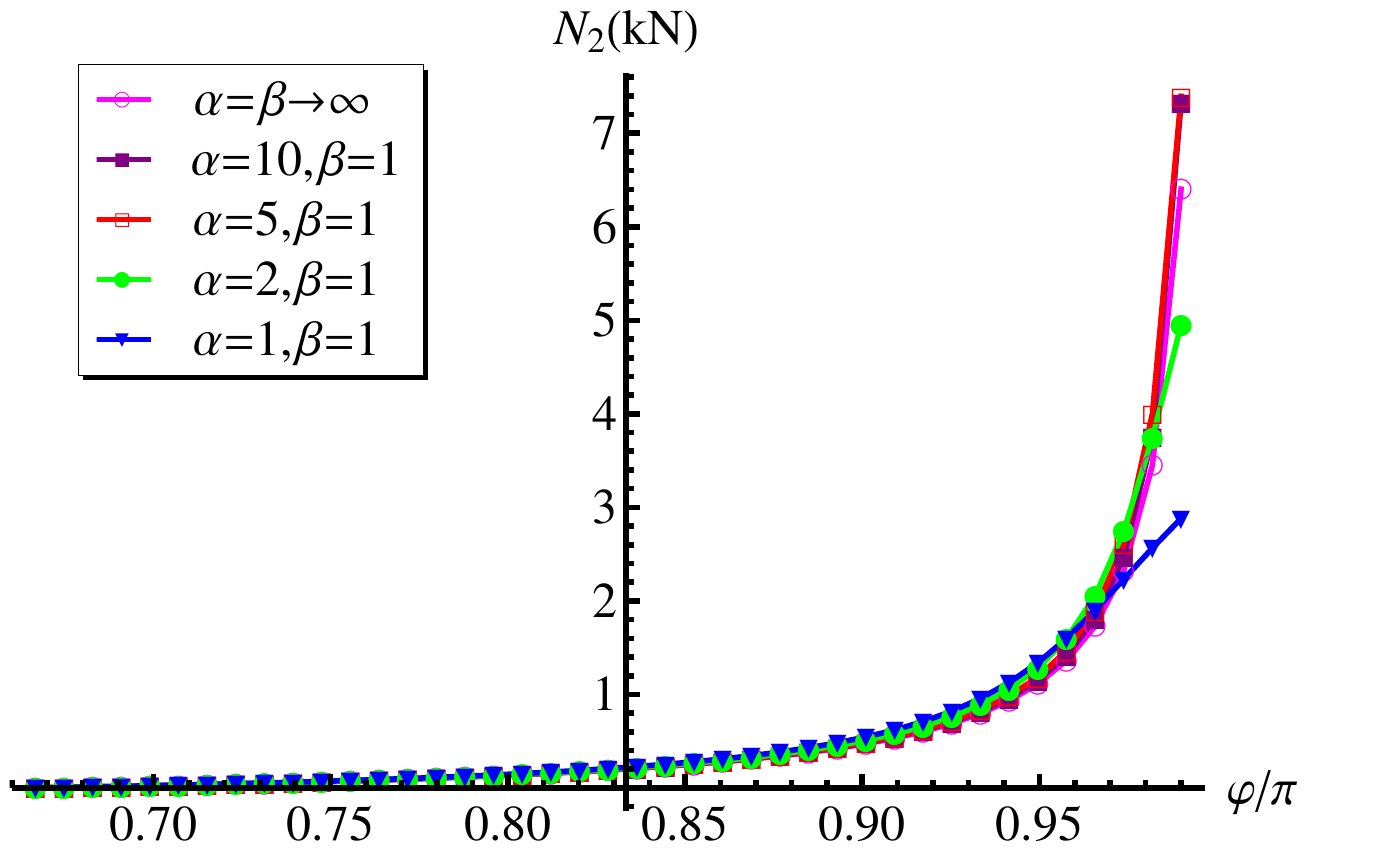,width=8.5cm}}\fi
\if\Images y\put(0,0){\psfig{figure=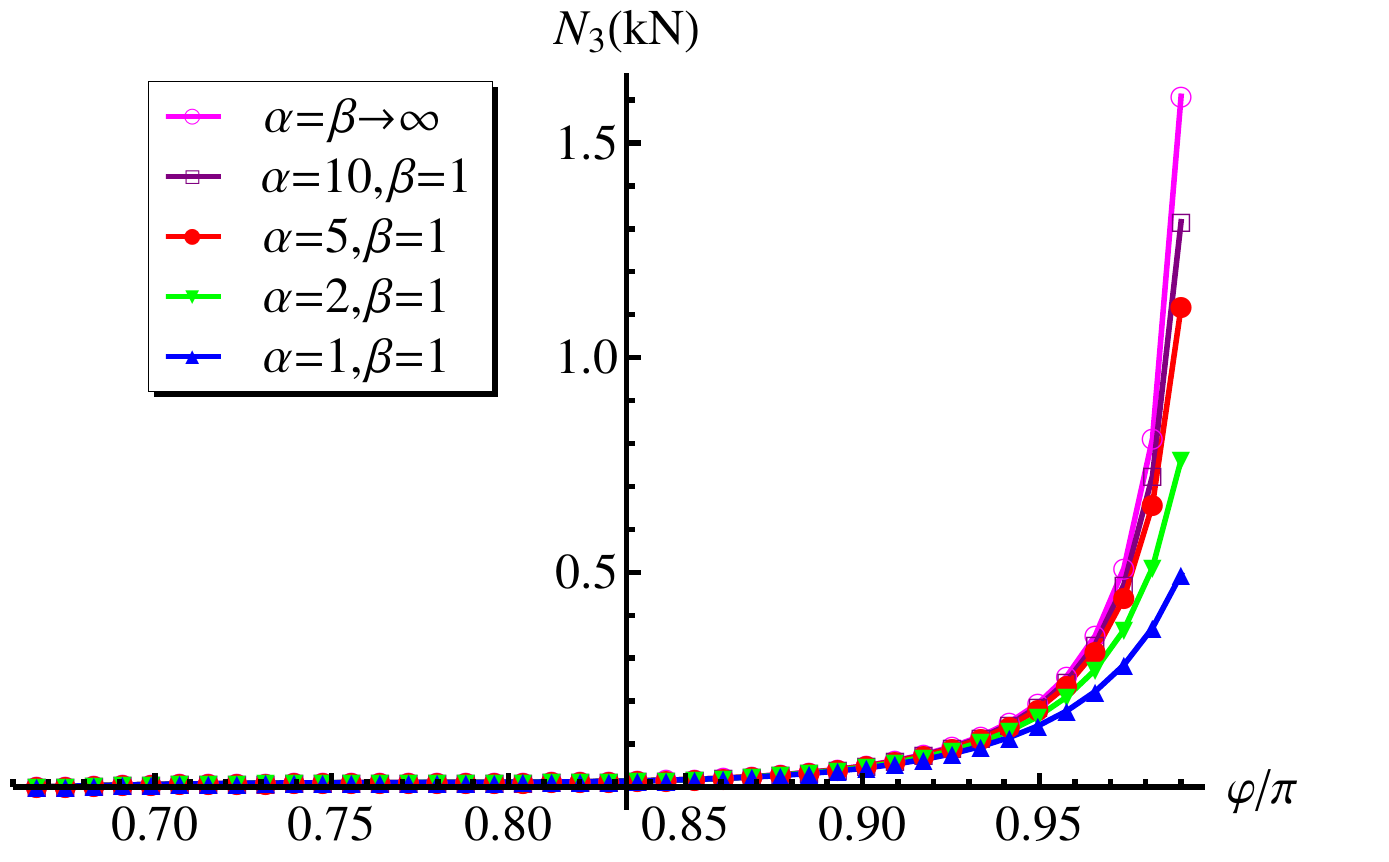,width=8.5cm}}\fi
\if\Images y\put(8,0){\psfig{figure=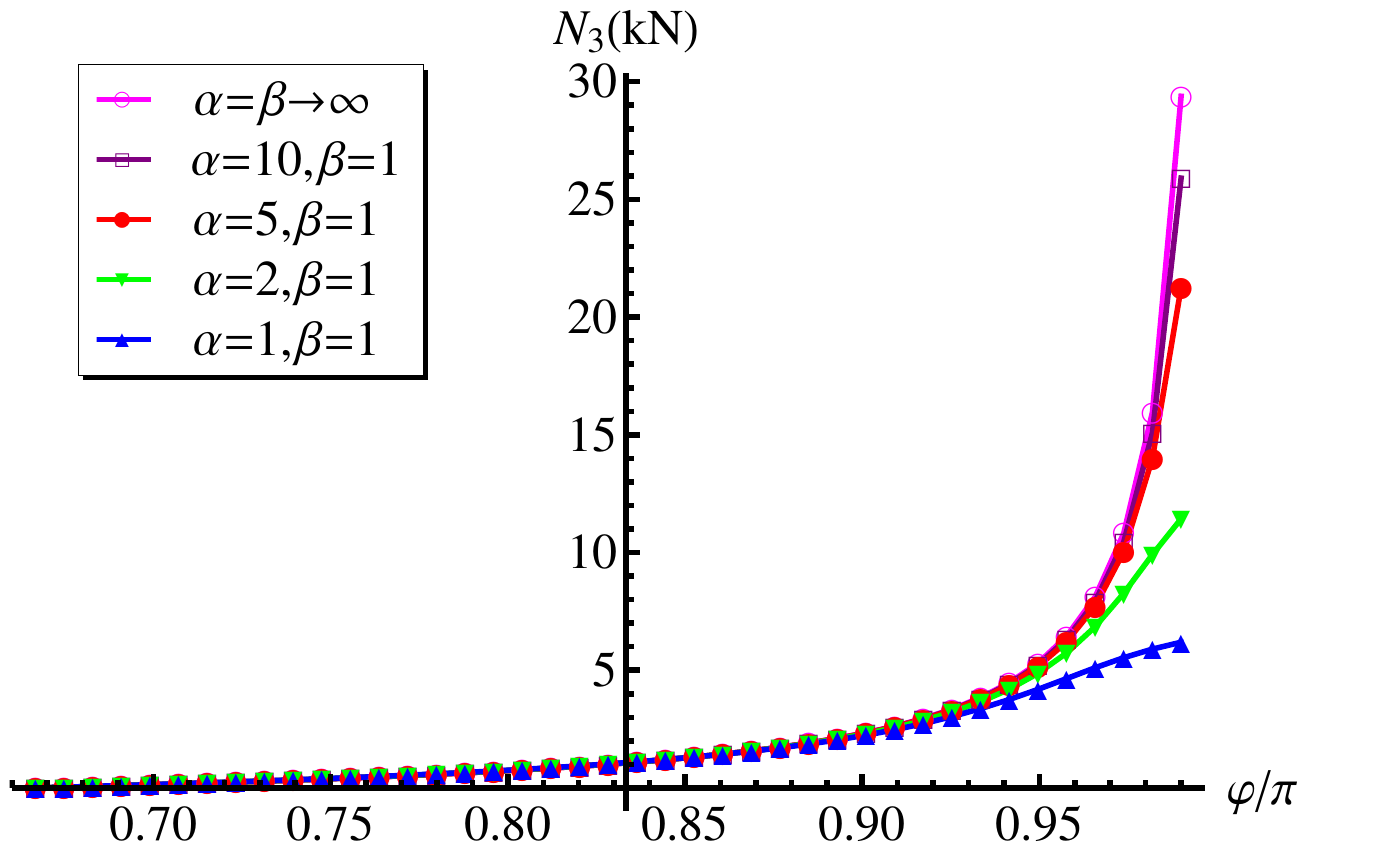,width=8.5cm}}\fi
\end{picture}
\caption{$K_h$ vs. $\varphi$ and $N_1,N_2,N_3$ vs. $\varphi$ curves of the slender prism model for $p_0$ = 0.005 (left), $p_0$ = 0.4 (right) and different values of $\alpha$ and $\beta$.
}
\label{NK-phi-ab-slender-panel}
\end{figure}

\begin{figure}[hbt]
\unitlength1cm
\begin{picture}(15,15)
\if\Images y\put(1.5,-0.5){\psfig{figure=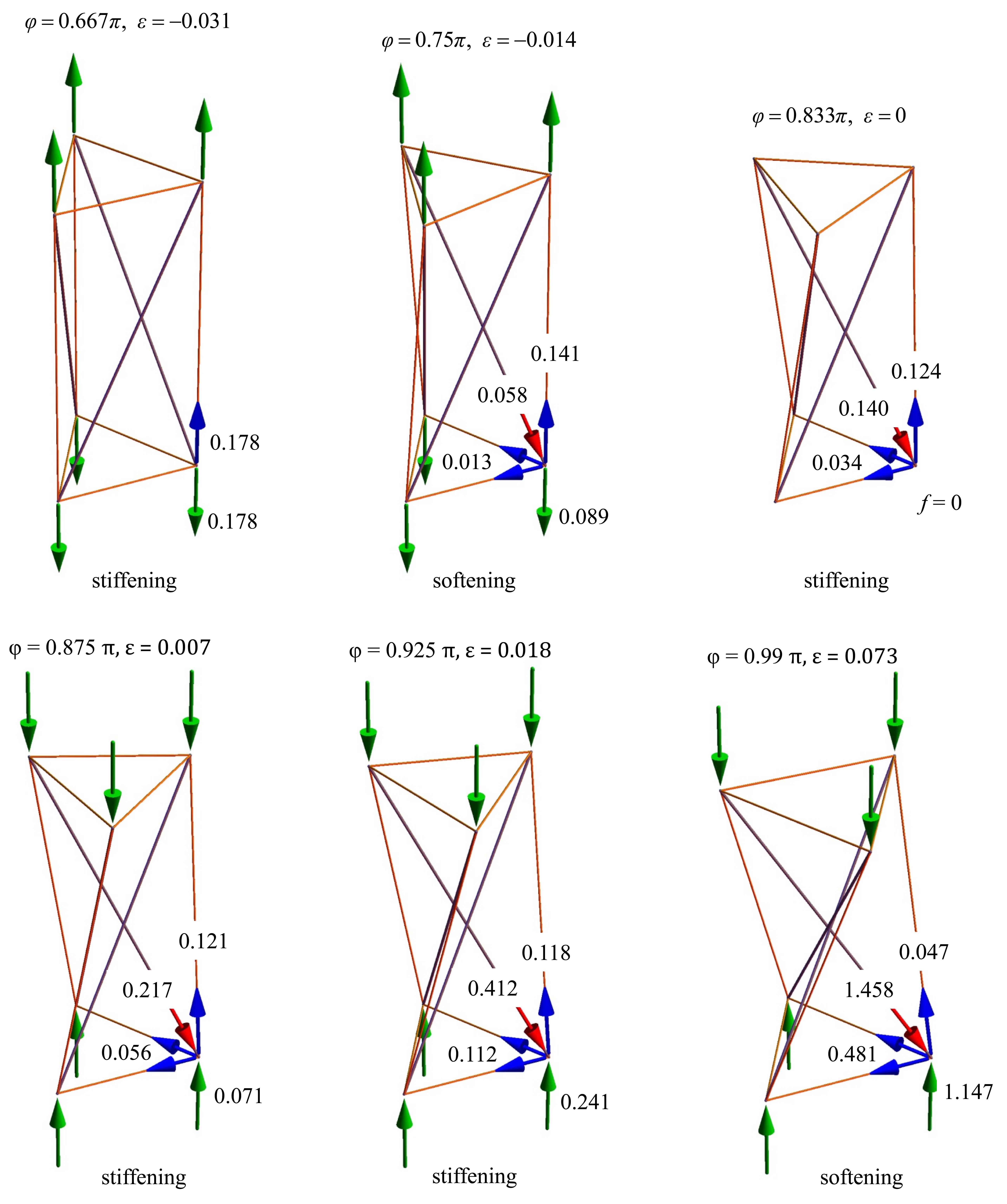,height=16cm}}\fi
\end{picture}
\caption{{\cFF{Member forces (kN) in different configurations of the slender prism, for $\alpha$ = $\beta$ = 1, and $p_0$ = 0.05.}}}
\label{Force_slender1}
\end{figure}

\begin{figure}[hbt]
\unitlength1cm
\begin{picture}(15,15)
\if\Images y\put(1.5,-0.5){\psfig{figure=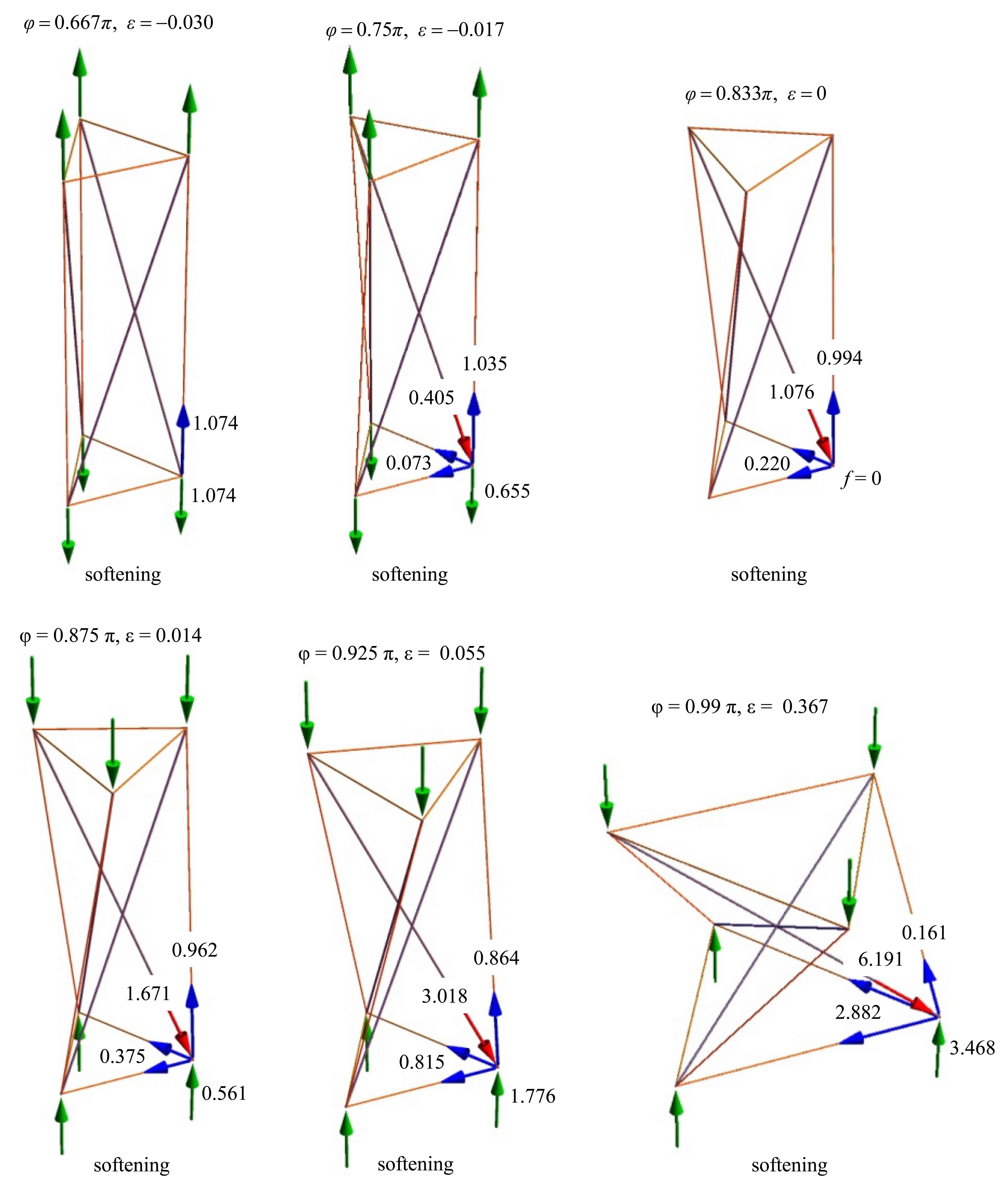,height=16cm}}\fi
\end{picture}
\caption{{\cFF{Member forces (kN) in different configurations of the slender prism, for $\alpha$ = $\beta$ = 1, and $p_0$ = 0.4.}}}
\label{Force_slender2}
\end{figure}

The `slender' prisms analyzed in the present study feature: $s_N = 0.162$ m, $\ell_N = 0.08$ m, and equilibrium height $h_0$ about twice the base side $\ell_0$ (cf. Table \ref{Sys2geom}).
{\cFF{
Figs. \ref{p-variab-system2} and \ref{Fr_rigidezza_variab_2}  show the force $F$ vs. $\delta$ curves of such prisms for different values of $p_0$, $\alpha$ and $\beta$, while  Figs. \ref{NK-phi-p0-slender-panel} and \ref{NK-phi-ab-slender-panel} provide the curves relating the axial stiffness $K_h$, the prism height $h$, and the axial forces $N_1$, $N_2$, $N_3$ with the angle of twist $\varphi$. Some snapshots of the deformation of the slender prism for $\varphi \in [2/3 \pi, \pi]$; $\alpha$ = $\beta$ = 1; $p_0$ = 0.05; and $p_0$ = 0.4 are illustrated in Figs. \ref{Force_slender1} and \ref{Force_slender2}.
}}

\begin{table}[htbp]
	\centering
	\scalebox{0.90}{
		\begin{tabular}{| c | c | c | c | c | c | c | c | c | c | c |}
    \hline
$\alpha$=$\beta$ & $p_0$ & $s_N$ (m) & $s_0$ (m) & $\ell_N$ (m) & $\ell_0$ (m) & $b_N$ (m) & $b_0$ (m) & $h_0$ (m) & $K_{h_0}$ (N/m) \\ \hline
1 & 0 & 0.1620 & 0.1620 & 0.080 & 0.0800 & 0.1834 &  0.1834 & 0.1602 & 0.0 \\ \hline
1 & 0.005 & 0.1620 & 0.1628 & 0.080 & 0.0801 & 0.1842 &  0.1842 & 0.1610 & 18552 \\ \hline
1 & 0.02 & 0.1620 & 0.1652 & 0.080 & 0.0804 & 0.1865 &  0.1865 & 0.1635 & 67597 \\ \hline
1 & 0.05 & 0.1620 & 0.1701 & 0.080 & 0.0811 & 0.1911 &  0.1911 & 0.1684 & 144450 \\ \hline
1 & 0.1 & 0.1620 & 0.1782 & 0.080 & 0.0821 & 0.1989 &  0.1989 & 0.1765 & 236357 \\ \hline
1 & 0.2 & 0.1620 & 0.1944 & 0.080 & 0.0840 & 0.2143 &  0.2143 & 0.1928 & 360901 \\ \hline
1 & 0.3 & 0.1620 & 0.2106 & 0.080 & 0.0856 & 0.2299 &  0.2299 & 0.2090 & 454877 \\ \hline
1 & 0.4 & 0.1620 & 0.2268 & 0.080 & 0.0871 & 0.2454 &  0.2454 & 0.2253 & 537673 \\ \hline
$\rightarrow \infty$ & 0 & 0.1620 & 0.1620 & 0.080 & 0.080 & 0.1834 &  0.1834 & 0.1602 & 0.0 \\ \hline
$\rightarrow \infty$ & 0.005 & 0.1620 & 0.1628 & 0.080 & 0.080 & 0.1841 &  0.1841 & 0.1610 & 19235 \\ \hline
$\rightarrow \infty$ & 0.02 & 0.1620 & 0.1652 & 0.080 & 0.080 & 0.1863 &  0.1863 & 0.1635 & 77745 \\ \hline
$\rightarrow \infty$ & 0.05 & 0.1620 & 0.1701 & 0.080 & 0.080 & 0.1906 &  0.1906 & 0.1684 & 196369 \\ \hline
$\rightarrow \infty$ & 0.1 & 0.1620 & 0.1782 & 0.080 & 0.080 & 0.1979 &  0.1979 & 0.1766 & 401754 \\ \hline
$\rightarrow \infty$ & 0.2 & 0.1620 & 0.1944 & 0.080 & 0.080 & 0.2126 &  0.2126 & 0.1929 & 840075 \\ \hline
$\rightarrow \infty$ & 0.3 & 0.1620 & 0.2106 & 0.080 & 0.080 & 0.2275 &  0.2275 & 0.2092 & 1315781 \\ \hline
$\rightarrow \infty$ & 0.4 & 0.1620 & 0.2268 & 0.080 & 0.080 & 0.2425 &  0.2425 & 0.2255 & 1829454 \\ \hline
		\end{tabular}	
		}	
\caption{Geometric variables and initial axial stiffness $K_{h_0}$ of the slender prism model for different values of the cross-string prestrain $p_0$; the fully elastic model ($\alpha=\beta=1$); and the rigid--elastic model ($\alpha=\beta \rightarrow +\infty$).
}	
	\label{Sys2geom}
\end{table}

\newpage

{\cFF{In the `el' model with $p_0 \le 0.3$, we observe that the compressive response first shows a stiffening branch, and next a softening branch (cf.
Figs. \ref{p-variab-system2}, \ref{NK-phi-p0-slender-panel}, and \ref{Force_slender1}). 
For $p_0 = 0.4$,  the compressive branch of the $F$ vs. $\delta$ (or $F$ vs. $\varphi$) response is instead always softening, and terminates with an unstable phase (Figs. \ref{p-variab-system2}, \ref{NK-phi-p0-slender-panel}, \ref{Force_slender2}). 
Figs. \ref{p-variab-system2} and \ref{NK-phi-p0-slender-panel} show that the tensile response of slender prisms is slightly softening for $p_0 \ge 0.1$.
In contrast, for $p_0 \le 0.05$ the same response is instead slightly stiffening.
It is worth noting that the above behaviors are markedly different from those exhibited by the thick prisms analyzed in Section \ref{thick prism}, since the latter feature unstable response in compression under low prestrains $p_0$, and always stiffening response in tension (Figs. \ref{p-variab-system1}, \ref{NK-phi-p0-thick-panel}, \ref{Force_thick}).
We now pass to examining the axial response of slender prisms for different values of the stiffness multipliers $\alpha$ and $\beta$, and $p_0=0.1$.
Fig. \ref{Fr_rigidezza_variab_2} shows that the compressive response for $p_0=0.1$ is almost linear when $\delta \rightarrow \delta^{max}$ in the `el' case, and tends to get infinitely stiff in the `rigel' limit. The tensile response is instead less sensitive to $\alpha$ and $\beta$, and always softening (Fig. \ref{Fr_rigidezza_variab_2}).
The individual responses of the prism members highlight that the softening response in compression is always associated with decreasing values of the force carried by the cross-strings (cf. Figs. \ref{NK-phi-p0-slender-panel} and \ref{NK-phi-ab-slender-panel}), as in the case of the thick prisms examined in the previous section. 
The deformations graphically illustrated in Fig. \ref{Force_slender2} highlight a marked stretching of the base-strings, in proximity to the locking configuration $\varphi = \pi$, when there results $\alpha$ = $\beta$ = 1, and $p_0$ = 0.4.
}}

\newpage

\newpage

\section{Experimental validation} \label{experiment}

\begin{figure}[hbt]
\unitlength1cm
\begin{picture}(11,6.5)
\if\Images y\put(3,-0.5){\psfig{figure=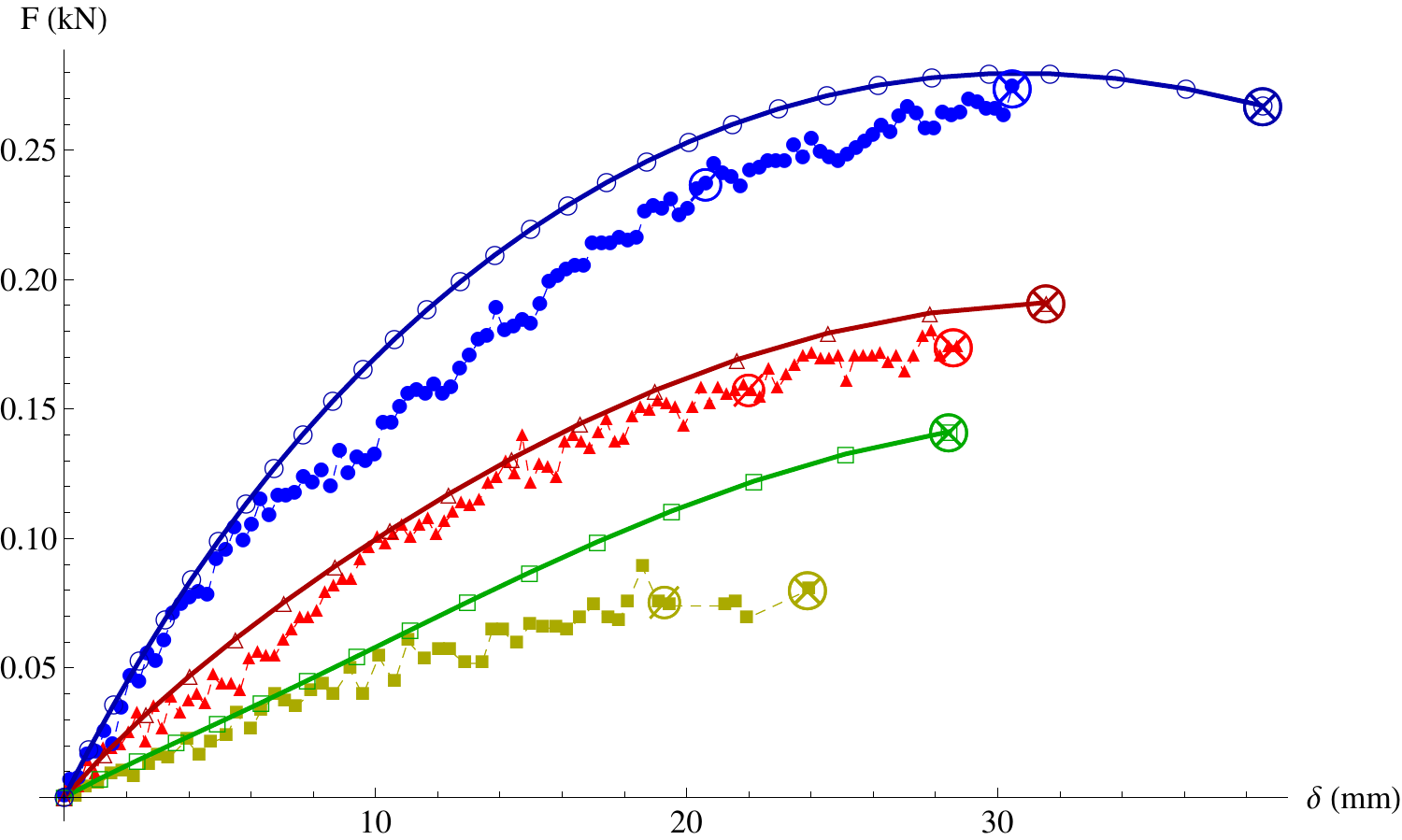,width=9.5cm}}\fi
\if\Images y\put(4.4,-2.8){\psfig{figure=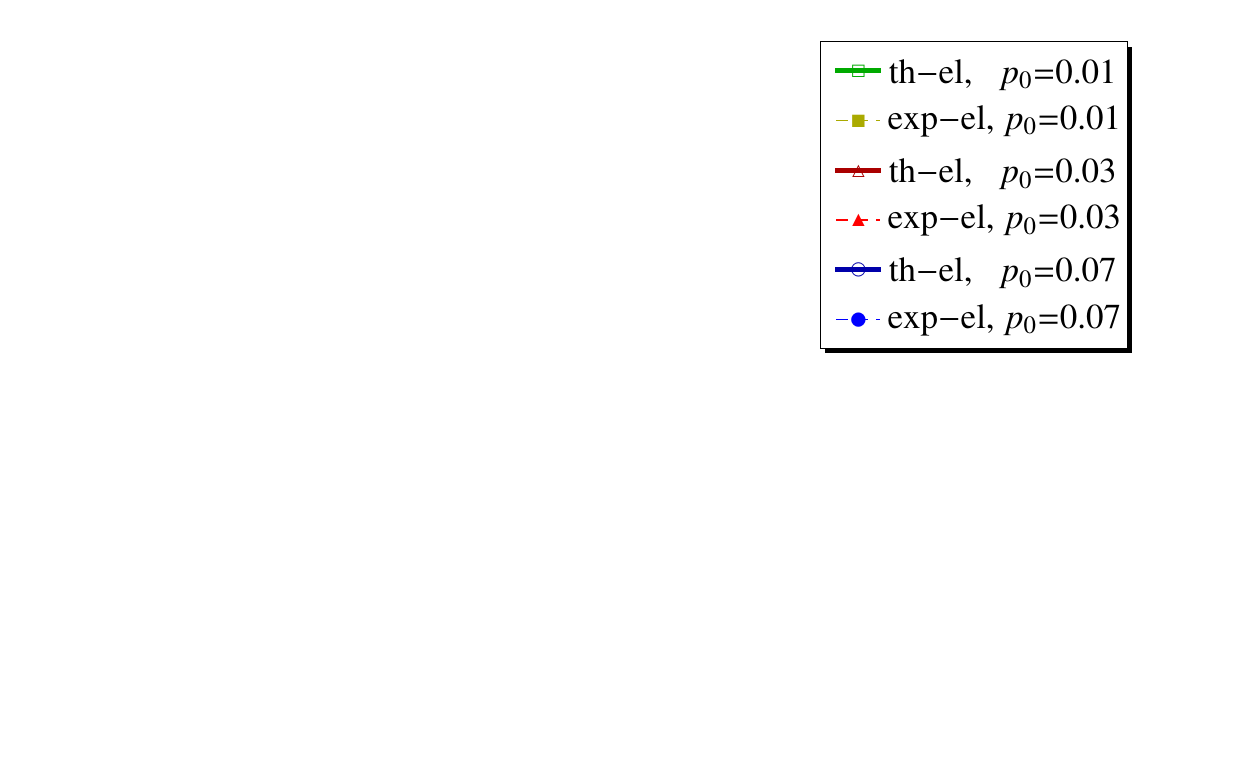,width=8cm}}\fi\end{picture}
\caption{Comparison of the theoretical and experimental responses of thick prisms with deformable bases.}
\label{Exp1}
\end{figure}

{\cFF{
The present section deals with an experimental validation of the models presented in Sections \ref{model} and \ref{results}, against the results of quasi-static compression tests on physical prism samples \citep{prot} (cf. also Section \ref{results}).
We first examine the experimental responses of the thick prism specimens described in Table \ref{db1_valori}, 
where $N_1^{(0)}$ denotes the axial force carried by the cross-strings in correspondence with the reference configuration.
Fig. \ref{Exp1} compares the theoretical ({`th-el'}) and experimental  (`exp-el') $F-\delta$ responses of such specimens, highlighting an overall good agreement between theory and experiments. We note a more compliant character of the experimental responses, as compared to those predicted by the fully-elastic model presented in Section \ref{model}, and oscillations of the experimental measurements.
Such theory vs. experiment mismatches are explained by signal noise; progressive damage to the nodes during loading; string damage due to the rubbing of Spectra\textsuperscript{\textregistered} fibers against the rivets placed at the nodes; and geometric imperfections (refer to \cite{prot} for detailed descriptions of such phenomena). 
In particular, geometric imperfections arising in the assembly phase prevent the three bars of the current prisms from simultaneously coming into contact with each other when the angle of twist approaches $\pi$.
The marker $\oslash$ in Fig. \ref{Exp1} indicates the first configuration at which two bars touch each other, while the marker $\otimes$ indicates the first configuration with all three bars interfering. It is worth noting that the full locking configuration (`$\otimes$') occurs at an angle of twist $\varphi$ appreciably lower than $\pi$,  due to geometric imperfections and the nonzero thickness of the bars.
Both the theoretical and experimental results shown in Fig. \ref{Exp1} indicate a clear softening character of the compressive response of the examined {thick} prisms.
}}

\begin{table}[htbp]
	\centering
				\begin{tabular}{| c | c | c | c | c | c | c | c |}
    \hline
type & $p_0$  & $s_N$ (m) & $s_0$ (m) &$N_1^{(0)}$(N)&$\ell_N$ (m) & $\ell_0$ (m) & $b_0$ (m) \\ \hline
\textit{el} & 0.01 & 0.080 & 0.081 & 30.9 & 0.132 & 0.134 &  0.165 \\ \hline
\textit{el} & 0.03 & 0.080 & 0.083 & 78.2 & 0.132 & 0.136 &  0.168  \\ \hline
\textit{el} & 0.07 & 0.080 & 0.085 & 170.0 & 0.132 & 0.140 &  0.174  \\ \hline

\end{tabular}
\caption{Geometric and mechanical properties of thick prism samples.}
	\label{}

		\label{db1_valori}
\end{table}

{\cFF{
We now pass to examining the experimental response of the slender prism models described in Table \ref{db2_valori}, which include two samples with deformable bases (`{el}' samples), and two samples aimed at reproducing the rigid--elastic model presented in Section \ref{rigelresp} (`{rigel}' samples). The latter were assembled by replacing the base-strings of the `{el}' systems with 12 mm thick aluminum plates (cf. Fig. \ref{rigel_prisms}, and \cite{prot}).
Fig. \ref{Exp2} illustrates a comparison between the theoretical and experimental responses of the `el' samples, which shows a rather good match between theory and experiment. 
In the present case, we observe reduced signal noise, as compared to the case of thick prisms, and all the bars getting simultaneously in touch at locking.
The main mismatch between the theoretical and experimental responses shown in Fig. \ref{Exp2} consists of an anticipated occurrence of prism locking in the physical models, which has already been observed and discussed in the case of the thick specimens.
It is interesting to note that both the theoretical and the experimental results shown in Fig. \ref{Exp2} indicate a slightly stiffening behavior of the `el' samples with a `slender' aspect ratio. 

The final experimental results presented in Fig. \ref{Exp3} are aimed at validating the rigid--elastic model presented in Section \ref{rigelresp} (`rigel' samples). One observes that the specimens endowed with nearly infinitely rigid bases feature a markedly stiff response in the proximity of the locking configuration, in line with the model presented in \cite{Oppenheim:2000}.
We observe a more compliant character of the experimental $F-\delta$ curves of `rigel' samples, as compared to the theoretical counterparts, which is explained by the not perfectly rigid behavior of the bases and the bars (physical samples), and the partial unthreading of the cross-strings from the lock washers placed at the the nodes \citep{prot}. The latter is induced by large tensile forces in the horizontal strings, when the system gets close to the locking configuration (cf. the theoretical results shown in Fig. \ref{NK-phi-ab-slender-panel}, for $\alpha=\beta \rightarrow \infty$).
%Both in the present and previous experiments, we didn't observe string yielding and/or bar buckling 
%up to the locking configuration.

\begin{table}[htbp]
	\centering
				\begin{tabular}{| c | c | c | c | c | c | c | c |}
    \hline
type & $p_0$ & $s_N$ (m) & $s_0$ (m) &$N_1^{(0)}$(N)&$\ell_N$ (m) & $\ell_0$ (m) & $b_0$ (m) \\ \hline
\textit{el} & 0.07 & 0.162 & 0.173 & 165.9 & 0.080 &0.081 &  0.194 \\ \hline
\textit{el}  & 0.09 & 0.162 & 0.176 & 219.9 &  0.080 & 0.082 &  0.197  \\ \hline
\textit{rigel}  & 0.06 & 0.162 & 0.172 & 150.0 &  0.080 & 0.080 &  0.192 \\ \hline
\textit{rigel}  & 0.11 & 0.162 & 0.181 & 286.0 &  0.080 & 0.080 &  0.200  \\ \hline
		\end{tabular}
\caption{Geometric and mechanical properties of slender prism samples.}
	\label{}

		\label{db2_valori}
\end{table}

\begin{figure}[hbt]
\unitlength1cm
\begin{picture}(13.0,9)	
\if\Images y\put(5,0){\psfig{figure=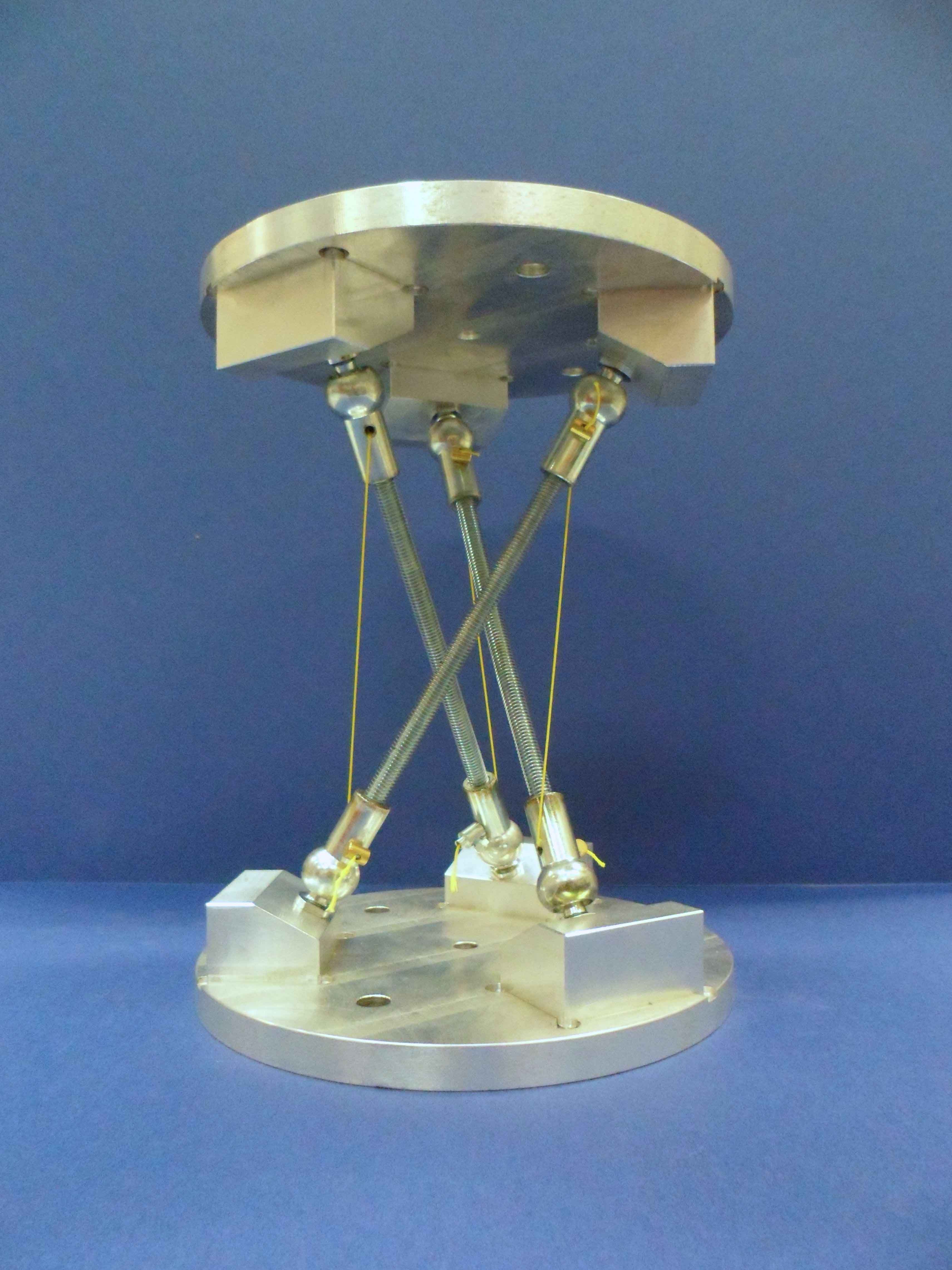,height=9cm}}\fi
\end{picture}
\caption{Photograph of a real-scale example of a slender prism endowed with thick aluminum bases \citep{prot}.}
\label{rigel_prisms}
\end{figure}

\begin{figure}[hbt]
\unitlength1cm
\begin{picture}(11,6.5)
\if\Images y\put(3,-0.5){\psfig{figure=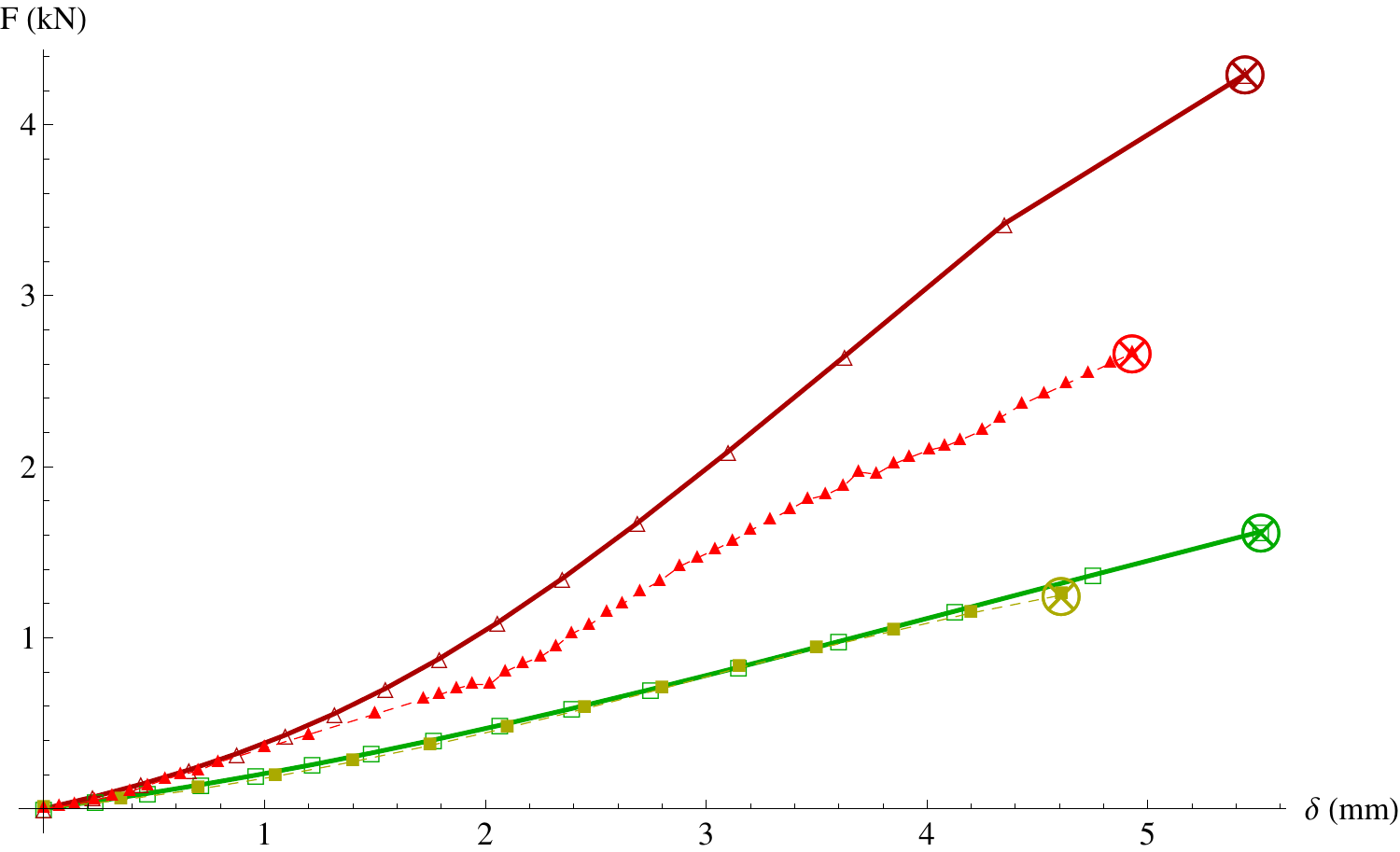,width=9.5cm}}\fi
\if\Images y\put(-1.9,-0.1){\psfig{figure=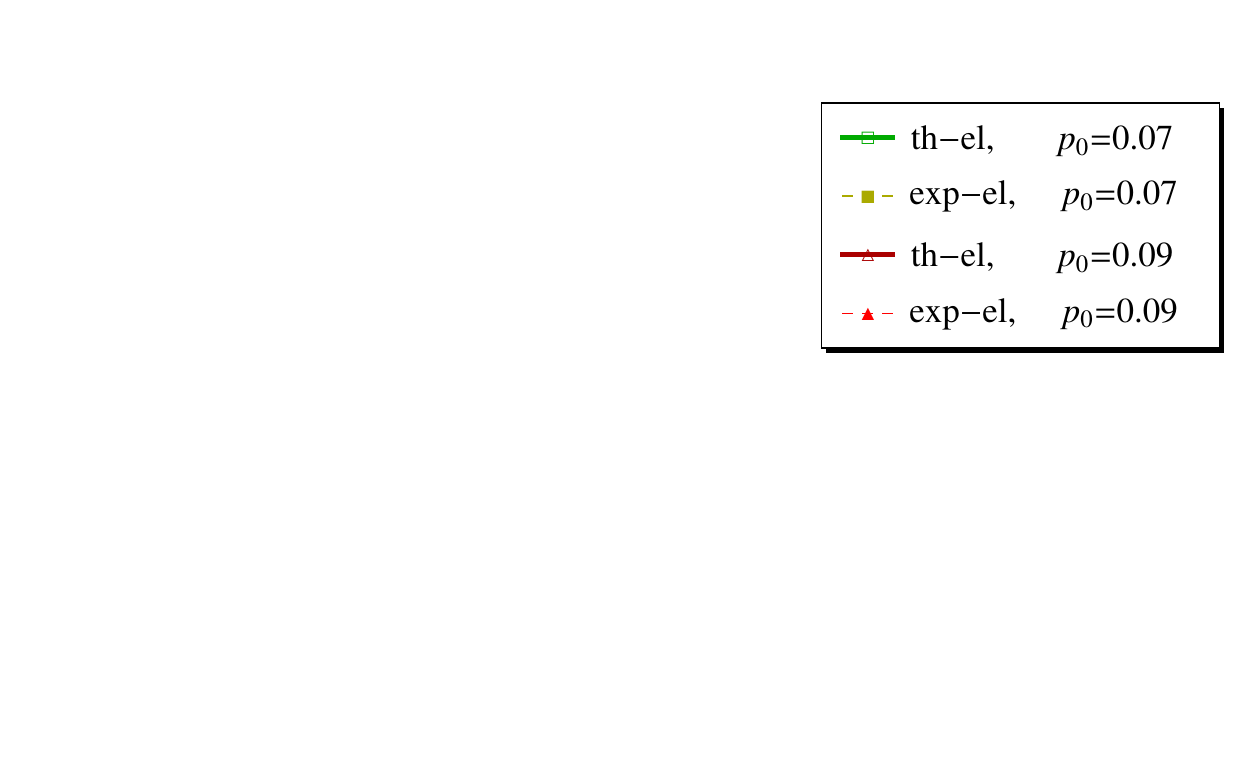,width=8.2cm}}\fi\end{picture}
\caption{Comparison of the theoretical and experimental responses of slender prisms with deformable bases.}
\label{Exp2}
\end{figure}

\begin{figure}[hbt]
\unitlength1cm
\begin{picture}(11,6.5)
\if\Images y\put(3,-0.5){\psfig{figure=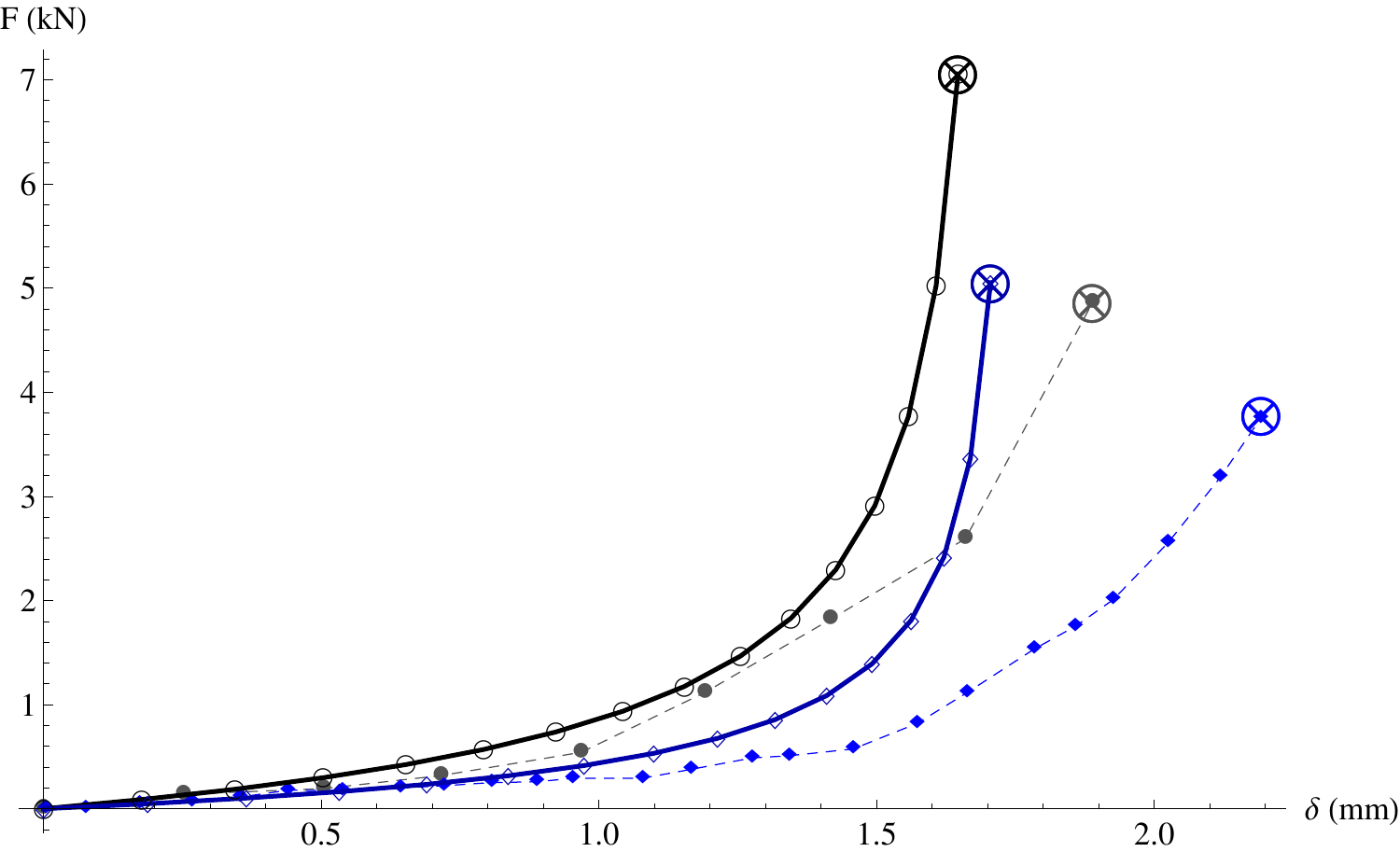,width=9.5cm}}\fi
\if\Images y\put(-1.9,-0.1){\psfig{figure=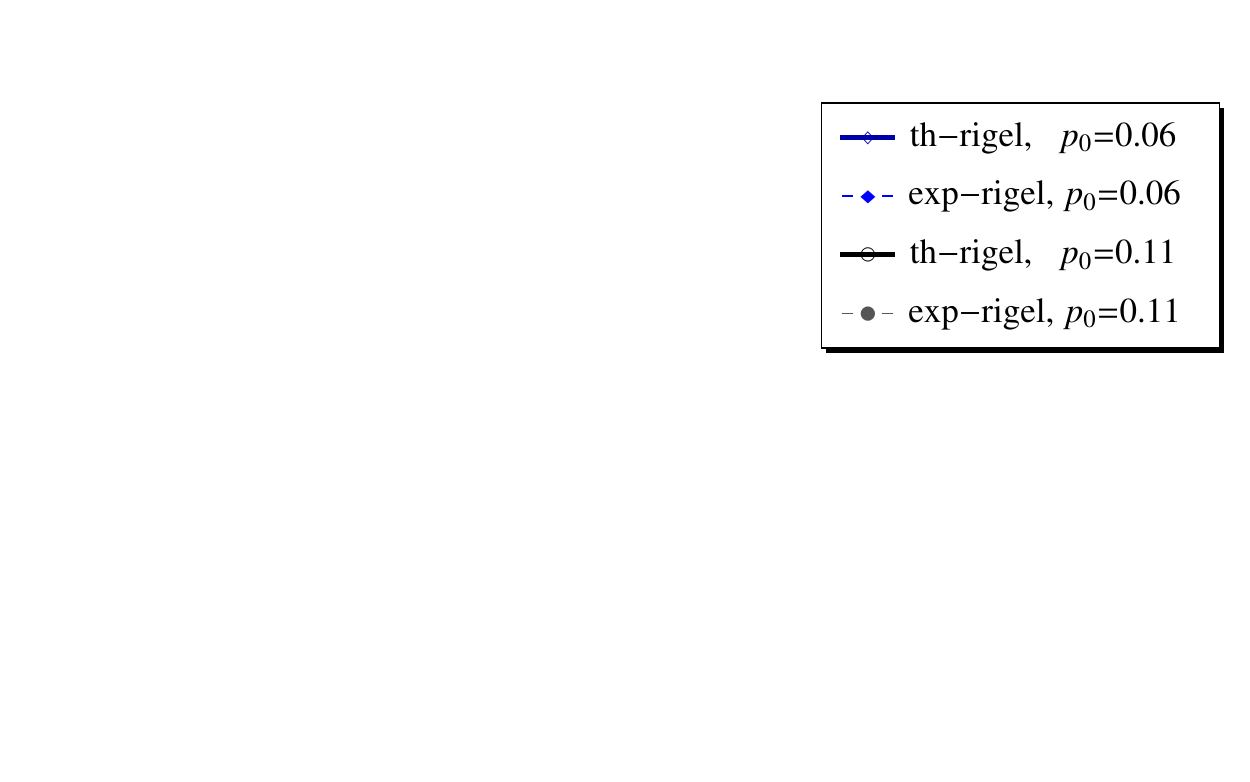,width=8.2cm}}\fi\end{picture}
\caption{Comparison of the theoretical and experimental responses of slender prisms with rigid bases.}
\label{Exp3}
\end{figure}

\section{Concluding remarks} \label{conclusions}

We have presented a fully elastic model of axially loaded tensegrity prisms, which generalizes previous models available in the literature \citep{Oppenheim:2000, FSD12}. 
The mechanical theory presented in  Section \ref{fullelresp} assumes that all the elements of a tensegrity prism respond as  elastic springs, and relaxes the rigidity constraints introduced in \cite{Oppenheim:2000}. On adopting the equilibrium approach to tensegrity systems described in \cite{Skelton2010}, we have written the equilibrium equations in the current configuration, thus developng a geometrical nonlinear model allowing for large displacements (Section \ref{fullelresp}). In addition, we have presented an incremental formulation of the equilibrium problem of axially loaded tensegrity prisms, which is particularly useful when using Netwon's iterative schemes in numerical simulations (Section \ref{rateproblem}).

The numerical results presented in Section \ref{results} highlight a rich variety of behaviors of tensegrity prisms under uniform axial loading and large displacements. The variegate mechanical response of such structures includes both extremely soft and markedly stiff deformation modes, depending on the geometry of the structure, the mechanical properties of the constituent elements, the magnitude of the cross-string prestrain $p_0$ (characterizing the whole state of self-stress), and the loading level (deformation-dependent behavior). We have found that `thick' prisms exhibit  softening response in compression under relatively low prestrains, and, on the contrary, stiffening response in tension over a large window of $p_0$ values (Figs. \ref{p-variab-system1}, \ref{NK-phi-p0-thick-panel}, \ref{Force_thick}). The softening response in compression of such structures is often associated with a snap buckling event, which might lead the prism to axial collapse (prism height tending to zero).
In contrast, we have noted that `slender' prisms need large cable prestrains to show softening response in compression, and relatively low prestrains
in order to feature softening response in tension (Fig. \ref{p-variab-system2}, \ref{NK-phi-p0-slender-panel}, \ref{Force_slender1}, \ref{Force_slender2}). 
By letting the base and bar rigidities tend to infinity, we have numerically observed that the compressive response of thick and slender prisms progressively switches to infinitely stiff in the proximity of the locking configuration  (Figs. \ref{Fr_rigidezza_variab}, \ref{NK-phi-ab-thick-panel}, \ref{Fr_rigidezza_variab_2}, \ref{NK-phi-ab-slender-panel}). In the rigid-elastic limit we have also noted that thick prisms exhibit stiffening response in tension (with the exception of cases characterized by extremely high values of $p_0$, cf. Fig. \ref{NK-phi-ab-thick-panel}), while slender prisms instead typically feature slightly softening response in tension (cf. Figs. \ref{Fr_rigidezza_variab_2} and \ref{NK-phi-ab-slender-panel}).
An experimental validation of the mechanical models presented in Section \ref{model} has been conducted against the results of quasi-static compression tests on physical samples \citep{prot}, with good agreement between theory and experiments.
The given experimental results have confirmed the switching from softening to stiffening of the compressive response of the tested samples, in relation to the prism aspect ratio, the magnitude of the applied prestress, and the rigidity of the terminal bases.

The outcomes of the present study significantly enlarge the known spectrum of behavior of tensegrity prisms under axial loading, as compared to the literature to date \citep{Oppenheim:2000, FSD12}, and pave the way to the fabrication of innovative periodic lattices and phononic crystals featuring extremal (softening/stiffening) responses.
It has been shown in \cite{FSD12} that 1D lattices of hard tensegrity prisms support extremely compact solitary waves. The `atomic scale localization'  of such  waves \citep{Friesecke:2002} may lead to create acoustic lenses capable of focusing pressure waves in very compact regions in space; to target tumors in hyperthermia applications; and to manufacture sensors/actuators for the nondestructive evaluation and monitoring of materials and structures \citep{Spadoni:2010, TensPatent2013}.
On the other hand, 
soft tensegrity lattices can be used to design acoustic metamaterials supporting special rarefaction waves, and innovative 
shock absorption devices \citep{HN12,Herbold:2013}.
Particularly challenging is the topology optimization of 3D tensegrity lattices showing soft and hard units (cf. the topologies shown in Fig. \ref{metamaterials}, which are obtained by stacking layers of tensegrity plates designed as in \cite{Skelton2010}), with the aim of designing anisotropic  systems featuring exceptional directional and band-gap properties (refer, e.g.., to \cite{Ruzzene:2005, FPD08, Porter:2009, DaraioPRL10, Ngo12, Leonard:2013, Manktelow:2013,Casadei:2013} and the references therein).
The results of the present study highlight that the self-stresses of the basic units are peculiar design variables of tensegrity metamaterials, which can be finely tuned in order to switch the local response from softening to stiffening, according to given anisotropy patterns.
Additional future extensions of the present study might involve the design of locally resonant materials incorporating tensegrity concepts, and the manufacture of tensegrity microstructures through Projection MicroStereoLithography \citep{Howon1, Howon2}, using swelling materials to create suitable self-stress states.
}}

\begin{figure}[hbt]
\unitlength1cm
\begin{picture}(9,12)
\if\Images y\put(-0.25,-0.5){\psfig{figure=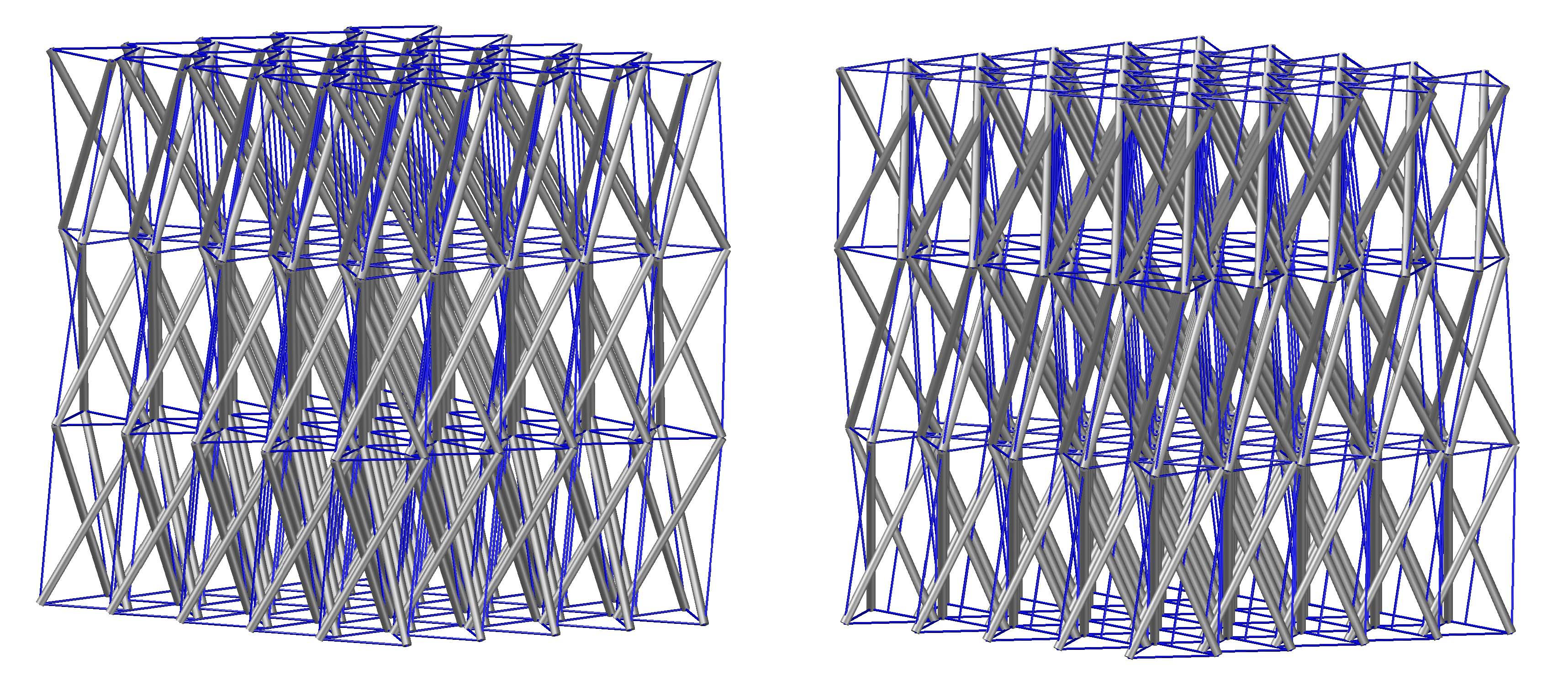,width=15cm}}\fi
\if\Images y\put(-0.25,5.5){\psfig{figure=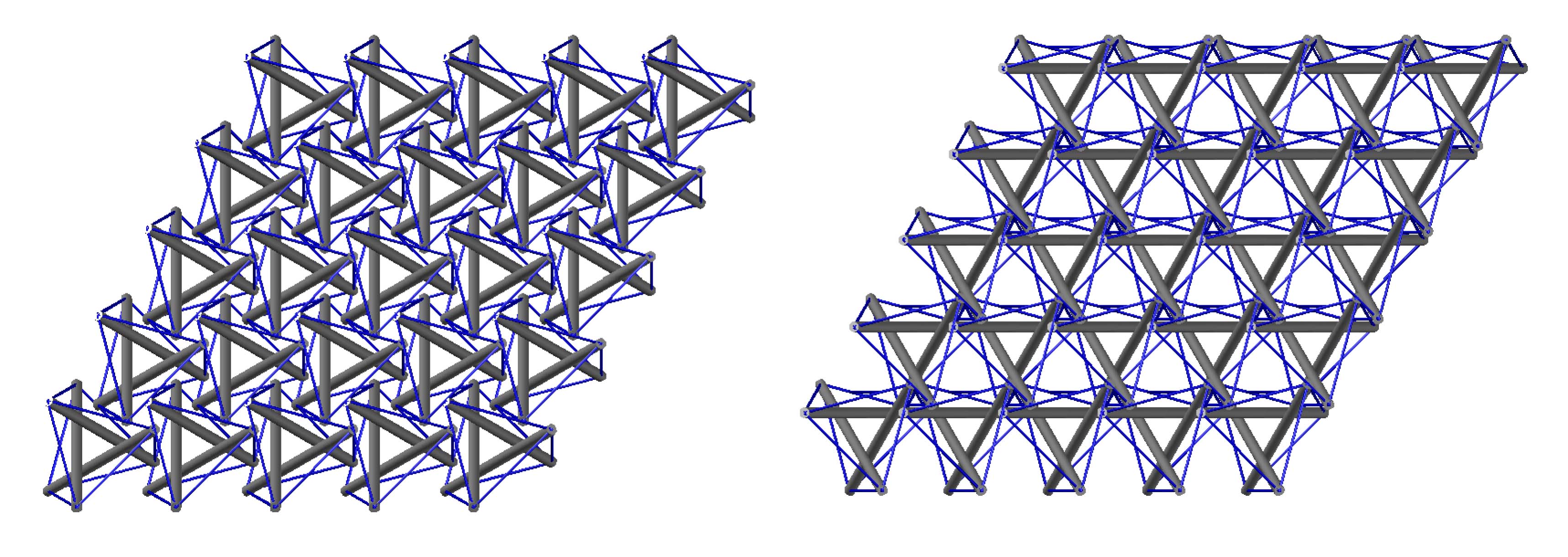,width=15cm}}\fi
\end{picture}
\caption{Different topologies of  3D tensegrity lattices obtained by stacking layers of tensegrity plates. Top: top views. Bottom: 3D views} 
\label{metamaterials}
\end{figure}
\medskip

\section*{Acknowledgements}
Support for this work was received from the
Italian Ministry of Foreign Affairs, Grant No. 00173/2014, Italy-USA  Scientific and Technological Cooperation 2014-2015
(`\textit{Lavoro realizzato con il contributo del Ministero degli Affari Esteri, Direzione Generale per la Promozione del Sistema Paese}').
The authors would like
to thank  Robert Skelton and Mauricio de Oliveira (University of California, San Diego) for many useful discussions and suggestions,
and Angelo Esposito (Department of Civil Engineering, University of Salerno) for 
his precious assistance in the preparation of the numerical simulations.

\newpage 

\section*{Appendix. Axial stiffness of a minimal regular tensegrity prism}\label{Appendix}
Let us examine the matrix $\bV$ introduced in Sect. \ref{fullelresp}. It is not difficult to verify that the entries of such a matrix have the following analytic expressions

\bea
V_{11} & = &\frac{1}{6} \left\{4 k_3 \sin ^2\left(\frac{\varphi }{2}\right) \left(\sqrt{3}-\frac{9 h^2 b_N}{\left(3 h^2-2 \ell ^2 \cos (\varphi )+2 \ell ^2\right)^{3/2}}\right)+\Bigl(k_1 \right.\Bigr.
\nonumber \\ 
& &   
\times\left(\sqrt{3 h^2-\sqrt{3} \ell ^2 \sin (\varphi )+\ell ^2 \cos (\varphi )+2 \ell ^2} \left(-9 h^2 \sin (\varphi )+\sqrt{3} \left(3 h^2+4 \ell ^2\right)  \right.\right.
\nonumber \\ 
& &   
\left.\times\cos (\varphi)+6 \sqrt{3} h^2-12 \ell ^2 \sin (\varphi )-3 \ell ^2 \sin (2 \varphi )-\sqrt{3} \ell ^2 \cos (2 \varphi )+6 \sqrt{3} \ell ^2\right)
\nonumber \\ 
& &   
\Bigl.\left.-9 h^2 s_N (-\sqrt{3} \sin (\varphi)+\cos (\varphi )+2)\right)\Bigr) /  \left(\left(3 h^2-\sqrt{3} \ell ^2 \sin (\varphi )+\ell ^2 \cos (\varphi )\right.\right.
\nonumber \\ 
& &   
\left.\left.\left.+2 \ell ^2\right)^{3/2}\right)+6 \sqrt{3} k_2 \right\} 
 \label{V11}
\eea

\bea
V_{12} & = & \frac{1}{6} \ell  \left(\frac{2 k_3 \sin (\varphi ) \left(\sqrt{3} \left(3 h^2-2 \ell ^2 \cos (\varphi )+2 \ell ^2\right)^{3/2}-3 b_N \left(3 h^2-\ell ^2 \cos (\varphi )+\ell^2\right)\right)}{\left(3 h^2-2 \ell ^2 \cos (\varphi )+2 \ell ^2\right)^{3/2}}   \right. 
\nonumber \\ 
& &   
 -\left(k_1 \left(2 \sqrt{3 h^2-\sqrt{3} \ell ^2 \sin (\varphi )+\ell ^2 \cos (\varphi )+2 \ell^2} \left(\left(9 h^2+6 \ell ^2\right) \cos (\varphi )+\sqrt{3}\right.\right.\right.
\nonumber \\ 
& &    
\left.\times
\sin(\varphi)\left(3 h^2-2 \ell ^2 \cos (\varphi)+2 \ell ^2\right)+3 \ell ^2 \cos (2 \varphi )\right)-3 s_N \left(2 \sqrt{3} \left(3 h^2+\ell ^2\right) \right.
\nonumber \\ 
& & 
\left.\left.\left.\times\cos (\varphi ) +2 \sin (\varphi )\left(3 h^2-\ell ^2 \cos (\varphi )+\ell ^2\right)+\sqrt{3} \ell ^2 \cos (2 \varphi)\right)\right)\right) /  \Bigl(2 \Bigl(3 h^2 \Bigr.\Bigr.
\nonumber \\ 
& & 
\left.\Bigl.\Bigl.-\sqrt{3}\ell ^2 \sin (\varphi )+\ell ^2 \cos (\varphi)+2 \ell ^2\Bigr)^{3/2}\Bigr)\right)
\label{V12}
\eea

\bea
V_{13} & = & \frac{3}{2} h \ell  \left(\frac{4 k_3 b_N \sin ^2\left(\frac{\varphi }{2}\right)}{\left(3 h^2-\ell ^2\left(2\cos (\varphi )+2\right)\right)^{3/2}}\right.
\nonumber \\ 
& & 
\left. +\frac{k_1 s_N \left(-\sqrt{3}
   \sin (\varphi )+\cos (\varphi )+2\right)}{\left(3 h^2-\ell ^2\left(\sqrt{3}\sin (\varphi )+ \cos (\varphi )+2 \right)\right)^{3/2}}\right)
\label{V13}
\eea

\bea
V_{21} & = & \frac{1}{6} \left(2 k_3 \sin (\varphi ) \left(\frac{9 h^2 b_N}{\left(3 h^2-2 \ell ^2 \cos (\varphi )+2 \ell ^2\right)^{3/2}}-\sqrt{3}\right)+\left(k_1 \right.\right.
\nonumber \\ 
& & 
\times\left(\sqrt{3 h^2-\sqrt{3}\ell ^2 \sin (\varphi )+\ell ^2 \cos (\varphi )+2 \ell ^2} \left(\left(9 h^2+6 \ell ^2\right) \cos (\varphi )+\sqrt{3} \sin (\varphi ) \right.\right.
\nonumber \\ 
& & 
\left.\left.\left.
\times \left(3 h^2 -2 \ell ^2 \cos (\varphi )+2\ell ^2\right)+3 \ell ^2 \cos (2 \varphi )\right)-9 h^2 s_N \left(\sin (\varphi )+\sqrt{3} \cos (\varphi )\right)\right)\right)
\nonumber \\ 
& & 
\left.
 / \left(\left(3 h^2-\sqrt{3} \ell ^2 \sin (\varphi )+\ell ^2\cos (\varphi )+2 \ell ^2\right)^{3/2}\right)\right)
\label{V21}
\eea

\bea
V_{22} & = & \frac{1}{6} \ell  \left(2 k_3 \left(-\frac{3 b_N \left(\ell ^2 (\cos (2 \varphi )+3)-2 \left(3 h^2+2 \ell ^2\right) \cos (\varphi )\right)}{2 \left(3 h^2-2 \ell ^2 \cos (\varphi
   )+2 \ell ^2\right)^{3/2}}-\sqrt{3} \cos (\varphi )\right)\right.
\nonumber \\ 
& & 
+\left(k_1 \left(3 s_N \left(6 \sqrt{3} h^2 \sin (\varphi )-2 \left(3 h^2+2 \ell ^2\right) \cos (\varphi )+4 \sqrt{3}
   \ell ^2 \sin (\varphi )+\sqrt{3} \ell ^2\right. \right.\right.
\nonumber \\ 
& & 
\left.
\times \sin (2 \varphi )+\ell ^2 \cos (2 \varphi )-6 \ell ^2\right)+2 \sqrt{3 h^2-\sqrt{3} \ell ^2 \sin (\varphi )+\ell ^2 \cos (\varphi )+2 \ell ^2} 
\nonumber \\ 
& &
\times\left(-9 h^2 \sin (\varphi )+\sqrt{3} \left(3 h^2+2 \ell ^2\right) \cos (\varphi )-6 \ell ^2 \sin (\varphi )-3 \ell ^2 \sin (2 \varphi )-\sqrt{3} \ell ^2 \right.
\nonumber \\ 
& &
\left.\left.\left.\left.
 \times\cos (2
   \varphi )+2 \sqrt{3} \ell ^2\right)\right)\right) / \left(2 \left(3 h^2-\sqrt{3} \ell ^2 \sin (\varphi )+\ell ^2 \cos (\varphi )+2 \ell ^2\right)^{3/2}\right)\right)
\label{V22}
\eea

\bea
V_{23} & = & \frac{3}{2} h \ell  \left(\frac{k_1 s_N \left(\sin (\varphi )+\sqrt{3} \cos (\varphi )\right)}{\left(3 h^2-\sqrt{3} \ell ^2 \sin (\varphi )+\ell ^2 \cos (\varphi )+2 \ell
   ^2\right)^{3/2}}\right.
\nonumber \\ 
& &
\left.
-\frac{2 k_3 b_N \sin (\varphi )}{\left(3 h^2-2 \ell ^2 \cos (\varphi )+2 \ell ^2\right)^{3/2}}\right)
\label{V23}
\eea

\bea
V_{31} & = & \frac{1}{2} \sqrt{3} h \left(-\frac{8 k_3 \ell  b_N \sin ^2\left(\frac{\varphi }{2}\right)}{\left(3 h^2-2 \ell ^2 \cos (\varphi )+2 \ell ^2\right)^{3/2}}
\right.
\nonumber \\ 
& &
\left.
-\frac{2 k_1 \ell  s_N
   \left(-\sqrt{3} \sin (\varphi )+\cos (\varphi )+2\right)}{\left(3 h^2-\sqrt{3} \ell ^2 \sin (\varphi )+\ell ^2 \cos (\varphi )+2 \ell ^2\right)^{3/2}}\right)
\label{V31}
\eea

\bea
V_{32} & = & \frac{1}{2} \sqrt{3} h \ell ^2 \left(\frac{k_1 s_N \left(\sin (\varphi )+\sqrt{3} \cos (\varphi )\right)}{\left(3 h^2-\sqrt{3} \ell ^2 \sin (\varphi )+\ell ^2 \cos (\varphi )+2
   \ell ^2\right)^{3/2}}
\right.
\nonumber \\ 
& &
\left.
-\frac{2 k_3 b_N \sin (\varphi )}{\left(3 h^2-2 \ell ^2 \cos (\varphi )+2 \ell ^2\right)^{3/2}}\right)
\label{V32}
\eea

\bea
V_{33} & = & k_3 \left(\frac{4 \sqrt{3} \ell ^2 b_N \sin ^2\left(\frac{\varphi }{2}\right)}{\left(3 h^2-2 \ell ^2 \cos (\varphi )+2 \ell ^2\right)^{3/2}}-1\right)
\nonumber \\ 
& &
+k_1 \left(-\frac{\sqrt{3}
   \ell ^2 s_N \left(\sqrt{3} \sin (\varphi )-\cos (\varphi )-2\right)}{\left(3 h^2-\sqrt{3} \ell ^2 \sin (\varphi )+\ell ^2 \cos (\varphi )+2 \ell ^2\right)^{3/2}}-1\right)
\label{V33}
\eea

By inserting the above results into Eqn. (\ref{Khel}) of Sect. \ref{rateproblem}, we easily obtain the axial stiffness $K_t^{el}$ of the fully-elastic model.
The reference value of such a quantity (for $\ell=\ell_0, \varphi=\varphi_0, h=h_0$) can be written as follows

\bea
K_{h_0}^{el} & = & \frac{p_0}{1+p_0} \left\{ 36 k_1 \eta_0 ^2 \left((3+2 \sqrt{3}+ \sqrt{3} \eta_0 ^2) k_1 k_2 + (-2+\sqrt{3}-\eta_0 ^2) k_1 k_1 \frac{p_0}{1+p_0}\right.\right.  
\nonumber \\ 
& & 
-6 k_2 k_1 \frac{p_0}{1+p_0}+k_3(2 \sqrt{3} k_1+ (-3+2 \sqrt{3}+ \sqrt{3} \eta_0 ^2)k_2  -(2+\sqrt{3}+ \eta_0 ^2)
\nonumber \\ 
& & 
\left. \left. \times k_1 \frac{p_0}{1+p_0})\right) \right\} / \left\{ 6 k_1 \frac{p_0}{1+p_0} \left(\sqrt{3}(1+8 \eta_0^2+2 \eta_0^4)k_2-2\eta_0^4 k_1 \frac{p_0}{1+p_0}\right) \right.
\nonumber \\ 
& & 
\Bigl. +k_1\Bigl(3(2+\sqrt{3}+\eta_0^2)k_2+ (-3+2\sqrt{3}+(-24+13\sqrt{3})\eta_0^2)k_1 \frac{p_0}{1+p_0}\Bigr) \Bigr.  
\nonumber \\ 
& & 
\left. +k_3\Bigl(6k_1+3(2-\sqrt{3}+\eta_0^2)k_2+(3+2\sqrt{3}+(24+13\sqrt{3})\eta_0^2)k_1 \frac{p_0}{1+p_0}\Bigr)\right\}  
\label{Kdb}
\eea

\noindent where:

\bea
\eta_0 = \ \frac{h_{0}}{a_{0}}
\label{rho}
\eea

\noindent For what concerns the rigid-elastic model presented in Sect. \ref{rigelresp}, we easily obtain

% NUOVA FORMULA

\bea
K_{t}^{rigel}(h) & = & \frac{1}{2 a^4 (2 \sqrt{3} a^2 \cos (\varphi +\frac{\pi
   }{6})+b^2)^{3/2}} \biggl\{3 k_1 \csc ^3(\varphi ) \left[\sqrt{3} \sin (\varphi +\frac{\pi }{6}) \left(a^4 \bigr. \right. \right.
\nonumber\\
& &
\left. \times (\cos (2 \varphi)+3) + 2 a^2 (b^2-2 a^2) \cos (\varphi )\right) (2 \sqrt{3} a^2 \cos (\varphi +\frac{\pi
   }{6})+b^2)
\nonumber\\
& & \times \Bigl( \sqrt{2 \sqrt{3} a^2 \cos (\varphi +\frac{\pi }{6})+b^2}-s_N \Bigr) - 2a^2 \sin (\varphi ) (2 a^2 \cos (\varphi )-2 a^2
\nonumber\\
& & 
+b^2)\biggl(\cos (\varphi +\frac{\pi }{6})(6 a^2 \cos (\varphi +\frac{\pi }{6})+\sqrt{3} b^2) \sqrt{2 \sqrt{3} a^2 \cos (\varphi +\frac{\pi }{6})+b^2} \biggr.
\nonumber\\
& &
\left. \biggl. \biggl. -s_N \left(3 a^2 \sin ^2(\varphi +\frac{\pi }{6})+6 a^2 \cos
   ^2(\varphi +\frac{\pi }{6})+\sqrt{3} b^2 \cos (\varphi +\frac{\pi
   }{6}) \right) \biggr)\right] \biggr\} \label{Krigel}
\eea

\medskip

\section*{References}

\bibliographystyle{apalike}

\end{document}